\documentclass[sigconf]{acmart}

\AtBeginDocument{%
  \providecommand\BibTeX{{%
    \normalfont B\kern-0.5em{\scshape i\kern-0.25em b}\kern-0.8em\TeX}}}

\copyrightyear{2024}
\acmYear{2024}
\setcopyright{rightsretained}
\acmConference[CHI '24]{Proceedings of the CHI Conference on Human Factors
in Computing Systems}{May 11--16, 2024}{Honolulu, HI, USA}
\acmBooktitle{Proceedings of the CHI Conference on Human Factors in
Computing Systems (CHI '24), May 11--16, 2024, Honolulu, HI, USA}
\acmDOI{10.1145/3613904.3642109}
\acmISBN{979-8-4007-0330-0/24/05}

\acmSubmissionID{3200}

\usepackage{xspace}
\usepackage{amsmath}
\usepackage{wrapfig}
\usepackage{enumitem}
\usepackage[noabbrev,capitalise,nameinlink]{cleveref}
\usepackage[utf8]{inputenc}

\usepackage{array}

\usepackage{tikz}

\usepackage{soul}

\usepackage{framed}

\usepackage{acmart-taps}

\definecolor{RoyalBlue}{HTML}{0071BC}

\definecolor{systemPurple}{HTML}{00BDB4}
\definecolor{systemBlue}{HTML}{007aff}
\definecolor{systemBlueTinted}{HTML}{99CAFF}
\definecolor{systemRed}{HTML}{ff3b30}

\usepackage{xspace}
\newcommand{\ie}{{i.e.,}\xspace}
\newcommand{\eg}{{e.g.,}\xspace}

\newcommand{\numParticipants}{{30}\xspace}

\newcommand{\participantsMLXP}{{7.1}\xspace}
\newcommand{\participantsEMLXP}{{4.1}\xspace}

\newcommand{\eml}{{efficient machine learning}\xspace}

\newcommand{\location}{{Apple}\xspace}

\newdimen\tempdim
\newcommand*{\ChartBox}[2]{%
  \begingroup
    \settoheight{\tempdim}{L}%
    \edef\tempheight{\the\tempdim}%
    \settodepth{\tempdim}{g}%
    \edef\tempdepth{\the\tempdim}%
    \tikz[
      baseline=0pt,
      inner sep=0pt,
    ]
    \node[
      fill={#2},
      rounded corners=1.5pt,
      anchor=base,
    ]{%
      \vphantom{g\"A}%
      \pgfmathsetlength{\tempdim}{#1}%
      \kern\tempdim\relax
    };%
  \endgroup
}

\newcommand{\tableSpaceAmount}{0mm}

\newcommand*{\chart}[2]{\ChartBox{3mm*#1}{#2}}

\newcommand{\pFifteen}{{E1}}
\newcommand{\pSixteen}{{E2}}
\newcommand{\pSix}{{E3}}
\newcommand{\pFourteen}{{E4}}
\newcommand{\pNine}{{E5}}
\newcommand{\pZero}{{E6}}
\newcommand{\pTwelve}{{E7}}
\newcommand{\pSeven}{{E8}}
\newcommand{\pTwo}{{E9}}
\newcommand{\pThirteen}{{E10}}
\newcommand{\pTen}{{E11}}
\newcommand{\pNineteen}{{E12}}
\newcommand{\pEleven}{{E13}}
\newcommand{\pTwentyeight}{{P1}}
\newcommand{\pTwentythree}{{P2}}
\newcommand{\pEight}{{P3}}

\newcommand{\pThirty}{{P5}}
\newcommand{\pTwentyone}{{P6}}
\newcommand{\pTwentytwo}{{P7}}
\newcommand{\pOne}{{P8}}
\newcommand{\pSeventeen}{{P9}}
\newcommand{\pTwentysix}{{P10}}
\newcommand{\pTwenty}{{P11}}
\newcommand{\pFive}{{T1}}
\newcommand{\pTwentyfive}{{T2}}
\newcommand{\pThree}{{T3}}
\newcommand{\pTwentynine}{{T4}}
\newcommand{\pFour}{{T5}}
\newcommand{\pTwentyseven}{{T6}}

\usepackage{graphicx,calc}

\newcommand*\inlinesfsymbol[1]{
    \raisebox{-0.5em}{\includegraphics[height=1.6em]{#1}}
}
\newcommand{\symbollowpower}{\inlinesfsymbol{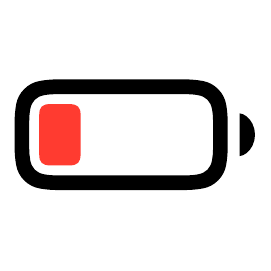}}
\newcommand{\symbolpower}{\inlinesfsymbol{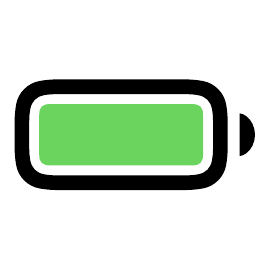}}
\newcommand{\symbolheat}{\inlinesfsymbol{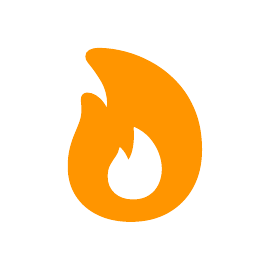}}
\newcommand{\symbolfps}{\inlinesfsymbol{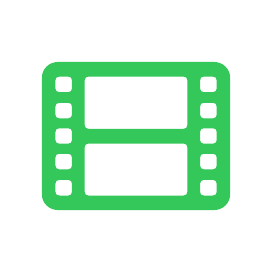}}
\newcommand{\symbollatency}{\inlinesfsymbol{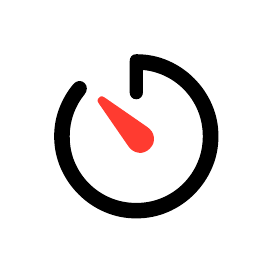}}
\newcommand{\symbolstorage}{\inlinesfsymbol{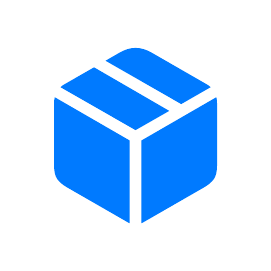}}
\newcommand{\symbolsleep}{\inlinesfsymbol{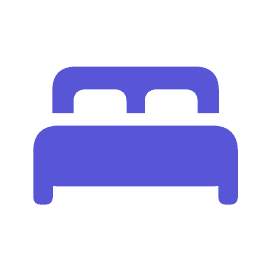}}

\newlength{\ParboxWidth}
\begin{document}

\title[Model Compression in Practice]{Model Compression in Practice: Lessons Learned from Practitioners Creating On-device Machine Learning Experiences}

\settopmatter{authorsperrow=4}

\author{Fred Hohman}
\affiliation{%
  \institution{Apple}
  \city{Seattle}
  \state{WA}
  \country{USA}
}
\email{fredhohman@apple.com}

\author{Mary Beth Kery}
\affiliation{%
  \institution{Apple}
  \city{Pittsburgh}
  \state{PA}
  \country{USA}
}
\email{mkery@apple.com}

\author{Donghao Ren}
\affiliation{%
  \institution{Apple}
  \city{Seattle}
  \state{WA}
  \country{USA}
}
\email{donghao@apple.com}

\author{Dominik Moritz}
\affiliation{%
  \institution{Apple}
  \city{Pittsburgh}
  \state{PA}
  \country{USA}
}
\email{domoritz@apple.com}

\begin{abstract}
On-device machine learning (ML) promises to improve the privacy, responsiveness, and proliferation of new, intelligent user experiences by moving ML computation onto everyday personal devices.
However, today's large ML models must be drastically compressed to run efficiently on-device, a hurtle that requires deep, yet currently niche expertise.
To engage the broader human-centered ML community in on-device ML experiences, we present the results from an interview study with \numParticipants experts at \location that specialize in producing efficient models.
We compile tacit knowledge that experts have developed through practical experience with model compression across different hardware platforms. 
Our findings offer pragmatic considerations missing from prior work, covering the design process, trade-offs, and technical strategies that go into creating efficient models.
Finally, we distill design recommendations for tooling to help ease the difficulty of this work and bring on-device ML into to more widespread practice.

\end{abstract}



\maketitle

\section{Introduction}
\label{sec:introduction}

Most modern machine learning (ML) models in production today occupy cloud servers of exceedingly more computational capacity and power than the device in your pocket.
Yet for delivering intelligent user experiences, there is good reason to move ML models onto personal computing devices people use every day.

\textit{On-device ML} is the practice of storing, training, and running ML models on an individual's device, such as a smartphone, tablet, or wearable.
However, as today's state-of-the-art models grow larger and larger in size (\eg into the billions of parameters~\cite{villalobos2022machine, owid2022artificialintelligence, zhao2023survey, stanford2023ai}), \textit{efficiency} remains the biggest barrier to on-device ML~\cite{coremltools,google2022why}.
An arbitrary ML model placed on a mobile device can easily consume every available resource of the device, whether it be compute, memory, or battery.
Creating efficient, on-device models brings new challenges to the ML development process.

In this paper, we argue that research in \textit{\eml} over the past decade has matured to the degree to where practitioners have an (albeit rapidly evolving) set of techniques to make on-device ML a reality.
The key idea is to shrink, optimize, and compress models, while simultaneously maintaining their accuracy.
To achieve this, practitioners develop strategies for how to best apply \textit{model compression} techniques to minimize the amount of computational resources needed.
Product designers, tool-builders, and ML practitioners today should be considering the benefits of on-device ML for their users:

First, on-device ML can be an enormous win for personal privacy.
While standard practice today is to send a user's (encrypted) data over a network to servers for ML inference, on-device ML cuts out this dependency such that a user's personal data never leaves their device.
Beyond inference, pre-trained models can be fine-tuned on-device to adapt to an individual user's preferences, while keeping those preferences private and local~\cite{konevcny2016federated, hu2020personalized, apple2023personalizing}.

Second, models on-device enable intelligent user experiences where they would not be possible otherwise.
For example, computational photography models within mobile cameras run inference at high frame rates (\eg 30 frames per second), which would not be possible if relying on the availability of a distant server.

Third, by going offline, on-device ML can create not only faster but also more portable experiences.
Since no data is sent over a network, users can still interact with ML-powered features in situations without internet access or cellular service.
This has the potential to broaden access of AI/ML features in more geographic regions, including rural and remote areas~\cite{tomavsev2020ai}.
By forgoing a network entirely, ML-powered features can run faster and remain responsive by alleviating network latency in an ``offline'' mode~\cite{google2022why}.

Lastly, removing the reliance on servers for ML inference has economic and environmental impact.
On-device ML cuts out the cost of a server, which may help individuals, non-profits, or smaller tech firms that previously could not afford server upkeep to now deliver ML-powered features to their users.
By reducing society's reliance on external servers, on-device ML may help reduce the carbon footprint of the cloud~\cite{monserrate2022cloud}.

In this paper, we seek to advocate for on-device ML by filling a crucial gap:
while efficient ML research and algorithmic advances have progressed tremendously~\cite{menghani2023efficient, zhao2022survey, gu2021server, zamzam2019resource, dhar2021survey,sehgal2019guidelines}, 
there currently exists little pragmatic guidance~\cite{nvidia2023docs} in literature, online, in books, or otherwise, for people wanting to create on-device ML user experiences.
On-device ML requires both clever algorithmic and user experience design, making this a fundamentally interdisciplinary problem that has received limited attention outside of ML venues.
Our work addresses:
\textbf{RQ: How should a broader audience of HCI and ML practitioners today optimize powerful models to design on-device, ML user experiences?}

To curate pragmatic guidance, we interviewed \numParticipants expert industry practitioners who are uniquely experienced with designing, developing, and deploying on-device ML at scale.
We capture the hard-won knowledge, insights, and tricks-of-the-trade that are traditionally omitted from technical ML publications. %
We draw connections between the design of ML user experiences and the compression strategies experts use to match design goals.
We contribute:

\begin{itemize}
\setlength\itemsep{0.5em}

    \item \textbf{Tacit knowledge that \numParticipants expert practitioners have developed} around the design processes, trade-offs, and strategies that go into deploying efficient on-device ML.
    Many decisions around optimization and compression strategy stem directly from user experience and product design.
    
    \item \textbf{A characterization of the key challenges practitioners face} when creating efficient models.
    Examples include the tension between optimizing performance against accuracy, and the necessity to work with hardware-level details.
    
    \item \textbf{Distilled design recommendations for interactive interfaces, tooling, and systems} that show promise to help practitioners optimize models and ultimately proliferate on-device ML experiences.
    
\end{itemize}

We conclude by discussing where the broader HCI + AI/ML research, design, and practitioner communities can engage in efficient, on-device ML.
We hope the results from this deep dive into a nascent area of ML user experience design helps spotlight its interdisciplinary importance and inspires others to contribute.

\section{Related Work in Human-Centered ML and Model Optimization}
\label{sec:related-work}
Our work attempts to bridge the Machine Learning, Ubiquitous Computing, Mobile Computing, and Human--Computer Interaction (HCI) communities, to build common language around an ML topic with broad, intersectional impact.
Our biggest challenge is in situating this research where no direct prior work exists.
In this section, we anchor our work in two main threads of HCI research.
First, overlapping with the HCI community, the Ubicomp and Mobile computing communities have long used model compression for on-device ML to make new user experiences possible~\cite{dhar2019device}.
Our work complements this literature by consolidating together strategies across many use cases, with the aim of broadening the audience of practitioners who can contribute to creating on-device ML systems (\cref{subsec:edge-computing}).
Second, our research continues a line of HCI work synthesizing lessons from real-world ML practice~\cite{amershi2019software, amershi2019guidelines, holstein2019improving, sambasivan2021everyone, liao2020questioning, bhatt2020explainable, yang2018investigating}, and we extend prior work into a previously unconsidered area of ML (\cref{subsec:hci}).
We situate our work in both of these directions, and end by covering related work on model efficiency for on-device ML itself (\cref{subsec:ml}).

\subsection{Ubiquitous Computing: Deploying ML on Mobile \& Edge Devices}
\label{subsec:edge-computing}
Challenges of on-device ML often occur in the context of mobile or edge computing~\cite{zhou2019edge, murshed2021machine, chen2019deep}.
Edge computing refers to small embedded hardware, wearables, or internet-of-things (IoT) devices where at least some computation is performed on-device (the ``edges'' of a network) rather than by a central server~\cite{murshed2021machine}. 
Example uses of compressed neural networks include facial recognition on mobile devices~\cite{zhen2021fast}, activity recognition on cameras~\cite{mishra2020teacher}, gesture recognition on smartwatches~\cite{xu2022enabling}, and respiratory monitoring on phones and smartwatches~\cite{chauhan2018performance, dai2021respwatch, liaqat2019wearbreathing}.
Since efficiency is critical for edge devices, the Ubiquitous computing, Mobile computing, and IoT research communities have contributed model compression advances around dynamic models~\cite{mishra2023designing, liu2021adaspring, wang2023genie}, structured sparsity~\cite{liberis2023differentiable}, and on-device training~\cite{yao2021context, jiang2021lightweight, li2023hierarchical}.
Our work differs from prior literature by not focusing on any single application or technique, and instead synthesizing practitioner strategies for model compression that can be used \textit{across} use cases and hardware types.
In our research, we interviewed ML practitioners deploying to a variety of consumer devices, including edge hardware.
We distinguish between device type only where advice differs for small embedded hardware, for example in \cref{subsubsec:section-edge-issues}.

\subsection{Human-Computer Interaction: Studying AI/ML Practitioners}
\label{subsec:hci}
Over the past decade, interview studies from HCI have played an important role in giving the public access to learn from AI/ML work that otherwise happens behind closed doors~\cite{amershi2019software, amershi2019guidelines, holstein2019improving, sambasivan2021everyone, liao2020questioning, bhatt2020explainable, yang2018investigating}. 
Interviews with practitioners are uniquely suited to capture the kinds of process-oriented insight, stories, and tricks-of-the-trade that are traditionally omitted from technical ML publications.
For instance, \citet{amershi2019software} interview and survey AI/ML practitioners at Microsoft to gather best practices for production ML development.
Researchers have studied how AI/ML teams collaborate across diverse technical roles~\cite{passi2018trust, zhang2020data, piorkowski2021ai}, or adopt new technology like AutoML~\cite{wang2019human, wang2021much, wang2021autods}.
\citet{sambasivan2021everyone} used interviews to profile organizational struggles between balancing modeling work with data quality work. 
\citet{holstein2019improving} found key mismatches between academic AI/ML fairness concepts and the kinds of real fairness issues product teams grapple with.
Many other aspects of AI fairness, accountability, transparency, and ethics have been found to be practitioner-driven, where design decisions and process have enormous impact~\cite{hong2020human, madaio2020co, modarres2018towards, delgado2021stakeholder, madaio2022assessing, robertson2021modeling, rakova2021responsible, hopkins2021machine}.
Like this prior work, we examine the production and practice side of on-device ML efficiency to illustrate the connection between ML compression choices and their impact on holistic ML user-experience design.

\subsubsection{User Experience Design for AI/ML}
\label{subsec:related-work-design}
Much of what the research community knows today about designing effective user experiences (UX) for AI/ML comes from interviews with seasoned product designers~\cite{yang2018investigating, zdanowska2022study, windl2022not, yildirim2022experienced}. 
Leading technology companies have distilled design wisdom from their own product teams into AI/ML design guidelines as public, educational resources~\cite{wright2020comparative}.
Examples include Apple's Human Interface Guidelines on Machine Learning~\cite{guidelines2019apple}, Microsoft's Human-AI Guidelines~\cite{amershi2019guidelines}, Google's People + AI Guidebook~\cite{guidebook2019google}, and IBM's Design for AI resources~\cite{design2019ibm}.
This body of work emphasizes that designing with machine learning requires new approaches from designers~\cite{yang2018machine, amershi2019guidelines, windl2022not, yildirim2022experienced, guidebook2019google, design2019ibm}.
Our paper contributes to this conversation by illustrating how power and performance issues of moving ML models on-device create real, tangible UX constraints for designers (for example, see Table \ref{tab:metrics}), and offers concrete design strategies for creating effective user experiences around those constraints (for example, see \cref{subsec:budget}).

\subsection{Machine Learning: On-Device ML \& Compression Techniques}
\label{subsec:ml}
The literature on \eml is broader than the scope of this paper, and for a comprehensive look we refer readers to the excellent survey papers cited here and below~\cite{menghani2023efficient, deng2020model, choudhary2020comprehensive, cheng2018model, treviso2023efficient}.
For neural networks, \citet{menghani2023efficient} breaks efficiency into 5 areas: (1) compression techniques, (2) learning techniques, (3) automation, (4) efficient architectures, and (5) infrastructure \& hardware.
Our interviews with practitioners contained more discussion of (1) and (5), and some discussion of (2) and (4).
We do not claim that this interview study is a comprehensive look at efficient machine learning.
Nonetheless, our work adds a fresh perspective to existing efficient ML literature: instead of detailing specific optimization techniques, here we profile higher-level strategies for how practitioners put techniques into practice to enable user-experience goals.

\subsubsection{On-device Inference vs. On-device Training}
\label{subsec:inference-v-training}
It is worthwhile to distinguish between on-device \textit{inference} and on-device \textit{training}.
On-device \textit{inference} refers to running an ML model on new input to get a prediction.
These models are typically pretrained on a server, then delivered to a device. %
On-device \textit{training} refers to either training a model from scratch or fine-tuning a model on a user's device.
While work in on-device ML encompasses both paradigms, in our paper we focus on the (currently) more common use case of on-device inference and leave on-device training for future work.
Training on-device is generally considered harder, since training usually requires far more resources~\cite{zhou2019edge}.
For a survey on the current challenges around on-device learning, see:~\cite{dhar2021survey, lim2020federated, zhou2019edge, murshed2021machine}.

\subsubsection{On-device Large Language Models and Foundation Models}
\label{subsubsec:llms}
In recent years, large language models (LLMs) and other generalized foundation models have raised the magnitude of size we expect from ML models~\cite{solaiman2023evaluating,movva2023large, owid2022artificialintelligence,villalobos2022machine}.
As models get bigger, so do the stakes for model compression and on-device efficiency~\cite{bommasani2021opportunities}.
Beyond speeding up inference~\cite{alizadeh2023llm}, efficiency methods for LLMs also aim to lower their enormous pretraining and fine-tuning costs.
Examples include low-rank adaption methods (\eg LRPD~\cite{zhao2016low} and LoRA~\cite{hu2021lora}), and parameter-efficient fine-tuning methods~\cite{ding2023parameter, ding2022delta, fu2023effectiveness, lialin2023scaling}.
The number of empirical works for compressing LLMs from 2023 alone would be difficult to capture, therefore we refer interested readers to the following surveys:~\cite{treviso2023efficient,zhou2023comprehensive,xu2023survey}.
We note that many of the techniques for compression that we discuss in~\cref{sec:background} are the same key ideas being applied to LLMs and foundation models.

\subsubsection{Publicly Available Compression Resources}
\label{subsec:existing-resources}

While most compression techniques originate in academic work~\cite{deng2020model, choudhary2020comprehensive, cheng2018model}, many resources for practitioners can be found within web tutorials and ML toolkits.
Examples include TensorFlow's quantization-aware training method~\cite{tf2020quantization,tf2018introducing}; PyTorch's experimental support for quantization~\cite{pytorch2023quantization}, sparsity~\cite{pytorch2023sparsity}, and it's accompanying examples~\cite{pytorch2023examples}; Google's quantization extension to Keras called QKeras~\cite{qkeras}; Microsoft's Neural Network Intelligence package and tool~\cite{ms2021nni}; Intel's Neural Compressor library~\cite{intel2020nc}; and Apple's MLX framework~\cite{apple2023mlx} and DNIKit~\cite{welsh2023dnikit}.
Other examples that target specific hardware include speeding up inference on FPGAs~\cite{fahim2021hls4ml} and compressing Core ML models to run on Apple platforms~\cite{coremltools}.
In terms of other community efforts, the appropriately named TinyML community has published a book~\cite{warden2019tinyml} and hosts community events and meetups.

\section{A Primer on Machine Learning Compression Techniques}
\label{sec:background}

\label{subsec:techniques}
To familiarize readers with model optimization, here we give a primer on common ML compression techniques.
This background will help ground the study results and provide context for the remainder of the paper.
\textit{Note we only cover common techniques mentioned by practitioners in our study---more exist.}

Model compression is a class of techniques used in on-device ML to reduce the computational resources a model consumes.
While it is not important to know every detail of every technique, it is useful to understand the variety of techniques at a high-level, and how they can be combined for bigger savings~\cite{han2015deep}.
This overview contains a brief description of each technique, and includes illustrations to build visual intuition.
Note that each technique below is truly a family of techniques, each with many nuanced variations.
For an in-depth review of the technical descriptions of compression techniques, see the following surveys:~\cite{menghani2023efficient, deng2020model, choudhary2020comprehensive, cheng2018model}.

\subsection{Quantization}
\label{subsec:quantization}
Convert the inputs, outputs, weights, and/or activations of a model from high-precision representations (\eg \texttt{fp32}) to lower-precision representations (\eg \texttt{fp16}, \texttt{int32}, \texttt{int16}, \texttt{int8}, and even \texttt{int2}). 
At a high-level, this coarsens a model.
For a detailed survey of quantization-specific techniques and their variations, see ~\cite{gholami2021survey}.

\begin{center}
    \includegraphics[width=0.6\linewidth]{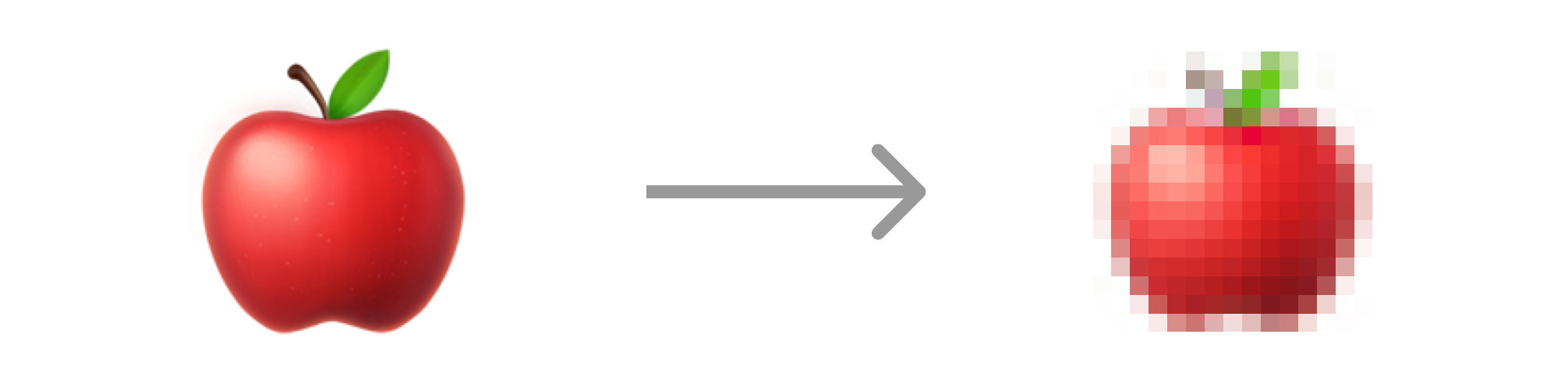}
\end{center}

\subsection{Palettization (Weight Clustering)}
\label{subsec:palletization}
Map the weights of a model to a discrete set of precomputed (or learned) values~\cite{cho2022differentiable, wu2018deep}.
Inspired by an artist's palette, the idea is to map many similar values to one average or approximate value, then use those new values for computing inference.
In this way, palettization is similar in spirit to algorithmic memoization, a dictionary, or a look-up table.
Palettization can make a model smaller but does not make a model faster since it incurs look-up time.

\begin{center}
    \includegraphics[width=0.6\linewidth]{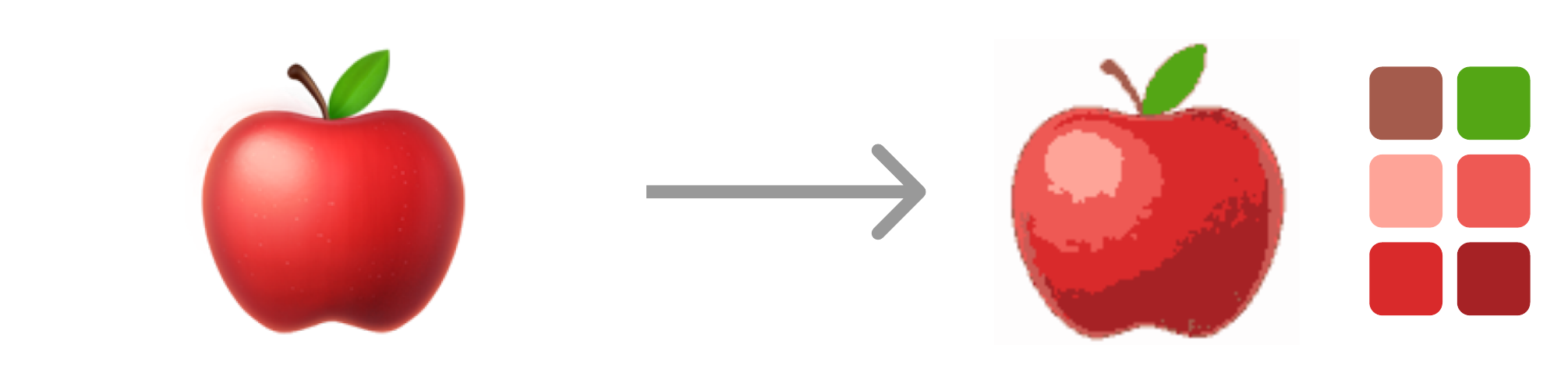}
\end{center}

\subsection{Pruning (Network Sparsity)}
\label{subsubsec:pruning}
Remove the least-important parts of a neural network to make it smaller~\cite{hoefler2021sparsity}.
The motivating idea is that modern neural networks are much more dense and overparameterized than is actually needed.
Since networks can contain billions of parameters, removing some or many parameters may not impact the final accuracy.
Pruning is a large family of techniques~\cite{hoefler2021sparsity}.

\subsubsection{Unstructured Pruning}
A model may have neurons or weights that do not much affect a model's decision~\cite{gholami2021survey}.
In unstructured pruning, the least important neurons or weights can be aggressively removed without affecting model accuracy~\cite{hoefler2021sparsity}.
Unstructured pruning has been shown to shrink a model by 10–100x its original size~\cite{hoefler2021sparsity}. 
The major downside is that this leads to \textit{sparse matrix operations}, which is a type of math that is slow on most hardware.
So although pruned models are small, they may be nearly as slow as the full-sized model~\cite{gholami2021survey}.

\begin{center}
    \includegraphics[width=0.6\linewidth]{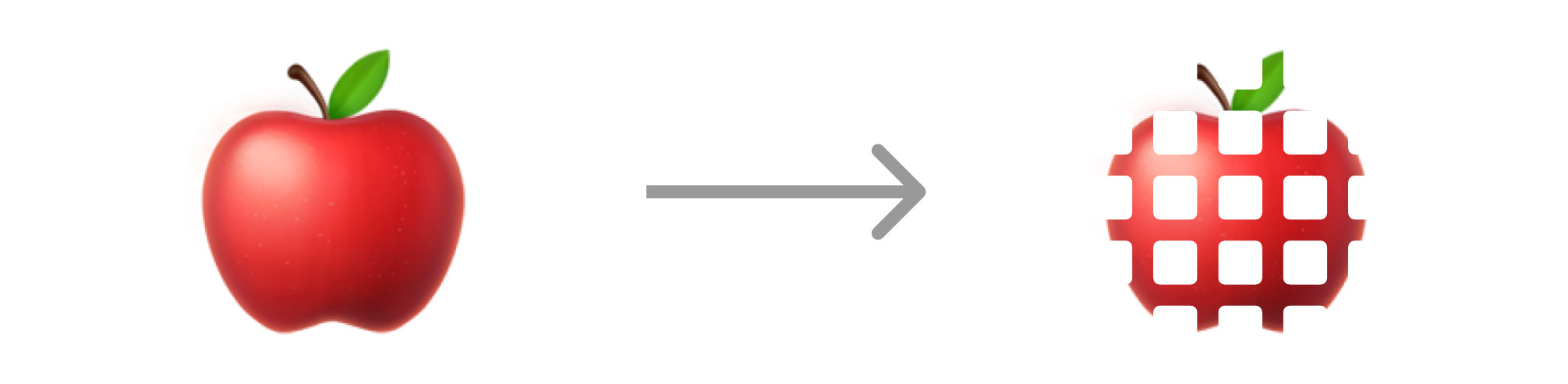}
\end{center}

\subsubsection{Structured Pruning}
Similar to unstructured pruning, the least important neurons or weights are identified, however, instead of removing (or zero-ing out) individual values, structured pruning removes entire structural elements, like channels or filters.
Since structured pruning removes much bigger chunks of a model, less pruning will be possible without serious degradation in model accuracy~\cite{gholami2021survey}.
However, by respecting the structure of the model, these pruned models maintain \textit{dense matrix operations}, which keeps the model fast on most hardware~\cite{hoefler2021sparsity}.

\begin{center}
    \includegraphics[width=0.6\linewidth]{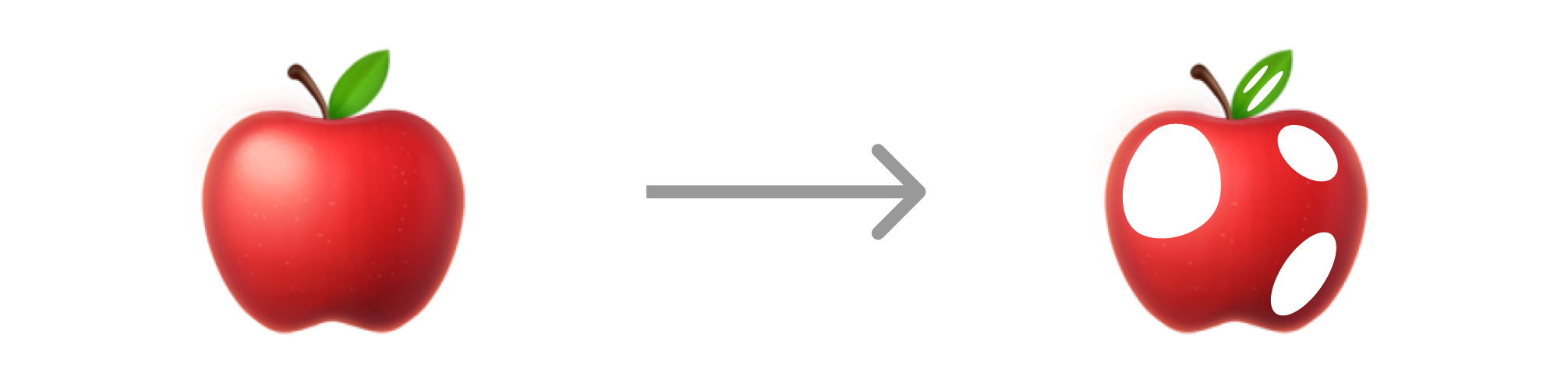}
\end{center}

\subsection{Distillation}
\label{subsubsec:distillation}
Train a smaller model to mimic a larger model.
Given a larger (teacher) model that is already highly accurate, transfer its learned features to a smaller (student) model by having the smaller model replicate the larger model's output.
One can then apply other compression techniques to the student model for further optimization~\cite{gholami2021survey, polino2018model}.
For a survey of distillation techniques, see~\cite{gou2021knowledge}.

\begin{center}
    \includegraphics[width=0.6\linewidth]{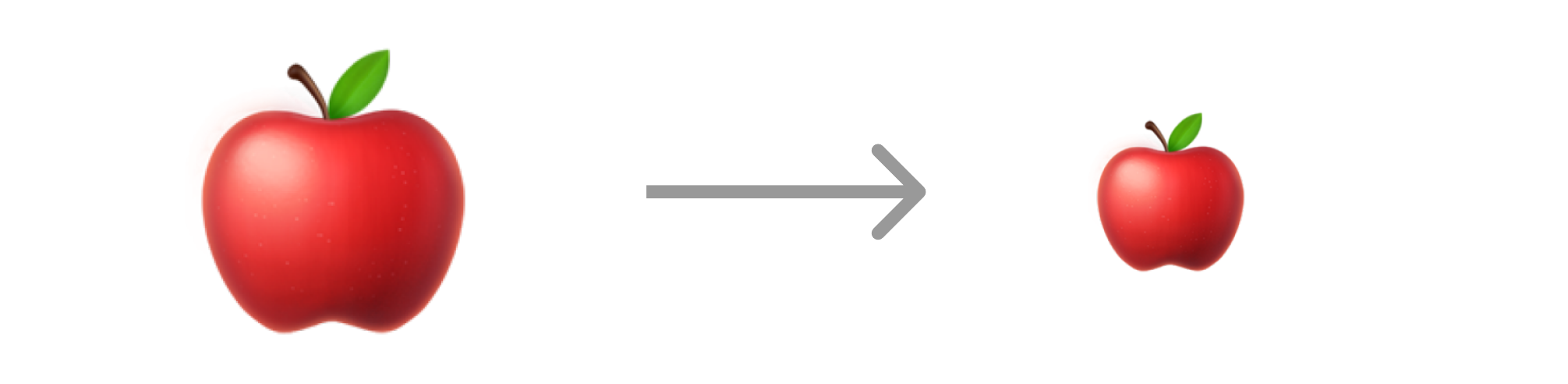}
\end{center}

\subsection{Efficient Neural Architectures}
\label{subsubsec:architecture}
Some model architectures are specifically designed to be small and efficient.
Prominent example architectures include MobileNets~\cite{howard2017mobilenets, sandler2018mobilenetv2}, ShuffleNets~\cite{zhang2018shufflenet}, and EfficientNets~\cite{tan2019efficientnet}.
All of these examples are designed to improve the efficiency of convolutions, which are vital to fitting computer vision models on-device~\cite{choudhary2020comprehensive}.
One can use other compression technqiues on efficient architectures to create even smaller models~\cite{choudhary2020comprehensive}.

\subsection{Dynamic Models}
\label{subsubsec:dynamic-models}
When thinking about ML applications, most people think of a single model with fixed inputs and outputs.
Dynamic models differ in that they take into account the differing prediction difficulty of each data input.
Depending on the input, a dynamic model may adapt its preprocessing steps, computational path, or model choice all together~\cite{zhu2021dynamic}.
While many types of dynamic models exist, we only discuss a selection of techniques that are mentioned later on.

\subsubsection{Early Exit Models}
\label{subsubsec:early-exit}
Allow for completing a prediction without the need to pass through the full model.
Motivated by the fact that some data points are easier to predict than others, early exit models run intermediate feature representations through additional output layers to check if the prediction is sufficiently correct.
For example, if a model is already confident about the current prediction, it finishes early rather than continuing to the next prediction layer.

\subsubsection{Gated Models}
\label{subsubsec:gated}
Train a smaller, approximate model whose prediction decides whether or not to invoke a larger model.
Similar to early exit models, some data points are easier to predict than others.
A fast, smaller model gates a larger model, for example by again using the confidence of the initial prediction.
At a high-level, these are systems of models that decide whether or not to spend extra compute if a prediction is still uncertain or unknown.

\section{Interview Study Methodology}
\label{sec:methodology}

\subsection{Study Protocol}
To capture emerging practices around \eml, we conducted semi-structured interviews~\cite{boyce2006conducting,knott2022interviews} with \numParticipants ML practitioners at \location to study how they approach model compression in their own work.
Our interview questions are outlined in \cref{appendix:questions}.
We gave participants ample time to flexibly speak to their specific work and express other topics beyond our question set that they felt were important to effectively optimizing models~\cite{knott2022interviews}.
The interviews took place between March and July 2022 with each conversation lasting from 45 minutes to 1 hour.
For all interviews, one researcher lead the interview questions, while another took detailed notes.
Where a participant approved, we also recorded conversations to refer back to during analysis.
No compensation was given, since all participants were salaried employees.
At the end of the study, we briefed participants and their teams on our results.
Our study protocol was approved by an internal IRB.

\subsection{Participants and Recruitment}
As discussed in \cref{sec:related-work}, in-depth interview studies with ML practitioners are uniquely suited to capture experts' tacit knowledge for the purpose of generating new, publicly-available guidance~\cite{amershi2019software, amershi2019guidelines, holstein2019improving, sambasivan2021everyone, liao2020questioning, bhatt2020explainable, yang2018investigating}. 
Our organization has a rare concentration of \eml experts, so we first reached out to known individuals.
We then used a snowball sampling strategy to reach a broader network of people involved in \eml.
To balance different perspectives, we sought practitioners working on ML for different domains and tasks (\eg vision, language, and sensing).
As the study evolved, we also chose new participants to help fill-in our largest areas of uncertainty.
Our participants, listed in \cref{tab:participants}, include ML research scientists, engineers, and managers spanning a wide breadth of application domains.
A natural saturation occurred when participants began to repeat know-how we had already recorded and began to \textit{only} suggest participants we had already included.

\subsection{Qualitative Data Analysis}
The first two authors conducted an iterative thematic analysis method to group common motivations, challenges, and best practices of model compression into categories~\cite{gibbs2007thematic}.
As we conducted more interviews, we continually discussed and updated these categories to reflect our new findings.
Each participant's data and transcripts were independently reviewed and manually coded by the first two authors at least twice using inductive coding~\cite{thomas2003general}.
In total, we spent 30+ hours interviewing participants and collected 23,500+ words of additional notes outside of the audio transcripts.
The final categories were split into two main sections: \cref{sec:results} ``Study Results'' and \cref{sec:opportunities} ``Design Opportunities for Efficient ML Tools and Interfaces.''
The composition of these two sections was formed by ordering major categories and hierarchically grouping within categories when relevant.

\subsection{Methodological Limitations}
While we found that learning from expert practitioners was tremendously valuable, any interview study has limits for generalizability.
Interview studies suffer from smaller population samples than other methods like a survey.
We chose to hold in-depth exploratory conversations with each participant---which a survey cannot support as well~\cite{knott2022interviews}.
On the other hand, we did not directly observe participants conducting their \eml work, which limits the specificity of our findings.
It was not possible to observe all participant's activity or model artifacts due to the sensitivity of their work.
Another concern is bias from a participant's role in the domain~\cite{boyce2006conducting}, as well as the power dynamics between interviewer and interviewees~\cite{kvale2006dominance}.
This study was conducted solely within one organization, therefore practitioners may hold organization-specific beliefs and practices~\cite{schein1990organizational}.
Despite our conscious efforts to recruit across ML application domains, we noticed a skew towards vision-based applications on images, video, and 3D scene data.
While interviews were conducted prior to the public rise of LLMs in late 2022~\cite{movva2023large}, many participants had NLP optimization experience, and as discussed in~\autoref{subsubsec:llms}, the same overall approaches for \eml are currently being applied to LLMs and other foundation model modalities.

Taking these limitations together, we caution readers to not consider the advice we detail from participants universally generalizable.
However it is with confidence that we share the rich pragmatic guidance these \numParticipants experts have to offer.

\section{Study Results}
\label{sec:results}
To answer our primary research question \textbf{RQ: How should a broader audience of HCI and ML practitioners today optimize powerful models to design on-device, ML user experiences?}, we first profile \textit{who} is working in model compression today, as emerged from our participant pool (\cref{subsec:persona}), and give a high-level state of \eml in practice (\cref{subsec:state-of-eml}).
We then break down our results along a typical AI/ML development process~\cite{mlworkflow2023google, tdsp2018microsoft, fayyad1996kdd, wirth2000crisp}, and use the machine learning workflow outlined by \citet{amershi2019software} as a reference point to map onto (in summary: \textit{model and data requirements}, \textit{model development and training}, and \textit{model evaluation}).
Similar to specifying \textit{model requirements} within ML~\cite{amershi2019software}, we characterize how practitioners design on-device machine learning experiences (\cref{subsec:on-device}) and plan model budgets (\cref{subsec:budget}).
Next, we discuss how compression affects the \textit{model development and training} process, and offer considerations for how people can strategize applying compression to their own work (\cref{subsec:invest}).
Lastly, we describe how practitioners \textit{evaluate compressed models} to balance accuracy versus performance---and avoid accidental compression artifacts (\cref{subsec:model-eval}).

Throughout the results we highlight actionable takeaways in
\aptLtoX[graphic=no,type=html]{
\begin{imageonly}{\setlength{\fboxsep}{3pt}\fbox{call-out boxes}}\end{imageonly}
}{
\setlength{\fboxsep}{3pt}
\fbox{call-out boxes}
}.
Note that some of these strategies are unique to model compression for on-device ML, while other strategies add a model compression twist to already-familiar software efficiency ideas.
We include both to balance domain-specific novelty with the key advice for on-device ML.

\subsection{Participants and Emergent Personas}
\label{subsec:persona}

Who does on-device ML efficiency?
Of the \numParticipants people we interviewed, participants had an average of \participantsMLXP (max 12) years experience with ML, and an average \participantsEMLXP (max 10) years experience with efficient ML.
Our participants had diverse breadth of expertise across domains, 
detailed in \cref{tab:participants}. 
Rather than their job title, we found that participant perspectives on \eml aligned more closely with their application focus.
For the purpose of understanding practitioners contributions, we define three distinct, emergent personas that best describe our participants:

\begin{itemize}
\setlength\itemsep{0.5em}

    \item \textbf{Compression \underline{E}xperts (E1--E13 in \cref{tab:participants}):}
    Seasoned research scientists and experienced engineers that lead on-device ML initiatives, develop new compression techniques, and consult on machine learning optimization efforts.
    Given the amount of experience these people have, they often manage or lead teams.
    \textit{Example: An ML research scientist who has a PhD in model compression and is developing novel techniques for model optimization.}
    
    \item \textbf{Machine Learning \underline{P}ractitioners (P1--P11 in \cref{tab:participants}):}
    ML engineers and data scientists that build and deploy models on-device where certain optimization budgets must be met.
    These people use compression as a means to an end rather than solely studying \eml.
    \textit{Example: An ML engineer optimizing a model's size to shrink it to 1MB while maintaining high accuracy.}
    
    \item \textbf{\underline{T}ooling Engineers (T1--T6 in \cref{tab:participants}):}
    Engineers and developers that focus on building frameworks, infrastructure, and tools for \eml.
    \textit{Example: A software engineer building and maintaining an organization-wide tool to help others compute efficiency metrics.}
    
\end{itemize}

\begin{table*}[htb]
\centering
\caption{
A summary of the \numParticipants participants from the interview study.
Participants are grouped by three emergent personas: Compression Experts (E), ML Practitioners (P), and Toolkit Engineers (T).
Participants indicated the number of years they have spent working on machine learning (light blue squares), and of those years how many have been spent working on model compression and optimization specifically (dark blue squares).
}

\begin{tabular}{lrlll}

\textbf{ID} & \multicolumn{2}{l}{\textbf{Experience (Compression / ML Years)}} & \textbf{Job Title (Manager \checkmark)} & \textbf{ML Application}\\
\hline
\\
\multicolumn{5}{l}{\textit{Compression Expert}}\\
\noalign{\vskip 1mm}
\hline
\noalign{\vskip 2mm}
E1&10/12&\chart{1}{systemBlue} \chart{1}{systemBlue} \chart{1}{systemBlue} \chart{1}{systemBlue} \chart{1}{systemBlue} \hspace{\tableSpaceAmount}\chart{1}{systemBlue} \chart{1}{systemBlue} \chart{1}{systemBlue} \chart{1}{systemBlue} \chart{1}{systemBlue} \hspace{\tableSpaceAmount}\chart{1}{systemBlueTinted} \chart{1}{systemBlueTinted} &ML Manager (\checkmark)&Efficient ML research\\
E2&7/12&\chart{1}{systemBlue} \chart{1}{systemBlue} \chart{1}{systemBlue} \chart{1}{systemBlue} \chart{1}{systemBlue} \hspace{\tableSpaceAmount}\chart{1}{systemBlue} \chart{1}{systemBlue} \chart{1}{systemBlueTinted} \chart{1}{systemBlueTinted} \chart{1}{systemBlueTinted} \hspace{\tableSpaceAmount}\chart{1}{systemBlueTinted} \chart{1}{systemBlueTinted} &ML Manager (\checkmark)&3D computer vision\\
E3&5/12&\chart{1}{systemBlue} \chart{1}{systemBlue} \chart{1}{systemBlue} \chart{1}{systemBlue} \chart{1}{systemBlue} \hspace{\tableSpaceAmount}\chart{1}{systemBlueTinted} \chart{1}{systemBlueTinted} \chart{1}{systemBlueTinted} \chart{1}{systemBlueTinted} \chart{1}{systemBlueTinted} \hspace{\tableSpaceAmount}\chart{1}{systemBlueTinted} \chart{1}{systemBlueTinted} &ML Manager (\checkmark)&Efficient ML research\\
E4&5/10&\chart{1}{systemBlue} \chart{1}{systemBlue} \chart{1}{systemBlue} \chart{1}{systemBlue} \chart{1}{systemBlue} \hspace{\tableSpaceAmount}\chart{1}{systemBlueTinted} \chart{1}{systemBlueTinted} \chart{1}{systemBlueTinted} \chart{1}{systemBlueTinted} \chart{1}{systemBlueTinted} \hspace{\tableSpaceAmount}&ML Engineer (\checkmark)&3D computer vision\\
E5&9/9&\chart{1}{systemBlue} \chart{1}{systemBlue} \chart{1}{systemBlue} \chart{1}{systemBlue} \chart{1}{systemBlue} \hspace{\tableSpaceAmount}\chart{1}{systemBlue} \chart{1}{systemBlue} \chart{1}{systemBlue} \chart{1}{systemBlue} &Research Scientist&Efficient ML research\\
E6&6/8&\chart{1}{systemBlue} \chart{1}{systemBlue} \chart{1}{systemBlue} \chart{1}{systemBlue} \chart{1}{systemBlue} \hspace{\tableSpaceAmount}\chart{1}{systemBlue} \chart{1}{systemBlueTinted} \chart{1}{systemBlueTinted} &ML Engineer (\checkmark)&Efficient computer vision\\
E7&5/8&\chart{1}{systemBlue} \chart{1}{systemBlue} \chart{1}{systemBlue} \chart{1}{systemBlue} \chart{1}{systemBlue} \hspace{\tableSpaceAmount}\chart{1}{systemBlueTinted} \chart{1}{systemBlueTinted} \chart{1}{systemBlueTinted} &ML Engineer (\checkmark)&3D computer vision\\
E8&2/8&\chart{1}{systemBlue} \chart{1}{systemBlue} \chart{1}{systemBlueTinted} \chart{1}{systemBlueTinted} \chart{1}{systemBlueTinted} \hspace{\tableSpaceAmount}\chart{1}{systemBlueTinted} \chart{1}{systemBlueTinted} \chart{1}{systemBlueTinted} &ML Manager (\checkmark)&Efficient ML research\\
E9&5/6&\chart{1}{systemBlue} \chart{1}{systemBlue} \chart{1}{systemBlue} \chart{1}{systemBlue} \chart{1}{systemBlue} \hspace{\tableSpaceAmount}\chart{1}{systemBlueTinted} &ML Engineer (\checkmark)&Efficient computer vision\\
E10&5/6&\chart{1}{systemBlue} \chart{1}{systemBlue} \chart{1}{systemBlue} \chart{1}{systemBlue} \chart{1}{systemBlue} \hspace{\tableSpaceAmount}\chart{1}{systemBlueTinted} &Research Scientist&Efficient ML research\\
E11&5/5&\chart{1}{systemBlue} \chart{1}{systemBlue} \chart{1}{systemBlue} \chart{1}{systemBlue} \chart{1}{systemBlue} \hspace{\tableSpaceAmount}&ML Engineer&Efficient ML research\\
E12&5/5&\chart{1}{systemBlue} \chart{1}{systemBlue} \chart{1}{systemBlue} \chart{1}{systemBlue} \chart{1}{systemBlue} \hspace{\tableSpaceAmount}&Research Scientist&3D computer vision\\
E13&3/5&\chart{1}{systemBlue} \chart{1}{systemBlue} \chart{1}{systemBlue} \chart{1}{systemBlueTinted} \chart{1}{systemBlueTinted} \hspace{\tableSpaceAmount}&ML Engineer&3D computer vision\\
\\
\multicolumn{5}{l}{\textit{ML Practitioner}}\\
\noalign{\vskip 1mm}
\hline
\noalign{\vskip 2mm}
P1&6/10&\chart{1}{systemBlue} \chart{1}{systemBlue} \chart{1}{systemBlue} \chart{1}{systemBlue} \chart{1}{systemBlue} \hspace{\tableSpaceAmount}\chart{1}{systemBlue} \chart{1}{systemBlueTinted} \chart{1}{systemBlueTinted} \chart{1}{systemBlueTinted} \chart{1}{systemBlueTinted} \hspace{\tableSpaceAmount}&ML Manager (\checkmark)&Efficient computer vision \\
P2&4/10&\chart{1}{systemBlue} \chart{1}{systemBlue} \chart{1}{systemBlue} \chart{1}{systemBlue} \chart{1}{systemBlueTinted} \hspace{\tableSpaceAmount}\chart{1}{systemBlueTinted} \chart{1}{systemBlueTinted} \chart{1}{systemBlueTinted} \chart{1}{systemBlueTinted} \chart{1}{systemBlueTinted} \hspace{\tableSpaceAmount}&ML Engineer&Machine translation\\
P3&3/10&\chart{1}{systemBlue} \chart{1}{systemBlue} \chart{1}{systemBlue} \chart{1}{systemBlueTinted} \chart{1}{systemBlueTinted} \hspace{\tableSpaceAmount}\chart{1}{systemBlueTinted} \chart{1}{systemBlueTinted} \chart{1}{systemBlueTinted} \chart{1}{systemBlueTinted} \chart{1}{systemBlueTinted} \hspace{\tableSpaceAmount}&Research Scientist (\checkmark)&ML sensing\\
P4&4/9&\chart{1}{systemBlue} \chart{1}{systemBlue} \chart{1}{systemBlue} \chart{1}{systemBlue} \chart{1}{systemBlueTinted} \hspace{\tableSpaceAmount}\chart{1}{systemBlueTinted} \chart{1}{systemBlueTinted} \chart{1}{systemBlueTinted} \chart{1}{systemBlueTinted} &ML Engineer&Machine translation\\
P5&5/8&\chart{1}{systemBlue} \chart{1}{systemBlue} \chart{1}{systemBlue} \chart{1}{systemBlue} \chart{1}{systemBlue} \hspace{\tableSpaceAmount}\chart{1}{systemBlueTinted} \chart{1}{systemBlueTinted} \chart{1}{systemBlueTinted} &Software Engineer&Computer vision\\
P6&1/6&\chart{1}{systemBlue} \chart{1}{systemBlueTinted} \chart{1}{systemBlueTinted} \chart{1}{systemBlueTinted} \chart{1}{systemBlueTinted} \hspace{\tableSpaceAmount}\chart{1}{systemBlueTinted} &ML Engineer&Multi-modal ML\\
P7&1/6&\chart{1}{systemBlue} \chart{1}{systemBlueTinted} \chart{1}{systemBlueTinted} \chart{1}{systemBlueTinted} \chart{1}{systemBlueTinted} \hspace{\tableSpaceAmount}\chart{1}{systemBlueTinted} &ML Engineer&Multi-modal ML\\
P8&2/5&\chart{1}{systemBlue} \chart{1}{systemBlue} \chart{1}{systemBlueTinted} \chart{1}{systemBlueTinted} \chart{1}{systemBlueTinted} \hspace{\tableSpaceAmount}&ML Engineer&ML fairness\\
P9&3/4&\chart{1}{systemBlue} \chart{1}{systemBlue} \chart{1}{systemBlue} \chart{1}{systemBlueTinted} &ML Engineer&ML hardware\\
P10&3/4&\chart{1}{systemBlue} \chart{1}{systemBlue} \chart{1}{systemBlue} \chart{1}{systemBlueTinted} &ML Engineer&3D computer vision\\
P11&1/4&\chart{1}{systemBlue} \chart{1}{systemBlueTinted} \chart{1}{systemBlueTinted} \chart{1}{systemBlueTinted} &Software Engineer (\checkmark)&Multi-modal ML\\
\\
\multicolumn{5}{l}{\textit{Tooling Engineer}}\\
\noalign{\vskip 1mm}
\hline
\noalign{\vskip 2mm}
T1&5/10&\chart{1}{systemBlue} \chart{1}{systemBlue} \chart{1}{systemBlue} \chart{1}{systemBlue} \chart{1}{systemBlue} \hspace{\tableSpaceAmount}\chart{1}{systemBlueTinted} \chart{1}{systemBlueTinted} \chart{1}{systemBlueTinted} \chart{1}{systemBlueTinted} \chart{1}{systemBlueTinted} \hspace{\tableSpaceAmount}&ML Manager (\checkmark)&Efficient ML tooling\\
T2&3/6&\chart{1}{systemBlue} \chart{1}{systemBlue} \chart{1}{systemBlue} \chart{1}{systemBlueTinted} \chart{1}{systemBlueTinted} \hspace{\tableSpaceAmount}\chart{1}{systemBlueTinted} &Software Engineer&Efficient ML tooling\\
T3&2/6&\chart{1}{systemBlue} \chart{1}{systemBlue} \chart{1}{systemBlueTinted} \chart{1}{systemBlueTinted} \chart{1}{systemBlueTinted} \hspace{\tableSpaceAmount}\chart{1}{systemBlueTinted} &Software Engineer&ML fairness\\
T4&4/4&\chart{1}{systemBlue} \chart{1}{systemBlue} \chart{1}{systemBlue} \chart{1}{systemBlue} &Software Engineer&Efficient ML algorithms\\
T5&3/3&\chart{1}{systemBlue} \chart{1}{systemBlue} \chart{1}{systemBlue} &Software Engineer&Efficient ML algorithms\\
T6&2/2&\chart{1}{systemBlue} \chart{1}{systemBlue} &Software Engineer&Efficient ML tooling\\

\end{tabular}
\label{tab:participants}
\end{table*}

Comparing the personas, we see clear differences between applications.
For example, compression experts are heavily (and nearly exclusively) focused on research, ML practitioners work across the most diverse set of domains (\eg vision, NLP, sensing, multi-modal models, fairness, and hardware), and tooling engineers focus on internal tools and compression algorithm implementation.
Throughout the paper, we label representative quotes from participants by their personas to illustrate the main findings from the study.
The persona labels offer additional context for situating a participant's background and perspective into the larger discussion. 

\subsection{The State of Efficient ML in Practice}
\label{subsec:state-of-eml}

Creating an efficient machine learning model is a \textit{``large design and constrained optimization problem''} [\pNine], where practitioners have to weigh decisions between many model performance metrics, such as model storage size, inference latency, power usage, device temperature, and model behavioral metrics such accuracy, precision, recall, and others.
To many readers, this characterization may sound like a job for an automated optimization algorithm.
However, a crucial finding of our results is that automation is not yet possible, do to the degree of ML, product design, and human expertise that goes into creating on-device applications.
Experts emphasized there is no single solution, \textit{``silver bullet''} [\pNineteen], \textit{``turnkey solution''} [\pNine], or \textit{``golden recipe''} [\pTwelve] for successfully building efficient machine learning models.
Partially, this is due to the rapidly moving-target of new state-of-the-art model architectures and new ML applications that may require a custom approach.

\begin{quote}
    \textit{``[Compression] works super well sometimes, but sometimes totally fails. Each project is too dependent, recipes tend to only work in some situations.''} --- \pTwentythree, \pTwentyone, \pTwentytwo
\end{quote}

The expertise for creating efficient ML models is currently held by a relatively small subset of ML practitioners, where \textit{``knowledge is handed-down from the few who know how to do it successfully,''} [\pSix].
These people are referred to as artisans and \textit{``based on earlier experience, know it's possible''} [\pTwentysix] to create highly efficient models.
Reducing a model's size by half, or by an order of magnitude, takes additional time and considerations [\pTwelve].
\textit{``It's like black magic or the dark arts, but you will get better with time,''} [\pTwelve].
Readers may recognize the ``dark arts'' or ``artisan'' language from earlier days of contemporary ML~\cite{patel2008investigating, hill2016trials}, before any real widespread knowledge base was established.
We hope these study results support growing such a knowledge-base for \eml, and it's role within a holistic ML process:

\begin{quote}
    ``\textit{[Efficient ML] means a lot of things to different people. Specific techniques on models weights help reduce size, but to get an efficient model comes from more careful design of the loss function, the system, which parts should and should not be modeled with ML.''} --- \pNine
\end{quote}

In practice, building efficient ML models describes the process of either building a new model from scratch or modifying an existing model to shrink it enough to fit performance constraints, such as model size, latency, power consumption, and even heat [\pEight].
Newcomers often gravitate towards the latest model compression techniques from literature as a first-step, while experts who consult on these projects encourage taking a step back to consider the entire data and model lifecycle:
\textit{``Don't do [model compression] blindly. Don't do [model compression] in a rush,''} [\pFive].
\textit{``Philosophically, [you] need to look at whole problem, not as an afterthought,''} [\pSix].

In addition to barriers of overall experience, efficient ML work differs perhaps most significantly in how much it requires a deep understanding of hardware details.
Low-level hardware details are not something that typical ML or software engineering usually needs to consider.
\textit{``[ML engineers] feel a bit intimated, and people don't feel like they understand this stuff well,''} [\pTwentyfive].

\subsection{The Metrics of Efficient, On-Device Models}
\label{subsec:on-device}

\begin{figure}[!tb]
    \centering
    \includegraphics[width=\columnwidth,alt={A diagram showing three ML models that shrink over time after ML developers apply (1) architecture search and (2) model compression to finally produce the most efficient model possible.}]{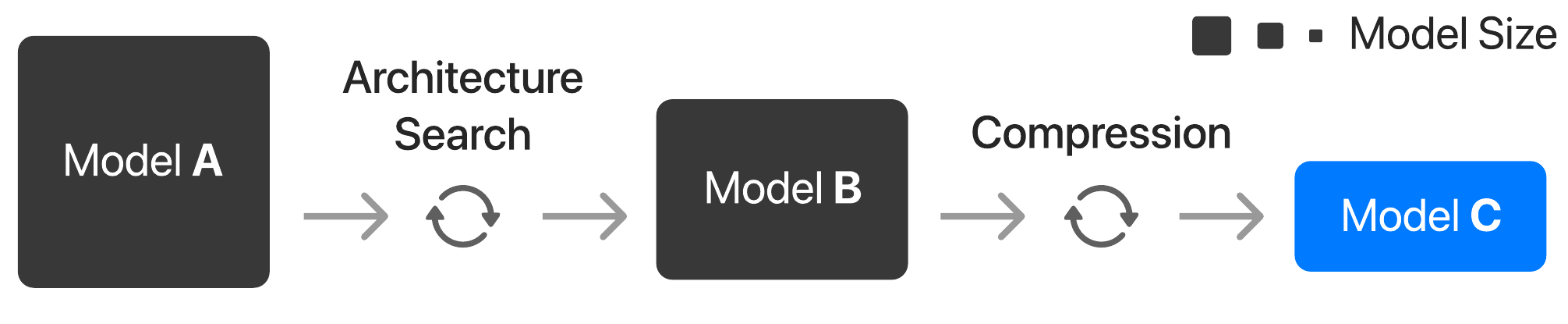} %
    \caption{
    Practitioners start with a feasibility model (A), which is a model of any size that demonstrates that the ML task works. Next, an architecture search looks for a smaller model that is equivalent to (A) and suitable for devices (B). A team decides a budget based on how much more efficient model (B) must be to deploy. Practitioners use techniques to compress their model (B) to reach the budget (C).
    }
    \label{fig:process}
\end{figure}

\begin{table*}
\renewcommand{\arraystretch}{1.5}%

\centering
\caption{Key aspects of on-device ML efficiency that impact user experience.}

\begin{tabular}{p{0.25\linewidth} | p{0.70\linewidth} } %
\textbf{User Perceivable Metrics} & \textbf{Negative User Impact}\\
\hline

\symbolstorage Model storage size & 
Modern deep learning models can easily grow to gigabytes. Devices have finite storage, so a model must not occupy so much space that it interferes with a user storing their own content.\\

\symbollowpower Power usage &
Users expect their devices to have long battery lives, but a resource-intensive model can quickly drain a battery. Addressing battery drain is particularly challenging for older devices with lower battery capacity and health.\\

\symbolheat Device temperature &
ML models that consume lots of computational power at once can heat up a user's device to where it is uncomfortably warm to the touch. Heat can also trigger thermal throttling, where the device slows down to avoid over-heating.\\

\symbollatency Inference latency &
Taking a long time to run inference can be frustrating for a user waiting on the result, and can make the experience feel unresponsive.\\

\symbolfps FPS (frames per second) &
A special case of inference latency. When processing live inputs (\eg video), this latency can create delays that makes the output appear choppy or unresponsive.\\

\hline

\textit{All metrics: model tested during high device resource load}
& Resources are shared. A user's device may be running many things simultaneously, including multiple ML models. A single model that demands too many resources will slow everything down.\\

\end{tabular}
\label{tab:metrics}
\end{table*}

For many practitioners we spoke to, \eml is of immediate importance to their work in developing new product features.
Efficiency is a deciding point on which features get approved to ship to users, so efficient ML enables user experiences that otherwise simply could not exist [\pFour, \pEight, \pNineteen, \pTwentythree, \pTwo].

\begin{quote}
    \textit{``Who cares if we can detect X if your model takes too much space? It will never make it to engineering requirements.''} --- \pEight
\end{quote}

As an umbrella term, \textit{efficiency} encompasses many metrics.
Common metrics that practitioners linked to user experience are summarized in \cref{tab:metrics}.
Some may be familiar with metrics commonly considered in ML workflows, such as latency and model storage size.
Other metrics may come as more foreign to an ML engineer accustomed to deploying their model server-side, \eg device temperature.
When a user is holding and carrying around a device all day, metrics like heat and battery life impact are crucial.

The prioritization of efficiency metrics depends on the model type and its intended usage.
While different practitioners we spoke to were focused on optimizing different aspects of efficiency, they all had a shared concept of a \textit{model budget}. 

\subsection{Deciding Model Budget}
\label{subsec:budget}

A model budget encompasses thresholds for speed, accuracy, size, or the amount of any specific resource a model is allowed to consume.
Model budgets are created individually per model, and are based as much in product and UX design as they are in device constraints. 

\subsubsection{Budgeting by Technical Feasibility}
\label{subsubsec:feasibility}
For new and novel ML applications, it is simply unknown on the onset how computational intensive a model might be.
Practitioners refer to reported metrics from related ML research and an organization's other models to sketch out an initial budget:

\begin{quote}
    \textit{``At the beginning, [we do a] `back of envelope calculation' where things need to be as honest as possible to what's realistic. At the beginning, you're more focused on accuracy. Over time there is refinement.''} --- \pTwelve
\end{quote}

A common refinement practice is outlined in \cref{fig:process}.
First, ML engineers will work on a ``feasibility'' model simply to see if the ML task ``\textit{is possible at all}'' [\pEight].
Once the feasibility model works at sufficient accuracy (\cref{fig:process}A), engineers work on an architecture search to find a smaller, more efficient model that achieves the accuracy goal (\cref{fig:process}B) [\pEight, \pTwelve].
Next, the budget is refined based on the potential to shrink the model using compression: ``\textit{given} [an] \textit{accuracy goal, what is the biggest percentage reduction possible?}'' [\pTen].
Some teams do this reduction estimation on their own, while others bring in ML compression experts [\pFour].
Since the compressed-model budget (\cref{fig:process}C) is often estimated \textit{before} engineers embark on model compression, the budget remains open to refinement, subject to product design constraints.
Ultimately, if an ML-powered feature requires a strict model budget, the model will not ship until engineers have found a way to reach those thresholds [\pEight].

\subsubsection{Budgeting by User's Experience of a Model}
Many aspects of budgeting come directly out of product feature design for where, when, and how often a model will be running.

\paragraph{\symbolfps One-Time, Real-Time, or All the Time?}
Latency budget is typically a per-feature UX decision dependent upon how a user perceives the model output and how often the model needs to run.
For example, in a photography app when tagging people in a new photo, a model needs to work fast enough so that users do not notice a delay.
Models on-device remove the wait on network response times, which make these latency budgets easier to achieve [\pTwentythree].

Models that only need to run once generally have a looser resource budget than continuous models.
For models running in real-time, only a few milliseconds per inference may be the maximum budget [\pTwentyeight].
This is most crucial in real-time scenarios where the user can perceive the model working, such as in live video applications [\pEleven].
A user will notice if their video feed appears choppy or a background filter appears delayed.
For this reason, latency budgets need to be as low as possible.
Similarly, strict latency budgets are given to models that need to run all the time, \ie ``always-on''.
Always-on models are continuously using the device's resources, so they are required to be as quick and lightweight as possible.
Some always-on models require a timely response to the user, for example, an always-on model that listens for a user to trigger a voice-assistant.

\paragraph{\symbollatency User Opt-In or Background Task?}
There are some applications where ML is the primary enabling technology, such as voice assistants or language translation apps.
If a user explicitly takes action to trigger a model, the user is in control of when the model runs (or does not run).
Product designers will allow models a larger budget for power and compute resources when a user explicitly opts-in [\pTen, \pTwelve, \pTwentythree].
It is also common to have supporting models running in the background.
Similar to any other background task, these models require a strict budget to stay unobtrusive:

\begin{quote}
    \textit{``If the user will never perceive the model, we still want the memory footprint to be inconspicuous so it doesn't interfere with them using their device.''} --- \pTwelve
\end{quote}

\paragraph{\symbolsleep Can it Wait for a User to Sleep?}
Many devices today will download and install routine updates at night, or whenever the device's user sleeps and has their device connected to power.
Updates are intensive computational tasks that can be disruptive to normal device function, thus it is conventional to wait until a user is not actively using their device to initiate.
By the same logic, scheduling a model to run overnight is a good option if the model's output is not something needed immediately.
If a user's device is plugged in, this also alleviates the concern of a model draining battery.
Disruption to the user is minimized, so models running overnight are also free to take a longer time with a bigger resource budget [\pEleven, \pEight]. 

\begin{framed}
\noindent\textbf{Strategy \#1}: First estimate the accuracy and latency of the original model architecture you want to run on-device. Then estimate how often the model needs to run (less often is ``easier'') and what it needs to deliver (less accurate is ``easier'') to budget the \textit{worst-case} performance that will be acceptable for your user experience. The gap between your starting estimate and worst-case budget will tell you how feasible your on-device ML plan is.
\end{framed}

\subsubsection{Top-down Budget by Application \& Device}
\begin{figure}[!t]
    \centering
    \includegraphics[width=\columnwidth,alt={A rectangular bar representing a fictional application's storage size that is broken up into various pieces. The piece that represents how much storage ML models take up is highlighted in blue.}]{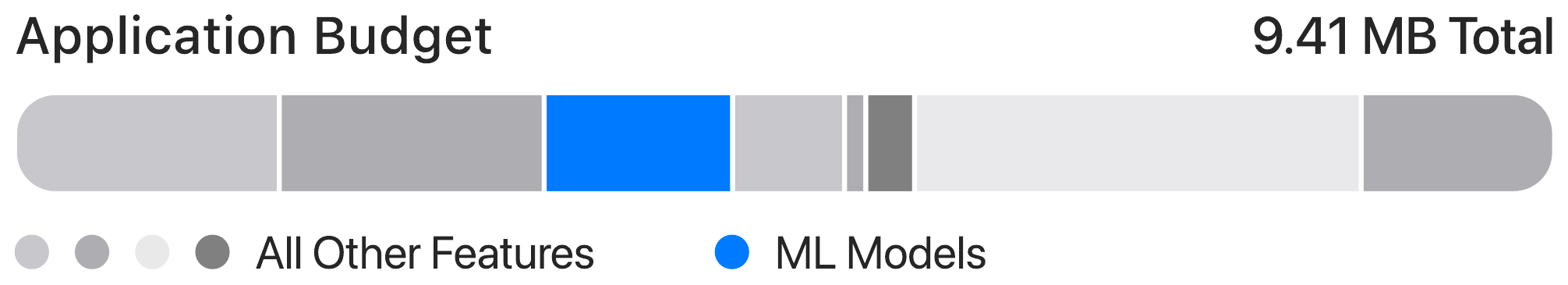} %
    \caption{Models needs to share space with the rest of an application. Model budget typically reflects how ``valuable'' a model is compared to all other features of the application. The above hypothetical example shows how much budget a model (blue) has in the overall system.}
    \label{fig:budget}
\end{figure}

As illustrated in \cref{fig:budget}, an application or device has finite resources---and ML models are only one component of a system.
For this reason, major budget allocation is typically decided by people with far reaching views and ownership in an organization [\pFifteen, \pTwentyeight, \pFive, \pSeventeen]:

\begin{quote}
    \textit{``\textup{[The product lead]} looks at budget for whole feature and allocates budget to individual models from there. There are dozens of algorithms. \textup{[Each model]} gets budget based on priority, practicality, and executive decisions on what is the most important algorithms to give space/time to.''} --- \pFive
\end{quote}

The difficulty of hitting budget can vary enormously depending on the model.
Thus, negotiation over budget allocations can continue and evolve until late in development [\pTwentyeight].

\paragraph{\symbolstorage Size Budget}
Any one application should not take up too much disk space on a device.
The challenge is that as more and more new features make use of ML, there are more models to fit in the same amount of storage:

\begin{quote}
    \textit{``Because of \textup{[neural networks]}, and the rising popularity of huge transformer models, compression becomes more and more important.''} --- \pTwo
\end{quote}

\paragraph{\symbolpower Power Budget}
A model's power budget is a measure of how much battery it is permitted to use.
Power consumption of a model can be tricky to estimate, and should be measured empirically on the target device for the most reliable estimate [\pZero, \pTwentynine, \pSeventeen].
Once again, real-time or always-on models require much stricter power budgets, because they continuously draw power.
Instead of measuring power at one moment of execution, these continuous models can be thought of in terms of the total percentage or total minutes of battery they consume in a day [\pZero, \pEight, \pTwelve].

\paragraph{\symbolheat Heat Budget}
Closely related to power consumption is device temperature.
In the words of \pTwelve{}: ``\textit{ship high-quality ML models that don't melt the user's device}.''
By melting, \pTwelve{} refers to both a device that is uncomfortably warm to the touch, and a device that has been forced into thermal mitigation.
Thermal mitigation, or thermal throttling, is when a device slows down processing to protect itself from overheating.
Hot and slow devices makes for a terrible user experience that should be avoided.
Thus, memory and power budgets must be set according to thermal measurements [\pZero, \pTwelve, \pFourteen, \pFifteen, \pSixteen]: ``\textit{Heat throttling is everyone's budget!}'' [\pFourteen].

\paragraph{Multiple Models in Concert}
A development team may measure the power, latency, or memory load of a single model as they develop it, but testing a single model in isolation is insufficient in situations where multiple models are powering an app or experience at the same time [\pZero, \pEleven, \pThirteen].
Multiple models running simultaneously will affect the overall memory pressure experienced by device at any moment, and thus the latency of each model as well as total power consumption.
For these situations, teams need to experimentally test models running in concert to get accurate measures of total consumption, and experimentally refine their budget for each individual model from there [\pEleven, \pThirteen]:

\begin{quote}
    \textit{``You could use too much memory, and that's a no-go. You don't know that until you test on device. [The memory bound] is not something you can compute ahead. There are multiple models happening at once, so until they are all put together you can't see exactly what the memory is going to be on a process.''} --- \pEleven
\end{quote}

\begin{framed}
\noindent\textbf{Strategy \#2}: Budget a model's resources by it's value-add to your overall experience. For a reliable estimate, run your model on the actual target devices early and often during model optimization. Measure compute usage, memory pressure, battery consumption, heat, and your model's storage size relative to the overall application. Run the model alone and in realistic, high-load scenarios where many process are running simultaneously to find out if your model is too resource greedy. Remember to not be a bad citizen: an ML feature that does not respect the resource needs of the overall device will result in a poor user experience.
\end{framed}

\subsubsection{Edge, IoT, and Low-power Devices} 
\label{subsubsec:section-edge-issues}
Devices like laptops, tablets, and mobile phones have far more memory and power resources than smaller, low-power ``edge'' devices, such as wearables or IoT devices.
For edge devices, budget is tight enough that even the bytes of the modeling code text itself can matter [\pTen].
For these devices, an expert recommended avoiding neural networks all together due to their high cost, and approximate the same accuracy as much as possible using the lowest profile conventional machine learning algorithms, such as decision trees [\pTen].

\subsubsection{Gating \& Variable-Budget Options}
There may be features where it is not (yet) technically possible to fit the ML component entirely on device.
One option is gating, where only \textit{some} inference is done on-device: a small simplified model lives on-device and is only highly accurate for common inputs.
If the small model detects complex or uncertain inputs, it invokes a high-powered ML model that uses more resources, or it sends the harder input to an external model on a server (see background \cref{subsubsec:gated}) [\pTen].

Another option is variable budgeting.
Product teams may be able to keep essential parts of a feature running smoothly under intermittent network availability by dynamically changing whether it uses higher-accuracy models or lower-accuracy models.
Similarly, product teams can also choose to lower ML accuracy when needed to keep a feature running under high-device memory load:

\begin{quote}
    \textit{``So people think about model compression as static resources, but the actual resources on device are dynamic based on what's running. So it's useful to on-the-fly change how much resources your model uses.''} --- \pThirteen
\end{quote}

For any model, there will be some size and efficiency budget at which the model can be shrunk without any noticeable difference to the user.
Below that threshold, accuracy degradation may negatively impact UX.
A budget must balance the best model accuracy with the lowest possible resource footprint.

\begin{framed}
\noindent\textbf{Strategy \#3}: On-device ML does not need to be all-or-nothing. If current compression techniques cannot fit your ML feature on-device, try breaking the feature down into subtasks. Delegate smaller subtask models to live on-device and delegate larger ML workloads to a server if appropriate. Prioritize on-device ML for feature subtasks that will help preserve your user's privacy and keep critical functionality responsive in the absence of a network connection.
\end{framed}

\subsection{Applying Compression during Development}
\label{subsec:invest}

Participants strongly emphasized that applying model compression \textit{is an investment}.
Compression techniques can take intensive engineering effort to apply---and may fail.
There are many circumstances that can cause one compression technique to generate enormous savings in one model, and completely fail to produce any savings in a different model, potentially after weeks of wasted engineering effort.
While this work is not an exact science, we detail general tips and guidance from practitioners.
Note that most of these techniques assume the model to compress is a neural network.

\subsubsection{Maximize Architecture Savings First}
\label{subsubsec:arch}
As shown in \cref{fig:process}, practitioners often start optimizing a model by finding a smaller architecture first.
\textit{``If you're running a common ResNet, look at strides, layer widths, and the number of layers''} [\pFour].
Participants also recommended neural architectures designed specifically for device efficiency [\pEleven, \pThirty, \pSix, \pTen, \pTwentyseven], such as MobileNet~\cite{howard2017mobilenets}.
Some applications do not need a heavy neural network; instead it might be possible, and even beneficial, to convert into a simpler decision tree or SVM [\pEight].
Architecture savings are seen as a more reliable first approach before attempting compression [\pEleven, \pFifteen]:

\begin{quote}
    \textit{``If you're trying to reduce the size, model compression is effective, but more like a last resort. If you have a model that is just overkill, [compression] can shave off 20-40\%, but just changing architecture to something smaller will have way bigger impact.''} --- \pTen
\end{quote}

Due to the high training cost of some models, participants discussed manual architecture selection (as opposed to auto-ML solutions that automate various stages of ML development) based on their expertise: \textit{``It's really trial and error by hand monitoring the hardware usage [and] identifying bottlenecks,''} [\pSixteen].

\subsubsection{Check the Architecture Against the Hardware}
\label{subsubsec:hardware-check}
A critical caveat to architecture optimization is that it is easy to be fooled.
A common mistake is to trust in the reproducibility of academic results: ``\textit{Don't expect what is ``efficient'' in a paper to exactly work in practice,}'' [\pTwelve].
The issue is that at the lowest level of computation, different hardware optimizes for different operations.
Reported efficiency measures are only reliable for the specific hardware set they were run on.
Even for specialized accelerator hardware designed for ML, the hardware will likely only support certain types of operations.
Since hardware cannot be altered at the same pace as software, the latest neural layer architectures from papers may not be possible to run efficiently on-device.
Practitioners emphasized the need to adapt their neural architecture to better fit hardware [\pFour, \pTwelve, \pSeventeen].

\subsubsection{Test Against a Range of Hardware}
\label{subsubsec:hardware-range}
In conventional machine learning work, and more generally much of modern software engineering, it is not required to know the details of the hardware that code will run on. 
In \eml, hardware may be one of the greatest challenges when working in on-device ML:

\begin{quote}
    \textit{``You must consider different layers of abstraction of a model from code to hardware. This can be hard for developers to understand.''} --- \pZero
\end{quote}

To further complicate matters, often practitioners are not considering a single specific hardware implementation, but rather a class of hardware.
Classes of hardware include platforms such as mobile phones, smartwatches, tablets, computers, and others.
Consider a practitioner building a model for a smartphone.
There are many types of smartphones, and versions of existing smartphones each with their own memory and compute details.
Hardware also changes over time, and practitioners may need to consider older versions of hardware that have already been shipped.
In the same way that front-end developers build responsive UIs and applications for multiple screensizes, so too do ML practitioners now need to optimize and test multiple hardware implementations specific to a set of devices [\pZero].

\begin{framed}
\noindent\textbf{Strategy \#4}: If starting from scratch, chose a model architecture specifically designed for mobile devices, but be aware that an architecture may not perform as-advertised due to implementation differences in hardware. It is crucial to profile models on your physical target hardware---and for every hardware version you aim to support. Older devices usually have fewer compute resources for your model.
\end{framed}

\subsubsection{Determine if a Model is Memory Bound or Compute Bound}
\label{subsec:memorybound}
Another useful strategy is to consider ahead which metrics are likely to be the biggest issue for a model type [\pFour].
A model is \textit{memory bound} if the size of the model or the size of the data is the biggest issue.
A model is \textit{compute bound} if the speed/latency of the model is the biggest issue.
For instance, \textit{``Computer vision tends to be memory bound and not compute bound. It's fast but the data sizes are massive''} [\pFour].
Participants also suggest going layer-by-layer to identify bottlenecks since individual layers can be either memory or compute bound individually [\pSixteen].

If memory-bound, architecture changes, sparsity techniques, or palettization techniques can help reduce model size.
On-disk size is considered the easiest memory savings to get from model compression [\pEleven].
If data transfer size is causing the memory issue, reducing data resolution for model processing can help, \eg reducing video data from 1080p to 720p [\pSeven].
If compute-bound, techniques like quantization can help speed up and simplify complex matrix math.
Quantization can also save memory since smaller numbers help keep computation in cache memory [\pNineteen].
Cache locality is a common latency bottleneck to check [\pZero, \pNineteen], that also helps with power problems:

\begin{quote}
    \textit{``Latency is easier than power to improve. You can check cache locality, moving data takes a lot of power so that's a big one to check.''} --- [\pZero]
\end{quote}

\subsubsection{Accuracy v. Effort Trade-Off}
Practitioners we spoke with had a sense of \textit{``accuracy v. effort''} [\pZero] or \textit{``risk v. reward''} [\pEleven] for some of the most common compression approaches.
As a general heuristic, compression techniques that need little-to-no retraining will be cheaper to apply---but at the cost of steeper model accuracy degradation.
Since some of these models take days, or more, to train, involving complex compression logic in the training process is costly.
Practitioners must consider cost in money, time, and effort that implementing a specific compression technique could take.

\paragraph{\textup{\textbf{(\$)}} Post-training  Quantization}
Quantizing a model's weights \textit{after} the model has been trained was widely considered the first go-to compression technique for many applications.

\begin{quote}
    \textit{``So quantization is a big one. You can usually quantize to 8-bit integers without losing accuracy. This isn't always the case. You do need to be careful where you apply quantization. But generally it's a pretty generic technique across different hardware. You can usually cut from \texttt{fp32} to \texttt{fp16} and can get speed up by going to integers.''} --- \pFourteen
\end{quote}

Since weight quantization can be done without additional training, it is considered cheap.
10/30 participants discussed using this approach [\pZero, \pOne, \pTwo, \pThree, \pFour, \pFive, \pEleven, \pTwelve, \pTwentyfive, \pTwentyfive].
Although post-training quantization is considered \textit{``easy''} [\pTwo] as far as ML compression techniques go, practitioners emphasized that it still often takes complex code to implement and there are many algorithm variations~\cite{gholami2021survey} to experiment with [\pFour].
For models that need high accuracy, post-training quantization may not be enough to hit budget without unacceptable accuracy degradation [\pTwo, \pFourteen, \pEleven, \pNine].

\paragraph{\textup{\textbf{(\$\$)}} Compression with Fine Tuning}
When model accuracy goes down after quantization or pruning, fine tuning can help recover the loss~\cite{gholami2021survey, hoefler2021sparsity}.
Fine tuning a model costs training time, but is necessary in many situations to recover accuracy.
Pruning techniques almost always require fine-tuning on the pruned model~\cite{hoefler2021sparsity}.
6/30 participants discussed successfully utilizing this approach [\pOne, \pThree, \pFive, \pSeven, \pEleven, \pThirteen].

\paragraph{\textup{\textbf{(\$\$\$+)}} Compression-aware Training}
Though it introduces additional complexity to model training, compression can be approached as yet another optimization that a model needs to learn during training.
11/30 participants discussed having experience with quantization learned during training [\pZero, \pTwo, \pFour, \pSix, \pNine, \pTwelve, \pThirteen, \pFifteen, \pNineteen, \pTwentyfive, \pTwentysix].
This approach is generally considered best-in-class and often required when designing always-on ML experiences that require low latency, \eg real-time computational photography models.
Practitioners recommended training-aware compression when a model needs to be compressed \textit{significantly} to meet budget while keeping high accuracy:

\begin{quote}
    \textit{``If you want to go to lower bit quantization, such as 4 or below, it's almost impossible to use post-training quantization because the difference in accuracy gets way too big. So for this level of compression you need to do training-aware compression.''} --- \pTwo
\end{quote}

Some practitioners had less experience with learned sparsity techniques and called them \textit{``high risk, high reward''} [\pEleven] because they are much harder to control compared to post-training sparsity techniques.
Although training-aware compression is considered the best form of optimization~\cite{gholami2021survey}, a major drawback is that is must be included in initial model training: \textit{``Not starting early with compression is a dead end,''} [\pSix].

Practitioners use their experience and early testing to judge how much compression a model will need.
For example, large vision models may start far exceeding their size and power budgets, and require heavy compression to meet budget.
In these scenarios, training-aware compression is likely necessary, which requires planning ahead: \textit{``It has to be done from day 0, not day 100,''} [\pTwelve].

For other applications, practitioners suggest estimating how much compression will be feasible with simple post-training quantization.
To estimate quantization savings \textit{before training a model}, first initialize the ML model architecture with random weights, then quantize, and test the model's speed and size on-device.
Even a coarse estimate may be worth getting a sense of the magnitude of savings from quantization [\pEleven]: 

\begin{quote}
    \textit{``[It] gives you a sense of how much quantization is going to help at all before you go through expensive training process.''} --- \pFour
\end{quote}

If post-training techniques produce results that are too far from budget goals, teams can then pivot to training-aware approaches.

\begin{framed}
\noindent\textbf{Strategy \#5}: Before you heavily invest in a particular compression technique, start with a cheap estimate of how much savings you might expect to gain. For example, initialize a mobile-friendly architecture with random weights, then quantize it. Profile this ``stand-in'' compressed model on-device and compare its efficiency to your budget. If it meets budget, then quantization may be enough. If it is far off budget (such is the case with many real-time computer vision models), consider more intensive training-aware compression techniques.
\end{framed}

\subsection{Evaluating Compressed Models: Efficiency, Accuracy, and Artifacts}
\label{subsec:model-eval}

The goal of model compression is to reduce a model's size while preserving its accuracy, or what some refer to as ``\textit{acceptable accuracy degradation}'' [\pSeven].
In general, past a certain point, shrinking a model will likely degrade its accuracy.
Conversely, in literature there are examples showing that compression can actually improve accuracy, by acting as a model regularizer and forcing the model to generalize better~\cite{kusupati2020soft, touvron2021training}.
This phenomenon is sometimes referred to as Occam's Hill~\cite{hoefler2021sparsity}: as light compression is applied, there can be an slight accuracy bump as the model is forced to generalize; however, as one increases the compression strength the model becomes \textit{too general} to properly work and accuracy quickly drops.
While these results have been observed on academic benchmarks, practitioners said that their models tend to generalize quite well already, so most often they see little improvement from the regularizer effect of compression [\pSeven, \pFifteen].
Thus, we focus this discussion on the much more common scenario where practitioners must wrestle with accuracy degradation.

\subsubsection{The Trade-off Curve Visualization}
\label{subsubsec:tradeoff}
\begin{figure}
    \centering
    \includegraphics[width=\columnwidth,alt={Two scatter plots where each dot represents a compressed ML model. The y-axis encodes model accuracy from 0\%--100\%. The x-axes of each chart encode latency (in ms) and power consumption (in mw) respectively. Each chart shows a slight curve from the bottom-left to the top-right, showing a trend that as a model gets more accurate it tends to have a higher latency and larger power consumption. The models that are chosen to be the best are the in knee of the curve and balance accuracy, latency, and power.}]{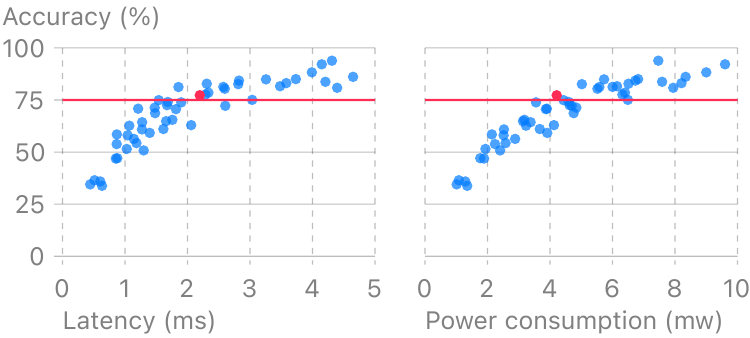} %
    \caption{
    A common chart used for model comparison and selection, where the y-axis is a model behavioral metric (\eg accuracy) and the x-axis is any number of performance metrics (\eg latency or power consumption).
    Each blue dot represents one model trained with different architectures or compression schemes, and the red line indicates the accuracy budget threshold.
    Practitioners look for the ``knee in the curve'' across charts: choosing a model above the accuracy threshold that minimizes latency and power consumption.
    In these charts, the selected model is colored red.
    }
    \label{fig:tuning-curves}
\end{figure}

To empirically compare multiple compressed models and select the one that satisfies budget requirements, practitioners typically plot a model behavioral metric (\eg accuracy) on the y-axis against any performance metric (\eg model size, model latency, or power consumption) on the x-axis~\cite{vasu2022mobileone, vasu2023fastvit}.
This is illustrated in \cref{fig:tuning-curves}.
Each dot represents a model architecture and compression pair.
These charts are called multiple names throughout different teams, including the trade-off curve, the accuracy Pareto curve, or the tuning curve.

\pEight{} described a scenario when optimizing over 6,000 small models.
In their application, the ML model was already on the order of kilobytes; therefore, they could retrain many new models quickly.
Comparing thousands of models in parallel, they plotted the F1-score on the y-axis against the model size on the x-axis.
This arrangement formed the usual curve.
Teams then filtered out models that do not satisfy their budgets, and ultimately selected a single model in the ``knee'' of the curve that has the best balance between what metrics they care about (\eg Pareto optimal).

\subsubsection{Compounding Changes Degrade Accuracy}
\label{subsubsec:compounding-error}
A common challenge with maintaining accuracy is that model compression techniques can compound in unintended ways.
In some scenarios, small optimizations throughout a model build on one another, so by the time a data input reaches the end of a network, its prediction is totally off.
In ML, this phenomenon is similar to the concept of exploding/vanishing gradients~\cite{bengio1994learning,pascanu2013difficulty}.
The effect of this compounding error is not obvious to spot in beginning [\pTwo].
For instance:

\begin{quote}
    \textit{``When you have addition that takes two quantized values to a quantized output, it's not easy to check that the range of the inputs are the same as the output ranges. You tend to lose resolution in the output where one branch dominates the range, and you lose the range of the lesser branch. You need additional care to check that they have the same range tracking during the forward quantization pass.''} --- \pZero
\end{quote}

For quantization, practitioners advise a ``gut check'' to ensure the quantized weights and activations match the ranges of the original model, or else accuracy will degrade [\pZero].

\subsubsection{Robust, End-to-end Data Evaluation}
\label{subsubsec:robust}
Since the amount of success in model optimization varies significantly by architecture and task, \textit{``without a clear evaluation strategy you won't know if you're making things better or worse,''} [\pTen].
Some ML-powered experiences are composed of multiple models working together.
Participants said they first test models individually to ensure they work standalone, then evaluate the multi-model systems to test compound effects [\pThirteen].
Practitioners said the curve visualizations (\cref{fig:tuning-curves}) are often the output of evaluation pipelines.

Many teams do extensive model evaluation, error, and failure analysis  [\pTwentyfive, \pNineteen, \pTwenty, \pTwentyone, \pTwentytwo, \pThirteen, \pTwentynine]. 
\textit{``We cannot assume compression doesn't change model behavior, so we look at confusion matrices and instances where the model gets it wrong''} [\pNineteen]. 
In one project that \pTwenty, \pTwentyone, and \pTwentytwo{} worked on where small mistakes could lead to a poor user experience, they built ``unit tests'' for different curated testing sets to monitor accuracy over time.
\pZero{} also emphasized using data unit tests to observe how single instances move through a model to both understand how much computational change occurs after compression and make the necessary adjustments to the model code. \textit{``Data so influential to hitting accuracy targets''} [\pSix].

\begin{framed}
\noindent\textbf{Strategy \#6}: Compression can degrade the accuracy of a model and change its behavior in unpredictable ways. It is essential to create a robust evaluation pipeline (\eg defining metrics, curating test sets) before you start optimizing your model, so that you can reliably observe shifts in model error afterwards. To prevent degradation from a failed optimization, compare optimized models with varying amounts of compression to your original model, inspecting the metrics, subpopulation behaviors, and internals, such as weights and activations, to ensure they are within expected ranges.
\end{framed}

\subsubsection{Model Compression Artifacts}
\label{subsubsec:artifacts}

\begin{figure}
    \centering
    \includegraphics[width=\columnwidth,alt={A grid of images showing the effect of too much compression which creates artifacts on images, audio, and ML models. The image artifacts show a reduction in colors and distort the image. The audio artifacts show missing frequencies from the audio's frequency range. The ML model is a small neural network with a question mark, indicating that we do not know the full scope of what ML compression artifacts look like.}]{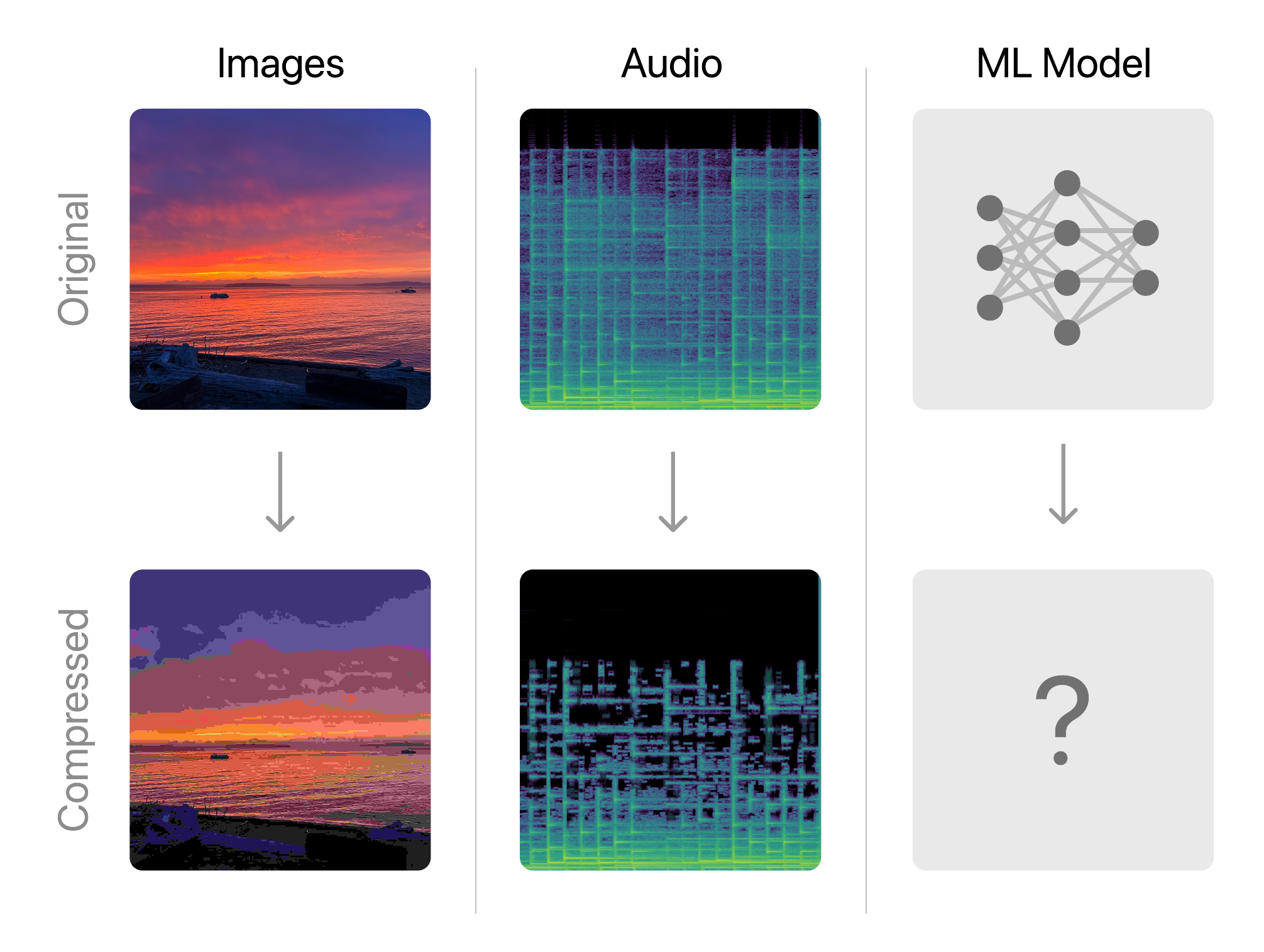} %
    \caption{Illustrative examples of compression artifacts in images and audio, but what do machine learning model artifacts look like and how do we identify them?}
    \label{fig:artifacts}
\end{figure}

As with other types of media, such as image, audio, and video data, if too much compression is applied it can produce \textit{compression artifacts:} noticeable distortions in the media (\cref{fig:artifacts}).
For example, compressing an image too much can produce blurry and blocky shapes over the subject matter; compressing audio can change the sound quality and waveform of the music.
Multiple participants describe scenarios where their model had artifacts, although they did not borrow this language from other types of media:

\begin{quote}
    \textit{``Compressing like 90\% can make things unstable with strange side effects. It's hard to figure out when you compress too much.''} --- \pEleven
\end{quote}

It is tempting to think of accuracy degradation as the only ML artifact; however, there are more subtle, even sinister implications depending on what a model is used for. 
For example, where some subpopulation of data is underrepresented in a dataset, \eg at the tail of a distribution, it could be that ML compression techniques remove this tail, amplifying existing biases.
Examples of this have been observed empirically in the few publications investigating robustness and evaluation of compressed models~\cite{hooker2020characterising, hooker2019compressed, ogueji2022intriguing, liebenwein2021lost, gale2019state}.

One specific compression artifact example story was told from multiple participants. 
In the case of an object detection model that needed to run at a high frame rate, during development teams noticed one day that the bounding boxes were jittering in their demo.
This finding was surprising, since the metrics from their most recent model iteration were reporting no changes, but \textit{``there were some weird side-effects''} [\pFour].
It was not until someone debugging the problem realized that they had applied quantization throughout the neural network, including the final prediction layer.
This coarsening of the data at the output produced correct bounding boxes, but resulted in a poor user experience.
Another participant had seen this before, and remarked if the output of a model is ``\textit{`fine grain,' don't quantize}'' [\pFive].

Generalizing from this example, one learning that teams have found is that when a model's prediction needs to be continuous, smooth, or is user-perceivable at a high frame rate, it is best not to compress the final output layer~\cite{deng2020model}.
Participants told us that other modeling tasks can be harder to optimize than classification [\pNine]. 
\textit{``Regression models have been an absolute nightmare to compress''} [\pTwelve].
Anytime the output needs to be a continuous or floating point value, compression techniques can produce artifacts [\pNineteen].

\subsubsection{Demoing User Experiences}
ML-powered experiences can be dependent upon multiple models working in concert, and the \textit{``best way to test compound [model] effects is having end-to-end evaluation''} [\pThirteen].
Another strategy to catch artifacts is testing models ``as close to the metal as possible,'' \ie loading the models on-device and building demo applications to run outside of lab environments.
\textit{``Since different hardware are not bit-wise accurate, metrics won't capture these changes''} [\pNineteen].
Another practitioner told us \textit{``user experience differences are the really important cases to isolate''} [\pThirty].
To find these user experience changes, teams have prioritized building demo applications where different models, \eg a baseline model and a compressed model, can be dynamically toggled back and forth for testing [\pTwentynine].
Some of these demos toggle between a model running on-device and a model running on a server [\pNineteen].
Interactive and live demos like these allow for ML teams to get feedback from other product stakeholders [\pThirteen].
They are particularly helpful for designers to get an overall ``feel'' of a model: interacting with a model in its intended environment to understand its capabilities and limitations [\pTwentynine]:

\begin{quote}
    \textit{``Get [the user experience] in the hands of people as fast as possible.''} --- \pTwentynine
\end{quote}

This notion of the ``feel'' of a model was described multiple times.
If an \textit{``ML engineer retrains [the model] and says it's better, we still need see if it feels better,''} [\pTwentynine].
To try and attribute the feel of a model to actionable development steps, practitioners show debugging modes on-device to observe live, real-time charts of a model's prediction and confidence, to help find edge cases and drill into specific errors that metrics may or may not capture [\pTwenty, \pTwentyone, \pTwentytwo].

\begin{framed}
\noindent\textbf{Strategy \#7}: Despite all the effort to create criteria and metrics to quantitatively measure and benchmark your model, your evaluation pipeline may not capture every aspect of your model's performance and behavior. Even if your model passes evaluation, move it on-device, in the intended environment in which it will run, and get it in the hands of users to demo. There is no other way to capture the feel of an ML-powered user experience.
\end{framed}

\section{Design Opportunities for Efficient ML Tools and Interfaces}
\label{sec:opportunities}

Given the unique challenges to efficient on-device ML, where can human-centered ML researchers, practitioners, and designers start to engage?
As \eml techniques are driven forward by advances in hardware engineering and ML research, there remains a major barrier in practically applying these techniques for real-world features and user experiences.

Tools influence what is possible: \textit{``there is a gap between what is possible with machine learning and what [tools] are being used,''} [\pFourteen].
Currently, the tooling for efficient on-device ML is underdeveloped, \textit{``homegrown and ad-hoc,''} [\pTwo]. 
Moreover, current tools focus on individual algorithms (\autoref{subsec:existing-resources}) instead of the holistic process: \textit{``this is a newer area, many tools are specific to a project and tend to be prototypes to prove feasibility,''} [\pTwo].
As we have demonstrated in this paper, there is considerable design planning, experimentation, and strategy that experts use to make on-device ML a reality.
Proper tools could help educate practitioners to develop these skills:

\begin{quote}
    \textit{``Tooling is education products disguised as software. Better tools make it easy to do correct things and harder to do incorrect things.''} --- \pTwentyfive
\end{quote}

We next share interdisciplinary HCI + ML research directions for supporting practitioners.
While the space of tooling opportunities is wide and will evolve over time, we provide a few actionable recommendations for future tools and interfaces that are ready for development today.

\subsection{Developing Intuition for Compression}

Although modern libraries for machine learning make it easier than ever develop models, that does not mean practitioners know how to best train, evaluate, and deploy models.
Machine learning is an inherently iterative and and empirical practice~\cite{patel2008investigating, amershi2019software}, and developing intuition for how models learn and behave is a major competitive advantage over blindly tweaking hyper-parameters.
Interactive playgrounds that provide a safe environment where practitioners can quickly build and test their ideas could be a great onboarding experience for learning about efficient ML and model compression.
There is already precedent for this within machine learning, where interactive and educational tools help learners develop intuition around how certain models train and make predictions.
Examples include include TensorFlow Playground~\cite{smilkov2017direct} for small neural networks, CNN Explainer~\cite{wang2021cnnexplainer} for convolutional neural networks, GANLab~\cite{kahng2018ganlab} for generative adversarial models, and Diffusion Explainer~\cite{lee2023diffusion} for stable diffusion models.
These interactives typically run fully in-browser, enabling learners to access and experiment with ML techniques without installing software to needing access to extra compute. 
Moreover, they have been incorporated into ML curricula to help people gain complementary, hands-on experience for specifics model categories.

Imagine a playground for efficient ML, where practitioners could learn and build intuition about different compression techniques [\pNineteen], model architectures [\pTwelve], and their effect on hardware [\pSeventeen].
Perhaps this playground could be web-based, where a small model is loaded in the browser running live inference.
Users could select different compression techniques, with varying degrees of strength, and see the impact on the model, its metrics, and its predictions.
This could help practitioners define realistic budgets (\cref{subsec:budget}), test different combinations of architectures and compression techniques (\cref{subsubsec:arch}), and inspect the difference between compute-bound or memory-bound models (\cref{subsec:memorybound}).

\subsection{Comparing Across Compression Schemes}

A common scenario in efficient model development is considering the trade-off between accuracy and performance (\cref{subsubsec:tradeoff}).
While conventional ML development requires model comparison between architectures and hyper-parameters, the optimization metrics that practitioners must also navigate add additional complexity when moving models on-device [\pFour].
Consider the scenario of having a well-performing model that does not hit a size budget.
What do you do next?
You can try every compression technique possible, but how do you ultimately select the best model that balances between the desired trade-offs?
In a similar process discussed in \cref{subsubsec:robust}, one practitioner mentioned that they \textit{``like to see an overview of a model's robustness before and after compression,''} [\pOne].

Better tooling to support model comparison could help practitioners compare a feasibility or baseline model (\cref{subsubsec:feasibility}) against different compressed models to select the best model possible.
There is rich opportunity for future work to investigate what to visualize and how.
For example, flexible interfaces should handle visualizing the similarities and differences between models, their internals, their outputs, and applied compression techniques, where each technique not only has a suite of hyperparameters but can be also be combined with one another.

This line of work could also draw from existing work in experiment tracking.
\pTwentysix{} described a project where the models they were developing were small, such that they could generate thousands of candidate models.
This participant always compared candidates against an uncompressed baseline model, and maintained documentation tracking their experimental history, with a few notes per model, so that other stakeholders could make an informed model selection.
However, this required diligent effort by the developer---experimental histories tend to exist across multiple documents and can result in long tables of values where patterns can be hard to discover.
Future opportunities for tooling point to model tracking systems that can generate interactive reports to help practitioners observe their model development history, while helping them select the best model (from potentially thousands of models) that passes their accuracy threshold and maximizes performance.

\subsection{Finding Model Bottlenecks for Targeted Compression}

In \cref{subsec:memorybound} and \cref{subsubsec:artifacts}, practitioners discuss analyzing a model layer-by-layer.
\textit{Targeted compression} is the practice of selectively optimizing specific components of a model, for example, individual layers in a neural network.
Currently, practitioners described the process as tedious, manual, and time-consuming, where \textit{``you must go layer by layer, operation by operation,''} [\pTwenty{}, \pTwentyone{}, \pTwentytwo{}].
There exists a space of interfaces that could help practitioners look inside their models, by mapping metrics to specific layers of a network [\pFour], finding the performance bottlenecks in a network [\pNine], comparing the input and output of these bottleneck layers [\pTwo], and selectively optimizing them.

Targeted compression tooling may also ease the difficulty of thinking at the hardware level.
\pZero{} gave an example: once an on-device model written in Python is compiled onto specialized hardware, the operations are expressed by what the hardware can support, and is likely not as readable as documented Python code.
To complicate matters, low-level hardware operations may not have a one-to-one mapping back to the original Python code, due to optimizations in the compilation process.
Better tools could help practitioners perform analytics on their models to find bottlenecks and easily traverse between different layers of abstraction (\eg model code and hardware operations).

\subsection{Evaluating Multi-model ML Systems}

It is often assumed that model evaluation is done on a single model, isolated in a ``lab environment.''
In practice, this is not the case.
Not only are models integrated into larger apps or codebases, many modern ML-powered experiences are the result of multiple models working together.
For example, models could be arranged in chains, where the outputs of one model are the inputs of another [\pThirteen].
If you compress one model, how does that impact downstream models?
Recalling the concept of compounding error discussed in \cref{subsubsec:compounding-error}, small errors introduced in earlier models can compound to produce bigger errors by the end of a model chain [\pThirteen].
\pSix{} emphasized this challenge, saying that the future of ML driven user experiences will be accomplished through multi-model systems. 
When multiple models are running simultaneously, each can be evaluated individually, but also must be considered as a whole, which makes it \textit{``super hard to reason about,''} [\pSix].
Future tools that generalize and can evaluate multi-model machine learning systems will have a big impact in helping practitioners build, debug, and make sense of large data-driven applications.

\subsection{Simplified Hardware Testing}
One recommendation repeated by practitioners in this study was to measure model performance on device (\cref{subsubsec:hardware-check}).
\pTwentynine{} emphasized that the performance of one model will like differ across different hardware (\cref{subsubsec:hardware-range}).
However, testing on real hardware can be a major barrier.
Due the variety of mobile devices and different versions of hardware, it can be hard to build for multiple hardware implementations simultaneously.
The average ML developer (outside of a hardware company) may not have access to all versions of all devices their users might have.
When it comes to future tooling opportunities, \pZero{} says it best, \textit{``the tools need to expose and take care of the cases for different hardware to maximize the hardware efficiency.''}
While there is undoubtedly room for ML-hardware innovation here, for the purposes of this work assume this problem can be reduced to looping over a set of hardware and evaluating a model.
With all these results, this problem could be cast as yet another comparison task, where tool designers need to enhance existing workflows to allow ML practitioners to track, organize, and see what hardware passed or failed certain criteria.

\subsection{Automating (Some) Compression Experiments}

While many of the practitioners we interviewed discussed creating efficient models as a human-in-the-loop iteration process, there are opportunities to automate applying different compression schemes and present results to developers. 
An apt analogy is what AutoML is to hyperparameter searching: instead of sequentially trying different compression schemes, perform a grid search and parallelize many different tests simultaneously. 
\pEight{} was particularly excited about compression automation.
Imagine mixed-initiative tools that take in a model and user-specified budgets, runs through a suite of compression techniques to find all possible experiments that satisfy the budgets, and finally presents a summary of recommended compression recipes for a practitioner to choose from.
Perhaps through tools like these we could even learn something about the behavior of compression itself.
However, we do not expect the process of creating efficient models to be completely automated.
Throughout our study, experts made it clear that successful model compression is an iterative development process (\cref{subsec:state-of-eml}), but leveraging the strengths of automation where appropriate suggests rich opportunity for mixed-initiative approaches for creating efficient models.

\section{Limitations \& Validity}
\label{sec:limtations}

Optimizing and creating efficient models to run on mobile devices is still relatively new, and best practices are evolving alongside the rapid pace of ML research.
Thus we expect not all the practices mentioned in this paper will stand the test of time.
As discussed in \cref{sec:methodology}, studying experts from a single organization naturally limits the perspective of this paper, therefore we also expect certain details of our findings may not generalize.
Nonetheless, we believe that the high-level strategies for on-device ML shared in this work---including budget design, strategies for investing in model optimization, and proper compression evaluation---will generalize to be useful for many different kinds of devices, domains, and ML user experiences.
We are eager to watch how this interdisciplinary area of research progresses in the future and reflect back on the practices outlined in this work to see which have remained, which have been removed, and which have been reinvented.

\section{Conclusion}
\label{sec:conclusion}

In this research we illustrated some of the pragmatic design considerations that go into efficient on-device machine learning.
While this paper is far less technical than related work from the ML literature, we feel the interdisciplinary bridge to UX and product design are critical to bringing on-device ML into more approachable and popular practice.
Through the results of this study, we are able to spotlight the holistic, end-to-end considerations that experts in model compression have developed from translating ML research into intelligent, on-device ML experiences.

\begin{acks}

We thank our participants at Apple for generously sharing their time and knowledge.
We also thank the many colleagues that have given their feedback on the paper.
Specifically, we thank Kayur Patel who sparked early ideas for this line of research, as well as Mohammad Rastegari, Sachin Mehta, and Yannick Assogba for their advisement on this work.

\end{acks}

\balance
\bibliographystyle{ACM-Reference-Format}

\bibliography{main}%


\begin{thebibliography}{132}


\ifx \showCODEN    \undefined \def \showCODEN     #1{\unskip}     \fi
\ifx \showDOI      \undefined \def \showDOI       #1{#1}\fi
\ifx \showISBNx    \undefined \def \showISBNx     #1{\unskip}     \fi
\ifx \showISBNxiii \undefined \def \showISBNxiii  #1{\unskip}     \fi
\ifx \showISSN     \undefined \def \showISSN      #1{\unskip}     \fi
\ifx \showLCCN     \undefined \def \showLCCN      #1{\unskip}     \fi
\ifx \shownote     \undefined \def \shownote      #1{#1}          \fi
\ifx \showarticletitle \undefined \def \showarticletitle #1{#1}   \fi
\ifx \showURL      \undefined \def \showURL       {\relax}        \fi
\providecommand\bibfield[2]{#2}
\providecommand\bibinfo[2]{#2}
\providecommand\natexlab[1]{#1}
\providecommand\showeprint[2][]{arXiv:#2}

\bibitem[tds(2018)]%
        {tdsp2018microsoft}
 \bibinfo{year}{2018}\natexlab{}.
\newblock \showarticletitle{What is the team data science process?}
\newblock \bibinfo{journal}{\emph{Microsoft}} (\bibinfo{year}{2018}).
\newblock
\urldef\tempurl%
\url{https://learn.microsoft.com/en-us/azure/architecture/data-science-process/overview}
\showURL{%
\tempurl}


\bibitem[des(2019)]%
        {design2019ibm}
 \bibinfo{year}{2019}\natexlab{}.
\newblock \showarticletitle{Design for AI}.
\newblock \bibinfo{journal}{\emph{IBM}} (\bibinfo{year}{2019}).
\newblock
\urldef\tempurl%
\url{https://www.ibm.com/design/ai/}
\showURL{%
\tempurl}


\bibitem[gui(2019a)]%
        {guidelines2019apple}
 \bibinfo{year}{2019}\natexlab{a}.
\newblock \showarticletitle{Human interface guidelines: Machine learning}.
\newblock \bibinfo{journal}{\emph{Apple Human Interface Guidelines}}
  (\bibinfo{year}{2019}).
\newblock
\urldef\tempurl%
\url{https://developer.apple.com/design/human-interface-guidelines/technologies/machine-learning/introduction}
\showURL{%
\tempurl}


\bibitem[gui(2019b)]%
        {guidebook2019google}
 \bibinfo{year}{2019}\natexlab{b}.
\newblock \showarticletitle{People + AI guidebook}.
\newblock \bibinfo{journal}{\emph{Google}} (\bibinfo{year}{2019}).
\newblock
\urldef\tempurl%
\url{https://pair.withgoogle.com/guidebook/}
\showURL{%
\tempurl}


\bibitem[mlw(2023)]%
        {mlworkflow2023google}
 \bibinfo{year}{2023}\natexlab{}.
\newblock \showarticletitle{Machine learning workflow}.
\newblock \bibinfo{journal}{\emph{Google}} (\bibinfo{year}{2023}).
\newblock
\urldef\tempurl%
\url{https://cloud.google.com/ai-platform/docs/ml-solutions-overview}
\showURL{%
\tempurl}


\bibitem[Alizadeh et~al\mbox{.}(2023)]%
        {alizadeh2023llm}
\bibfield{author}{\bibinfo{person}{Keivan Alizadeh}, \bibinfo{person}{Iman
  Mirzadeh}, \bibinfo{person}{Dmitry Belenko}, \bibinfo{person}{Karen
  Khatamifard}, \bibinfo{person}{Minsik Cho}, \bibinfo{person}{Carlo~C
  Del~Mundo}, \bibinfo{person}{Mohammad Rastegari}, {and}
  \bibinfo{person}{Mehrdad Farajtabar}.} \bibinfo{year}{2023}\natexlab{}.
\newblock \showarticletitle{Llm in a flash: Efficient large language model
  inference with limited memory}.
\newblock \bibinfo{journal}{\emph{arXiv preprint arXiv:2312.11514}}
  (\bibinfo{year}{2023}).
\newblock


\bibitem[Amershi et~al\mbox{.}(2019a)]%
        {amershi2019software}
\bibfield{author}{\bibinfo{person}{Saleema Amershi}, \bibinfo{person}{Andrew
  Begel}, \bibinfo{person}{Christian Bird}, \bibinfo{person}{Robert DeLine},
  \bibinfo{person}{Harald Gall}, \bibinfo{person}{Ece Kamar},
  \bibinfo{person}{Nachiappan Nagappan}, \bibinfo{person}{Besmira Nushi}, {and}
  \bibinfo{person}{Thomas Zimmermann}.} \bibinfo{year}{2019}\natexlab{a}.
\newblock \showarticletitle{Software engineering for machine learning: A case
  study}. In \bibinfo{booktitle}{\emph{2019 IEEE/ACM 41st International
  Conference on Software Engineering: Software Engineering in Practice}}. IEEE,
  \bibinfo{pages}{291--300}.
\newblock
\urldef\tempurl%
\url{https://doi.org/10.1109/icse-seip.2019.00042}
\showDOI{\tempurl}


\bibitem[Amershi et~al\mbox{.}(2019b)]%
        {amershi2019guidelines}
\bibfield{author}{\bibinfo{person}{Saleema Amershi}, \bibinfo{person}{Dan
  Weld}, \bibinfo{person}{Mihaela Vorvoreanu}, \bibinfo{person}{Adam Fourney},
  \bibinfo{person}{Besmira Nushi}, \bibinfo{person}{Penny Collisson},
  \bibinfo{person}{Jina Suh}, \bibinfo{person}{Shamsi Iqbal},
  \bibinfo{person}{Paul~N Bennett}, \bibinfo{person}{Kori Inkpen},
  {et~al\mbox{.}}} \bibinfo{year}{2019}\natexlab{b}.
\newblock \showarticletitle{Guidelines for human-AI interaction}. In
  \bibinfo{booktitle}{\emph{Proceedings of the 2019 CHI Conference on Human
  Factors in Computing Systems}}. \bibinfo{pages}{1--13}.
\newblock
\urldef\tempurl%
\url{https://doi.org/10.1145/3290605.3300233}
\showDOI{\tempurl}


\bibitem[Apple(2023a)]%
        {coremltools}
\bibfield{author}{\bibinfo{person}{Apple}.} \bibinfo{year}{2023}\natexlab{a}.
\newblock \bibinfo{booktitle}{\emph{Optimizing models - Core ML Tools
  overview}}.
\newblock
\urldef\tempurl%
\url{https://coremltools.readme.io/docs}
\showURL{%
\tempurl}


\bibitem[Apple(2023b)]%
        {apple2023personalizing}
\bibfield{author}{\bibinfo{person}{Apple}.} \bibinfo{year}{2023}\natexlab{b}.
\newblock \bibinfo{booktitle}{\emph{Personalizing a model with on-device
  updates}}.
\newblock
\urldef\tempurl%
\url{https://developer.apple.com/documentation/coreml/model_personalization/personalizing_a_model_with_on-device_updates}
\showURL{%
\tempurl}


\bibitem[Bengio et~al\mbox{.}(1994)]%
        {bengio1994learning}
\bibfield{author}{\bibinfo{person}{Yoshua Bengio}, \bibinfo{person}{Patrice
  Simard}, {and} \bibinfo{person}{Paolo Frasconi}.}
  \bibinfo{year}{1994}\natexlab{}.
\newblock \showarticletitle{Learning long-term dependencies with gradient
  descent is difficult}.
\newblock \bibinfo{journal}{\emph{IEEE Transactions on Neural Networks}}
  \bibinfo{volume}{5}, \bibinfo{number}{2} (\bibinfo{year}{1994}),
  \bibinfo{pages}{157--166}.
\newblock


\bibitem[Bhatt et~al\mbox{.}(2020)]%
        {bhatt2020explainable}
\bibfield{author}{\bibinfo{person}{Umang Bhatt}, \bibinfo{person}{Alice Xiang},
  \bibinfo{person}{Shubham Sharma}, \bibinfo{person}{Adrian Weller},
  \bibinfo{person}{Ankur Taly}, \bibinfo{person}{Yunhan Jia},
  \bibinfo{person}{Joydeep Ghosh}, \bibinfo{person}{Ruchir Puri},
  \bibinfo{person}{Jos{\'e}~MF Moura}, {and} \bibinfo{person}{Peter
  Eckersley}.} \bibinfo{year}{2020}\natexlab{}.
\newblock \showarticletitle{Explainable machine learning in deployment}. In
  \bibinfo{booktitle}{\emph{Proceedings of the 2020 Conference on Fairness,
  Accountability, and Transparency}}. \bibinfo{pages}{648--657}.
\newblock
\urldef\tempurl%
\url{https://doi.org/10.1145/3351095.3375624}
\showDOI{\tempurl}


\bibitem[Bommasani et~al\mbox{.}(2021)]%
        {bommasani2021opportunities}
\bibfield{author}{\bibinfo{person}{Rishi Bommasani}, \bibinfo{person}{Drew~A
  Hudson}, \bibinfo{person}{Ehsan Adeli}, \bibinfo{person}{Russ Altman},
  \bibinfo{person}{Simran Arora}, \bibinfo{person}{Sydney von Arx},
  \bibinfo{person}{Michael~S Bernstein}, \bibinfo{person}{Jeannette Bohg},
  \bibinfo{person}{Antoine Bosselut}, \bibinfo{person}{Emma Brunskill},
  {et~al\mbox{.}}} \bibinfo{year}{2021}\natexlab{}.
\newblock \showarticletitle{On the opportunities and risks of foundation
  models}.
\newblock \bibinfo{journal}{\emph{arXiv preprint arXiv:2108.07258}}
  (\bibinfo{year}{2021}).
\newblock


\bibitem[Boyce and Neale(2006)]%
        {boyce2006conducting}
\bibfield{author}{\bibinfo{person}{Carolyn Boyce} {and} \bibinfo{person}{Palena
  Neale}.} \bibinfo{year}{2006}\natexlab{}.
\newblock \bibinfo{booktitle}{\emph{Conducting in-depth interviews: A guide for
  designing and conducting in-depth interviews for evaluation input}}.
  Vol.~\bibinfo{volume}{2}.
\newblock \bibinfo{publisher}{Pathfinder International Watertown, MA}.
\newblock


\bibitem[Chauhan et~al\mbox{.}(2018)]%
        {chauhan2018performance}
\bibfield{author}{\bibinfo{person}{Jagmohan Chauhan},
  \bibinfo{person}{Jathushan Rajasegaran}, \bibinfo{person}{Suranga
  Seneviratne}, \bibinfo{person}{Archan Misra}, \bibinfo{person}{Aruna
  Seneviratne}, {and} \bibinfo{person}{Youngki Lee}.}
  \bibinfo{year}{2018}\natexlab{}.
\newblock \showarticletitle{Performance characterization of deep learning
  models for breathing-based authentication on resource-constrained devices}.
\newblock \bibinfo{journal}{\emph{Proceedings of the ACM on Interactive,
  Mobile, Wearable and Ubiquitous Technologies}} \bibinfo{volume}{2},
  \bibinfo{number}{4} (\bibinfo{year}{2018}), \bibinfo{pages}{1--24}.
\newblock


\bibitem[Chen and Ran(2019)]%
        {chen2019deep}
\bibfield{author}{\bibinfo{person}{Jiasi Chen} {and} \bibinfo{person}{Xukan
  Ran}.} \bibinfo{year}{2019}\natexlab{}.
\newblock \showarticletitle{Deep learning with edge computing: A review}.
\newblock \bibinfo{journal}{\emph{Proc. IEEE}} \bibinfo{volume}{107},
  \bibinfo{number}{8} (\bibinfo{year}{2019}), \bibinfo{pages}{1655--1674}.
\newblock
\urldef\tempurl%
\url{https://doi.org/10.1109/jproc.2019.2921977}
\showDOI{\tempurl}


\bibitem[Cheng et~al\mbox{.}(2018)]%
        {cheng2018model}
\bibfield{author}{\bibinfo{person}{Yu Cheng}, \bibinfo{person}{Duo Wang},
  \bibinfo{person}{Pan Zhou}, {and} \bibinfo{person}{Tao Zhang}.}
  \bibinfo{year}{2018}\natexlab{}.
\newblock \showarticletitle{Model compression and acceleration for deep neural
  networks: The principles, progress, and challenges}.
\newblock \bibinfo{journal}{\emph{IEEE Signal Processing Magazine}}
  \bibinfo{volume}{35}, \bibinfo{number}{1} (\bibinfo{year}{2018}),
  \bibinfo{pages}{126--136}.
\newblock
\urldef\tempurl%
\url{https://doi.org/10.1109/msp.2017.2765695}
\showDOI{\tempurl}


\bibitem[Cho et~al\mbox{.}(2022)]%
        {cho2022differentiable}
\bibfield{author}{\bibinfo{person}{Minsik Cho}, \bibinfo{person}{Keivan~A.
  Vahid}, \bibinfo{person}{Saurabh Adya}, {and} \bibinfo{person}{Mohammad
  Rastegari}.} \bibinfo{year}{2022}\natexlab{}.
\newblock \showarticletitle{Differentiable k-means clustering layer for neural
  network compression}. In \bibinfo{booktitle}{\emph{International Conference
  on Learning Representations}}.
\newblock
\urldef\tempurl%
\url{https://arxiv.org/abs/2108.12659}
\showURL{%
\tempurl}


\bibitem[Choudhary et~al\mbox{.}(2020)]%
        {choudhary2020comprehensive}
\bibfield{author}{\bibinfo{person}{Tejalal Choudhary}, \bibinfo{person}{Vipul
  Mishra}, \bibinfo{person}{Anurag Goswami}, {and} \bibinfo{person}{Jagannathan
  Sarangapani}.} \bibinfo{year}{2020}\natexlab{}.
\newblock \showarticletitle{A comprehensive survey on model compression and
  acceleration}.
\newblock \bibinfo{journal}{\emph{Artificial Intelligence Review}}
  \bibinfo{volume}{53}, \bibinfo{number}{7} (\bibinfo{year}{2020}),
  \bibinfo{pages}{5113--5155}.
\newblock
\urldef\tempurl%
\url{https://doi.org/10.1007/s10462-020-09816-7}
\showDOI{\tempurl}


\bibitem[Dai et~al\mbox{.}(2021)]%
        {dai2021respwatch}
\bibfield{author}{\bibinfo{person}{Ruixuan Dai}, \bibinfo{person}{Chenyang Lu},
  \bibinfo{person}{Michael Avidan}, {and} \bibinfo{person}{Thomas
  Kannampallil}.} \bibinfo{year}{2021}\natexlab{}.
\newblock \showarticletitle{Respwatch: Robust measurement of respiratory rate
  on smartwatches with photoplethysmography}. In
  \bibinfo{booktitle}{\emph{Proceedings of the International Conference on
  Internet-of-Things Design and Implementation}}. \bibinfo{pages}{208--220}.
\newblock


\bibitem[Delgado et~al\mbox{.}(2021)]%
        {delgado2021stakeholder}
\bibfield{author}{\bibinfo{person}{Fernando Delgado}, \bibinfo{person}{Stephen
  Yang}, \bibinfo{person}{Michael Madaio}, {and} \bibinfo{person}{Qian Yang}.}
  \bibinfo{year}{2021}\natexlab{}.
\newblock \showarticletitle{Stakeholder participation in AI: Beyond ``add
  diverse stakeholders and stir''}.
\newblock \bibinfo{journal}{\emph{arXiv}} (\bibinfo{year}{2021}).
\newblock


\bibitem[Deng et~al\mbox{.}(2020)]%
        {deng2020model}
\bibfield{author}{\bibinfo{person}{Lei Deng}, \bibinfo{person}{Guoqi Li},
  \bibinfo{person}{Song Han}, \bibinfo{person}{Luping Shi}, {and}
  \bibinfo{person}{Yuan Xie}.} \bibinfo{year}{2020}\natexlab{}.
\newblock \showarticletitle{Model compression and hardware acceleration for
  neural networks: A comprehensive survey}.
\newblock \bibinfo{journal}{\emph{Proc. IEEE}} \bibinfo{volume}{108},
  \bibinfo{number}{4} (\bibinfo{year}{2020}), \bibinfo{pages}{485--532}.
\newblock
\urldef\tempurl%
\url{https://doi.org/10.1109/jproc.2020.2976475}
\showDOI{\tempurl}


\bibitem[Developers(2022)]%
        {google2022why}
\bibfield{author}{\bibinfo{person}{Google Developers}.} \bibinfo{year}{Accessed
  2022}\natexlab{}.
\newblock \bibinfo{title}{Why on-device machine learning?}
\newblock
\newblock
\urldef\tempurl%
\url{https://developers.google.com/learn/topics/on-device-ml/learn-more}
\showURL{%
\tempurl}


\bibitem[Dhar et~al\mbox{.}(2021a)]%
        {dhar2019device}
\bibfield{author}{\bibinfo{person}{Sauptik Dhar}, \bibinfo{person}{Junyao Guo},
  \bibinfo{person}{Jiayi Liu}, \bibinfo{person}{Samarth Tripathi},
  \bibinfo{person}{Unmesh Kurup}, {and} \bibinfo{person}{Mohak Shah}.}
  \bibinfo{year}{2021}\natexlab{a}.
\newblock \showarticletitle{On-device machine learning: An algorithms and
  learning theory perspective}.
\newblock \bibinfo{journal}{\emph{ACM Transactions on Internet Things}}
  (\bibinfo{year}{2021}).
\newblock
\urldef\tempurl%
\url{https://doi.org/10.1145/3450494}
\showDOI{\tempurl}


\bibitem[Dhar et~al\mbox{.}(2021b)]%
        {dhar2021survey}
\bibfield{author}{\bibinfo{person}{Sauptik Dhar}, \bibinfo{person}{Junyao Guo},
  \bibinfo{person}{Jiayi Liu}, \bibinfo{person}{Samarth Tripathi},
  \bibinfo{person}{Unmesh Kurup}, {and} \bibinfo{person}{Mohak Shah}.}
  \bibinfo{year}{2021}\natexlab{b}.
\newblock \showarticletitle{A survey of on-device machine learning: An
  algorithms and learning theory perspective}.
\newblock \bibinfo{journal}{\emph{ACM Transactions on Internet of Things}}
  \bibinfo{volume}{2}, \bibinfo{number}{3} (\bibinfo{year}{2021}),
  \bibinfo{pages}{1--49}.
\newblock
\urldef\tempurl%
\url{https://doi.org/10.1145/3450494}
\showDOI{\tempurl}


\bibitem[Ding et~al\mbox{.}(2022)]%
        {ding2022delta}
\bibfield{author}{\bibinfo{person}{Ning Ding}, \bibinfo{person}{Yujia Qin},
  \bibinfo{person}{Guang Yang}, \bibinfo{person}{Fuchao Wei},
  \bibinfo{person}{Zonghan Yang}, \bibinfo{person}{Yusheng Su},
  \bibinfo{person}{Shengding Hu}, \bibinfo{person}{Yulin Chen},
  \bibinfo{person}{Chi-Min Chan}, \bibinfo{person}{Weize Chen},
  {et~al\mbox{.}}} \bibinfo{year}{2022}\natexlab{}.
\newblock \showarticletitle{Delta tuning: A comprehensive study of parameter
  efficient methods for pre-trained language models}.
\newblock \bibinfo{journal}{\emph{arXiv preprint arXiv:2203.06904}}
  (\bibinfo{year}{2022}).
\newblock


\bibitem[Ding et~al\mbox{.}(2023)]%
        {ding2023parameter}
\bibfield{author}{\bibinfo{person}{Ning Ding}, \bibinfo{person}{Yujia Qin},
  \bibinfo{person}{Guang Yang}, \bibinfo{person}{Fuchao Wei},
  \bibinfo{person}{Zonghan Yang}, \bibinfo{person}{Yusheng Su},
  \bibinfo{person}{Shengding Hu}, \bibinfo{person}{Yulin Chen},
  \bibinfo{person}{Chi-Min Chan}, \bibinfo{person}{Weize Chen},
  {et~al\mbox{.}}} \bibinfo{year}{2023}\natexlab{}.
\newblock \showarticletitle{Parameter-efficient fine-tuning of large-scale
  pre-trained language models}.
\newblock \bibinfo{journal}{\emph{Nature Machine Intelligence}}
  \bibinfo{volume}{5}, \bibinfo{number}{3} (\bibinfo{year}{2023}),
  \bibinfo{pages}{220--235}.
\newblock


\bibitem[Fahim et~al\mbox{.}(2021)]%
        {fahim2021hls4ml}
\bibfield{author}{\bibinfo{person}{Farah Fahim}, \bibinfo{person}{Benjamin
  Hawks}, \bibinfo{person}{Christian Herwig}, \bibinfo{person}{James
  Hirschauer}, \bibinfo{person}{Sergo Jindariani}, \bibinfo{person}{Nhan Tran},
  \bibinfo{person}{Luca~P Carloni}, \bibinfo{person}{Giuseppe Di~Guglielmo},
  \bibinfo{person}{Philip Harris}, \bibinfo{person}{Jeffrey Krupa},
  {et~al\mbox{.}}} \bibinfo{year}{2021}\natexlab{}.
\newblock \showarticletitle{hls4ml: An open-source codesign workflow to empower
  scientific low-power machine learning devices}.
\newblock  (\bibinfo{year}{2021}).
\newblock
\showeprint[arXiv]{2103.05579}


\bibitem[Fayyad et~al\mbox{.}(1996)]%
        {fayyad1996kdd}
\bibfield{author}{\bibinfo{person}{Usama Fayyad}, \bibinfo{person}{Gregory
  Piatetsky-Shapiro}, {and} \bibinfo{person}{Padhraic Smyth}.}
  \bibinfo{year}{1996}\natexlab{}.
\newblock \showarticletitle{The KDD process for extracting useful knowledge
  from volumes of data}.
\newblock \bibinfo{journal}{\emph{Commun. ACM}} \bibinfo{volume}{39},
  \bibinfo{number}{11} (\bibinfo{year}{1996}), \bibinfo{pages}{27--34}.
\newblock


\bibitem[Fu et~al\mbox{.}(2023)]%
        {fu2023effectiveness}
\bibfield{author}{\bibinfo{person}{Zihao Fu}, \bibinfo{person}{Haoran Yang},
  \bibinfo{person}{Anthony Man-Cho So}, \bibinfo{person}{Wai Lam},
  \bibinfo{person}{Lidong Bing}, {and} \bibinfo{person}{Nigel Collier}.}
  \bibinfo{year}{2023}\natexlab{}.
\newblock \showarticletitle{On the effectiveness of parameter-efficient
  fine-tuning}. In \bibinfo{booktitle}{\emph{Proceedings of the AAAI Conference
  on Artificial Intelligence}}, Vol.~\bibinfo{volume}{37}.
  \bibinfo{pages}{12799--12807}.
\newblock


\bibitem[Gale et~al\mbox{.}(2019)]%
        {gale2019state}
\bibfield{author}{\bibinfo{person}{Trevor Gale}, \bibinfo{person}{Erich Elsen},
  {and} \bibinfo{person}{Sara Hooker}.} \bibinfo{year}{2019}\natexlab{}.
\newblock \showarticletitle{The state of sparsity in deep neural networks}.
\newblock \bibinfo{journal}{\emph{arXiv preprint arXiv:1902.09574}}
  (\bibinfo{year}{2019}).
\newblock


\bibitem[Gholami et~al\mbox{.}(2021)]%
        {gholami2021survey}
\bibfield{author}{\bibinfo{person}{Amir Gholami}, \bibinfo{person}{Sehoon Kim},
  \bibinfo{person}{Zhen Dong}, \bibinfo{person}{Zhewei Yao},
  \bibinfo{person}{Michael~W Mahoney}, {and} \bibinfo{person}{Kurt Keutzer}.}
  \bibinfo{year}{2021}\natexlab{}.
\newblock \showarticletitle{A survey of quantization methods for efficient
  neural network inference}.
\newblock \bibinfo{journal}{\emph{arXiv}} (\bibinfo{year}{2021}).
\newblock
\showeprint[arXiv]{2103.13630}


\bibitem[Giattino et~al\mbox{.}(2022)]%
        {owid2022artificialintelligence}
\bibfield{author}{\bibinfo{person}{Charlie Giattino}, \bibinfo{person}{Edouard
  Mathieu}, \bibinfo{person}{Veronika Samborska}, \bibinfo{person}{Julia
  Broden}, {and} \bibinfo{person}{Max Roser}.} \bibinfo{year}{2022}\natexlab{}.
\newblock \showarticletitle{Artificial intelligence}.
\newblock \bibinfo{journal}{\emph{Our World in Data}} (\bibinfo{year}{2022}).
\newblock
\newblock
\shownote{https://ourworldindata.org/artificial-intelligence}.


\bibitem[Gibbs(2007)]%
        {gibbs2007thematic}
\bibfield{author}{\bibinfo{person}{Graham~R Gibbs}.}
  \bibinfo{year}{2007}\natexlab{}.
\newblock \showarticletitle{Thematic coding and categorizing}.
\newblock \bibinfo{journal}{\emph{Analyzing Qualitative Data}}
  \bibinfo{volume}{703} (\bibinfo{year}{2007}), \bibinfo{pages}{38--56}.
\newblock


\bibitem[Google(2019)]%
        {qkeras}
\bibfield{author}{\bibinfo{person}{Google}.} \bibinfo{year}{2019}\natexlab{}.
\newblock \bibinfo{booktitle}{\emph{QKeras}}.
\newblock
\urldef\tempurl%
\url{https://github.com/google/qkeras}
\showURL{%
\tempurl}


\bibitem[Gou et~al\mbox{.}(2021)]%
        {gou2021knowledge}
\bibfield{author}{\bibinfo{person}{Jianping Gou}, \bibinfo{person}{Baosheng
  Yu}, \bibinfo{person}{Stephen~J Maybank}, {and} \bibinfo{person}{Dacheng
  Tao}.} \bibinfo{year}{2021}\natexlab{}.
\newblock \showarticletitle{Knowledge distillation: A survey}.
\newblock \bibinfo{journal}{\emph{International Journal of Computer Vision}}
  \bibinfo{volume}{129}, \bibinfo{number}{6} (\bibinfo{year}{2021}),
  \bibinfo{pages}{1789--1819}.
\newblock
\urldef\tempurl%
\url{https://doi.org/10.1007/s11263-021-01453-z}
\showDOI{\tempurl}


\bibitem[Gu et~al\mbox{.}(2021)]%
        {gu2021server}
\bibfield{author}{\bibinfo{person}{Renjie Gu}, \bibinfo{person}{Chaoyue Niu},
  \bibinfo{person}{Fan Wu}, \bibinfo{person}{Guihai Chen},
  \bibinfo{person}{Chun Hu}, \bibinfo{person}{Chengfei Lyu}, {and}
  \bibinfo{person}{Zhihua Wu}.} \bibinfo{year}{2021}\natexlab{}.
\newblock \showarticletitle{From server-based to client-based machine learning:
  A comprehensive survey}.
\newblock \bibinfo{journal}{\emph{Comput. Surveys}} \bibinfo{volume}{54},
  \bibinfo{number}{1} (\bibinfo{year}{2021}), \bibinfo{pages}{1--36}.
\newblock
\urldef\tempurl%
\url{https://doi.org/10.1145/3424660}
\showDOI{\tempurl}


\bibitem[Han et~al\mbox{.}(2016)]%
        {han2015deep}
\bibfield{author}{\bibinfo{person}{Song Han}, \bibinfo{person}{Huizi Mao},
  {and} \bibinfo{person}{William~J Dally}.} \bibinfo{year}{2016}\natexlab{}.
\newblock \showarticletitle{Deep compression: Compressing deep neural networks
  with pruning, trained quantization and huffman coding}.
\newblock  (\bibinfo{year}{2016}).
\newblock


\bibitem[Hannun et~al\mbox{.}(2023)]%
        {apple2023mlx}
\bibfield{author}{\bibinfo{person}{Awni Hannun}, \bibinfo{person}{Jagrit
  Digani}, \bibinfo{person}{Angelos Katharopoulos}, {and}
  \bibinfo{person}{Ronan Collobert}.} \bibinfo{year}{2023}\natexlab{}.
\newblock \bibinfo{booktitle}{\emph{{MLX}: Efficient and flexible machine
  learning on Apple silicon}}.
\newblock
\urldef\tempurl%
\url{https://github.com/ml-explore}
\showURL{%
\tempurl}


\bibitem[Hill et~al\mbox{.}(2016)]%
        {hill2016trials}
\bibfield{author}{\bibinfo{person}{Charles Hill}, \bibinfo{person}{Rachel
  Bellamy}, \bibinfo{person}{Thomas Erickson}, {and} \bibinfo{person}{Margaret
  Burnett}.} \bibinfo{year}{2016}\natexlab{}.
\newblock \showarticletitle{Trials and tribulations of developers of
  intelligent systems: A field study}. In \bibinfo{booktitle}{\emph{2016 IEEE
  Symposium on Visual Languages and Human-Centric Computing}}. IEEE,
  \bibinfo{pages}{162--170}.
\newblock
\urldef\tempurl%
\url{https://doi.org/10.1109/vlhcc.2016.7739680}
\showDOI{\tempurl}


\bibitem[Hoefler et~al\mbox{.}(2021)]%
        {hoefler2021sparsity}
\bibfield{author}{\bibinfo{person}{Torsten Hoefler}, \bibinfo{person}{Dan
  Alistarh}, \bibinfo{person}{Tal Ben-Nun}, \bibinfo{person}{Nikoli Dryden},
  {and} \bibinfo{person}{Alexandra Peste}.} \bibinfo{year}{2021}\natexlab{}.
\newblock \showarticletitle{Sparsity in deep learning: Pruning and growth for
  efficient inference and training in neural networks}.
\newblock \bibinfo{journal}{\emph{Journal of Machine Learning Research}}
  \bibinfo{volume}{22}, \bibinfo{number}{241} (\bibinfo{year}{2021}),
  \bibinfo{pages}{1--124}.
\newblock


\bibitem[Holstein et~al\mbox{.}(2019)]%
        {holstein2019improving}
\bibfield{author}{\bibinfo{person}{Kenneth Holstein}, \bibinfo{person}{Jennifer
  Wortman~Vaughan}, \bibinfo{person}{Hal Daum{\'e}~III}, \bibinfo{person}{Miro
  Dudik}, {and} \bibinfo{person}{Hanna Wallach}.}
  \bibinfo{year}{2019}\natexlab{}.
\newblock \showarticletitle{Improving fairness in machine learning systems:
  What do industry practitioners need?}. In
  \bibinfo{booktitle}{\emph{Proceedings of the 2019 CHI Conference on Human
  Factors in Computing Systems}}. \bibinfo{pages}{1--16}.
\newblock
\urldef\tempurl%
\url{https://doi.org/10.1145/3290605.3300830}
\showDOI{\tempurl}


\bibitem[Hong et~al\mbox{.}(2020)]%
        {hong2020human}
\bibfield{author}{\bibinfo{person}{Sungsoo~Ray Hong}, \bibinfo{person}{Jessica
  Hullman}, {and} \bibinfo{person}{Enrico Bertini}.}
  \bibinfo{year}{2020}\natexlab{}.
\newblock \showarticletitle{Human factors in model interpretability: Industry
  practices, challenges, and needs}.
\newblock \bibinfo{journal}{\emph{Proceedings of the ACM on Human-Computer
  Interaction}} \bibinfo{volume}{4}, \bibinfo{number}{CSCW1}
  (\bibinfo{year}{2020}), \bibinfo{pages}{1--26}.
\newblock


\bibitem[Hooker et~al\mbox{.}(2019)]%
        {hooker2019compressed}
\bibfield{author}{\bibinfo{person}{Sara Hooker}, \bibinfo{person}{Aaron
  Courville}, \bibinfo{person}{Gregory Clark}, \bibinfo{person}{Yann Dauphin},
  {and} \bibinfo{person}{Andrea Frome}.} \bibinfo{year}{2019}\natexlab{}.
\newblock \showarticletitle{What do compressed deep neural networks forget?}
\newblock \bibinfo{journal}{\emph{arXiv preprint arXiv:1911.05248}}
  (\bibinfo{year}{2019}).
\newblock


\bibitem[Hooker et~al\mbox{.}(2020)]%
        {hooker2020characterising}
\bibfield{author}{\bibinfo{person}{Sara Hooker}, \bibinfo{person}{Nyalleng
  Moorosi}, \bibinfo{person}{Gregory Clark}, \bibinfo{person}{Samy Bengio},
  {and} \bibinfo{person}{Emily Denton}.} \bibinfo{year}{2020}\natexlab{}.
\newblock \showarticletitle{Characterising bias in compressed models}.
\newblock \bibinfo{journal}{\emph{arXiv}} (\bibinfo{year}{2020}).
\newblock
\showeprint{2010.03058}


\bibitem[Hopkins and Booth(2021)]%
        {hopkins2021machine}
\bibfield{author}{\bibinfo{person}{Aspen Hopkins} {and} \bibinfo{person}{Serena
  Booth}.} \bibinfo{year}{2021}\natexlab{}.
\newblock \showarticletitle{Machine learning practices outside big tech: How
  resource constraints challenge responsible development}. In
  \bibinfo{booktitle}{\emph{Proceedings of the 2021 AAAI/ACM Conference on AI,
  Ethics, and Society}}. \bibinfo{pages}{134--145}.
\newblock


\bibitem[Howard et~al\mbox{.}(2017)]%
        {howard2017mobilenets}
\bibfield{author}{\bibinfo{person}{Andrew~G Howard}, \bibinfo{person}{Menglong
  Zhu}, \bibinfo{person}{Bo Chen}, \bibinfo{person}{Dmitry Kalenichenko},
  \bibinfo{person}{Weijun Wang}, \bibinfo{person}{Tobias Weyand},
  \bibinfo{person}{Marco Andreetto}, {and} \bibinfo{person}{Hartwig Adam}.}
  \bibinfo{year}{2017}\natexlab{}.
\newblock \showarticletitle{Mobilenets: Efficient convolutional neural networks
  for mobile vision applications}.
\newblock \bibinfo{journal}{\emph{arXiv}}  \bibinfo{volume}{abs/1704.04861}
  (\bibinfo{year}{2017}).
\newblock
\showeprint[arXiv]{1704.04861}


\bibitem[Hu et~al\mbox{.}(2022)]%
        {hu2021lora}
\bibfield{author}{\bibinfo{person}{Edward~J Hu}, \bibinfo{person}{Yelong Shen},
  \bibinfo{person}{Phillip Wallis}, \bibinfo{person}{Zeyuan Allen-Zhu},
  \bibinfo{person}{Yuanzhi Li}, \bibinfo{person}{Shean Wang},
  \bibinfo{person}{Lu Wang}, {and} \bibinfo{person}{Weizhu Chen}.}
  \bibinfo{year}{2022}\natexlab{}.
\newblock \showarticletitle{Lora: Low-rank adaptation of large language
  models}.
\newblock \bibinfo{journal}{\emph{International Conference on Learning
  Representations}} (\bibinfo{year}{2022}).
\newblock


\bibitem[Hu et~al\mbox{.}(2020)]%
        {hu2020personalized}
\bibfield{author}{\bibinfo{person}{Rui Hu}, \bibinfo{person}{Yuanxiong Guo},
  \bibinfo{person}{Hongning Li}, \bibinfo{person}{Qingqi Pei}, {and}
  \bibinfo{person}{Yanmin Gong}.} \bibinfo{year}{2020}\natexlab{}.
\newblock \showarticletitle{Personalized federated learning with differential
  privacy}.
\newblock \bibinfo{journal}{\emph{IEEE Internet of Things Journal}}
  \bibinfo{volume}{7}, \bibinfo{number}{10} (\bibinfo{year}{2020}),
  \bibinfo{pages}{9530--9539}.
\newblock


\bibitem[Intel(2020)]%
        {intel2020nc}
\bibfield{author}{\bibinfo{person}{Intel}.} \bibinfo{year}{2020}\natexlab{}.
\newblock \bibinfo{booktitle}{\emph{Neural Compressor}}.
\newblock
\urldef\tempurl%
\url{https://github.com/intel/neural-compressor}
\showURL{%
\tempurl}


\bibitem[Jiang et~al\mbox{.}(2021)]%
        {jiang2021lightweight}
\bibfield{author}{\bibinfo{person}{Linshan Jiang}, \bibinfo{person}{Rui Tan},
  \bibinfo{person}{Xin Lou}, {and} \bibinfo{person}{Guosheng Lin}.}
  \bibinfo{year}{2021}\natexlab{}.
\newblock \showarticletitle{On lightweight privacy-preserving collaborative
  learning for Internet of Things by independent random projections}.
\newblock \bibinfo{journal}{\emph{ACM Transactions on Internet of Things}}
  \bibinfo{volume}{2}, \bibinfo{number}{2} (\bibinfo{year}{2021}),
  \bibinfo{pages}{1--32}.
\newblock


\bibitem[Kahng et~al\mbox{.}(2018)]%
        {kahng2018ganlab}
\bibfield{author}{\bibinfo{person}{Minsuk Kahng}, \bibinfo{person}{Nikhil
  Thorat}, \bibinfo{person}{Duen Horng~Polo Chau}, \bibinfo{person}{Fernanda~B
  Viégas}, {and} \bibinfo{person}{Martin Wattenberg}.}
  \bibinfo{year}{2018}\natexlab{}.
\newblock \showarticletitle{Gan lab: Understanding complex deep generative
  models using interactive visual experimentation}.
\newblock \bibinfo{journal}{\emph{IEEE Transactions on Visualization and
  Computer Graphics}} \bibinfo{volume}{25}, \bibinfo{number}{1}
  (\bibinfo{year}{2018}), \bibinfo{pages}{1--11}.
\newblock
\urldef\tempurl%
\url{https://poloclub.github.io/ganlab/}
\showURL{%
\tempurl}


\bibitem[Knott et~al\mbox{.}(2022)]%
        {knott2022interviews}
\bibfield{author}{\bibinfo{person}{Eleanor Knott}, \bibinfo{person}{Aliya~Hamid
  Rao}, \bibinfo{person}{Kate Summers}, {and} \bibinfo{person}{Chana Teeger}.}
  \bibinfo{year}{2022}\natexlab{}.
\newblock \showarticletitle{Interviews in the social sciences}.
\newblock \bibinfo{journal}{\emph{Nature Reviews Methods Primers}}
  \bibinfo{volume}{2}, \bibinfo{number}{1} (\bibinfo{year}{2022}),
  \bibinfo{pages}{1--15}.
\newblock


\bibitem[Kone{\v{c}}n{\`y} et~al\mbox{.}(2016)]%
        {konevcny2016federated}
\bibfield{author}{\bibinfo{person}{Jakub Kone{\v{c}}n{\`y}},
  \bibinfo{person}{H~Brendan McMahan}, \bibinfo{person}{Felix~X Yu},
  \bibinfo{person}{Peter Richt{\'a}rik}, \bibinfo{person}{Ananda~Theertha
  Suresh}, {and} \bibinfo{person}{Dave Bacon}.}
  \bibinfo{year}{2016}\natexlab{}.
\newblock \showarticletitle{Federated learning: Strategies for improving
  communication efficiency}.
\newblock \bibinfo{journal}{\emph{arXiv preprint arXiv:1610.05492}}
  (\bibinfo{year}{2016}).
\newblock


\bibitem[Kusupati et~al\mbox{.}(2020)]%
        {kusupati2020soft}
\bibfield{author}{\bibinfo{person}{Aditya Kusupati}, \bibinfo{person}{Vivek
  Ramanujan}, \bibinfo{person}{Raghav Somani}, \bibinfo{person}{Mitchell
  Wortsman}, \bibinfo{person}{Prateek Jain}, \bibinfo{person}{Sham Kakade},
  {and} \bibinfo{person}{Ali Farhadi}.} \bibinfo{year}{2020}\natexlab{}.
\newblock \showarticletitle{Soft threshold weight reparameterization for
  learnable sparsity}. In \bibinfo{booktitle}{\emph{International Conference on
  Machine Learning}}. PMLR, \bibinfo{pages}{5544--5555}.
\newblock


\bibitem[Kvale(2006)]%
        {kvale2006dominance}
\bibfield{author}{\bibinfo{person}{Steinar Kvale}.}
  \bibinfo{year}{2006}\natexlab{}.
\newblock \showarticletitle{Dominance through interviews and dialogues}.
\newblock \bibinfo{journal}{\emph{Qualitative Inquiry}} \bibinfo{volume}{12},
  \bibinfo{number}{3} (\bibinfo{year}{2006}), \bibinfo{pages}{480--500}.
\newblock


\bibitem[Lee et~al\mbox{.}(2023)]%
        {lee2023diffusion}
\bibfield{author}{\bibinfo{person}{Seongmin Lee}, \bibinfo{person}{Benjamin
  Hoover}, \bibinfo{person}{Hendrik Strobelt}, \bibinfo{person}{Zijie~J Wang},
  \bibinfo{person}{ShengYun Peng}, \bibinfo{person}{Austin Wright},
  \bibinfo{person}{Kevin Li}, \bibinfo{person}{Haekyu Park},
  \bibinfo{person}{Haoyang Yang}, {and} \bibinfo{person}{Duen~Horng Chau}.}
  \bibinfo{year}{2023}\natexlab{}.
\newblock \showarticletitle{Diffusion explainer: Visual explanation for
  text-to-image stable diffusion}.
\newblock \bibinfo{journal}{\emph{arXiv preprint arXiv:2305.03509}}
  (\bibinfo{year}{2023}).
\newblock


\bibitem[Li et~al\mbox{.}(2023)]%
        {li2023hierarchical}
\bibfield{author}{\bibinfo{person}{Youpeng Li}, \bibinfo{person}{Xuyu Wang},
  {and} \bibinfo{person}{Lingling An}.} \bibinfo{year}{2023}\natexlab{}.
\newblock \showarticletitle{Hierarchical clustering-based personalized
  federated learning for robust and fair human activity recognition}.
\newblock \bibinfo{journal}{\emph{Proceedings of the ACM on Interactive,
  Mobile, Wearable and Ubiquitous Technologies}} \bibinfo{volume}{7},
  \bibinfo{number}{1} (\bibinfo{year}{2023}), \bibinfo{pages}{1--38}.
\newblock


\bibitem[Lialin et~al\mbox{.}(2023)]%
        {lialin2023scaling}
\bibfield{author}{\bibinfo{person}{Vladislav Lialin}, \bibinfo{person}{Vijeta
  Deshpande}, {and} \bibinfo{person}{Anna Rumshisky}.}
  \bibinfo{year}{2023}\natexlab{}.
\newblock \showarticletitle{Scaling down to scale up: A guide to
  parameter-efficient fine-tuning}.
\newblock \bibinfo{journal}{\emph{arXiv preprint arXiv:2303.15647}}
  (\bibinfo{year}{2023}).
\newblock


\bibitem[Liao et~al\mbox{.}(2020)]%
        {liao2020questioning}
\bibfield{author}{\bibinfo{person}{Q~Vera Liao}, \bibinfo{person}{Daniel
  Gruen}, {and} \bibinfo{person}{Sarah Miller}.}
  \bibinfo{year}{2020}\natexlab{}.
\newblock \showarticletitle{Questioning the AI: Informing design practices for
  explainable AI user experiences}. In \bibinfo{booktitle}{\emph{Proceedings of
  the 2020 CHI Conference on Human Factors in Computing Systems}}.
  \bibinfo{pages}{1--15}.
\newblock
\urldef\tempurl%
\url{https://doi.org/10.1145/3313831.3376590}
\showDOI{\tempurl}


\bibitem[Liaqat et~al\mbox{.}(2019)]%
        {liaqat2019wearbreathing}
\bibfield{author}{\bibinfo{person}{Daniyal Liaqat}, \bibinfo{person}{Mohamed
  Abdalla}, \bibinfo{person}{Pegah Abed-Esfahani}, \bibinfo{person}{Moshe
  Gabel}, \bibinfo{person}{Tatiana Son}, \bibinfo{person}{Robert Wu},
  \bibinfo{person}{Andrea Gershon}, \bibinfo{person}{Frank Rudzicz}, {and}
  \bibinfo{person}{Eyal~De Lara}.} \bibinfo{year}{2019}\natexlab{}.
\newblock \showarticletitle{WearBreathing: Real world respiratory rate
  monitoring using smartwatches}.
\newblock \bibinfo{journal}{\emph{Proceedings of the ACM on Interactive,
  Mobile, Wearable and Ubiquitous Technologies}} \bibinfo{volume}{3},
  \bibinfo{number}{2} (\bibinfo{year}{2019}), \bibinfo{pages}{1--22}.
\newblock


\bibitem[Liberis and Lane(2023)]%
        {liberis2023differentiable}
\bibfield{author}{\bibinfo{person}{Edgar Liberis} {and}
  \bibinfo{person}{Nicholas~D Lane}.} \bibinfo{year}{2023}\natexlab{}.
\newblock \showarticletitle{Differentiable neural network pruning to enable
  smart applications on microcontrollers}.
\newblock \bibinfo{journal}{\emph{Proceedings of the ACM on Interactive,
  Mobile, Wearable and Ubiquitous Technologies}} \bibinfo{volume}{6},
  \bibinfo{number}{4} (\bibinfo{year}{2023}), \bibinfo{pages}{1--19}.
\newblock


\bibitem[Liebenwein et~al\mbox{.}(2021)]%
        {liebenwein2021lost}
\bibfield{author}{\bibinfo{person}{Lucas Liebenwein}, \bibinfo{person}{Cenk
  Baykal}, \bibinfo{person}{Brandon Carter}, \bibinfo{person}{David Gifford},
  {and} \bibinfo{person}{Daniela Rus}.} \bibinfo{year}{2021}\natexlab{}.
\newblock \showarticletitle{Lost in pruning: The effects of pruning neural
  networks beyond test accuracy}.
\newblock \bibinfo{journal}{\emph{Proceedings of Machine Learning and Systems}}
   \bibinfo{volume}{3} (\bibinfo{year}{2021}), \bibinfo{pages}{93--138}.
\newblock


\bibitem[Lim et~al\mbox{.}(2020)]%
        {lim2020federated}
\bibfield{author}{\bibinfo{person}{Wei Yang~Bryan Lim},
  \bibinfo{person}{Nguyen~Cong Luong}, \bibinfo{person}{Dinh~Thai Hoang},
  \bibinfo{person}{Yutao Jiao}, \bibinfo{person}{Ying-Chang Liang},
  \bibinfo{person}{Qiang Yang}, \bibinfo{person}{Dusit Niyato}, {and}
  \bibinfo{person}{Chunyan Miao}.} \bibinfo{year}{2020}\natexlab{}.
\newblock \showarticletitle{Federated learning in mobile edge networks: A
  comprehensive survey}.
\newblock \bibinfo{journal}{\emph{IEEE Communications Surveys \& Tutorials}}
  \bibinfo{volume}{22}, \bibinfo{number}{3} (\bibinfo{year}{2020}),
  \bibinfo{pages}{2031--2063}.
\newblock
\urldef\tempurl%
\url{https://doi.org/10.1109/comst.2020.2986024}
\showDOI{\tempurl}


\bibitem[Liu et~al\mbox{.}(2021)]%
        {liu2021adaspring}
\bibfield{author}{\bibinfo{person}{Sicong Liu}, \bibinfo{person}{Bin Guo},
  \bibinfo{person}{Ke Ma}, \bibinfo{person}{Zhiwen Yu}, {and}
  \bibinfo{person}{Junzhao Du}.} \bibinfo{year}{2021}\natexlab{}.
\newblock \showarticletitle{AdaSpring: Context-adaptive and
  runtime-evolutionary deep model compression for mobile applications}.
\newblock \bibinfo{journal}{\emph{Proceedings of the ACM on Interactive,
  Mobile, Wearable and Ubiquitous Technologies}} \bibinfo{volume}{5},
  \bibinfo{number}{1} (\bibinfo{year}{2021}), \bibinfo{pages}{1--22}.
\newblock


\bibitem[Madaio et~al\mbox{.}(2022)]%
        {madaio2022assessing}
\bibfield{author}{\bibinfo{person}{Michael Madaio}, \bibinfo{person}{Lisa
  Egede}, \bibinfo{person}{Hariharan Subramonyam}, \bibinfo{person}{Jennifer
  Wortman~Vaughan}, {and} \bibinfo{person}{Hanna Wallach}.}
  \bibinfo{year}{2022}\natexlab{}.
\newblock \showarticletitle{Assessing the fairness of AI systems: AI
  practitioners' processes, challenges, and needs for support}.
\newblock \bibinfo{journal}{\emph{Proceedings of the ACM on Human-Computer
  Interaction}} \bibinfo{volume}{6}, \bibinfo{number}{CSCW1}
  (\bibinfo{year}{2022}), \bibinfo{pages}{1--26}.
\newblock


\bibitem[Madaio et~al\mbox{.}(2020)]%
        {madaio2020co}
\bibfield{author}{\bibinfo{person}{Michael~A Madaio}, \bibinfo{person}{Luke
  Stark}, \bibinfo{person}{Jennifer Wortman~Vaughan}, {and}
  \bibinfo{person}{Hanna Wallach}.} \bibinfo{year}{2020}\natexlab{}.
\newblock \showarticletitle{Co-designing checklists to understand
  organizational challenges and opportunities around fairness in AI}. In
  \bibinfo{booktitle}{\emph{Proceedings of the 2020 CHI Conference on Human
  Factors in Computing Systems}}. \bibinfo{pages}{1--14}.
\newblock


\bibitem[Menghani(2023)]%
        {menghani2023efficient}
\bibfield{author}{\bibinfo{person}{Gaurav Menghani}.}
  \bibinfo{year}{2023}\natexlab{}.
\newblock \showarticletitle{Efficient deep learning: A survey on making deep
  learning models smaller, faster, and better}.
\newblock \bibinfo{journal}{\emph{Comput. Surveys}} \bibinfo{volume}{55},
  \bibinfo{number}{12} (\bibinfo{year}{2023}), \bibinfo{pages}{1--37}.
\newblock


\bibitem[Microsoft(2021)]%
        {ms2021nni}
\bibfield{author}{\bibinfo{person}{Microsoft}.}
  \bibinfo{year}{2021}\natexlab{}.
\newblock \bibinfo{booktitle}{\emph{Neural network intelligence}}.
\newblock
\urldef\tempurl%
\url{https://github.com/microsoft/nni}
\showURL{%
\tempurl}


\bibitem[Mishra and Gupta(2023)]%
        {mishra2023designing}
\bibfield{author}{\bibinfo{person}{Rahul Mishra} {and}
  \bibinfo{person}{Hari~Prabhat Gupta}.} \bibinfo{year}{2023}\natexlab{}.
\newblock \showarticletitle{Designing and training of lightweight neural
  networks on edge devices using early halting in knowledge distillation}.
\newblock \bibinfo{journal}{\emph{IEEE Transactions on Mobile Computing}}
  (\bibinfo{year}{2023}).
\newblock


\bibitem[Mishra et~al\mbox{.}(2020)]%
        {mishra2020teacher}
\bibfield{author}{\bibinfo{person}{Rahul Mishra}, \bibinfo{person}{Hari~Prabhat
  Gupta}, {and} \bibinfo{person}{Tanima Dutta}.}
  \bibinfo{year}{2020}\natexlab{}.
\newblock \showarticletitle{Teacher, trainee, and student based knowledge
  distillation technique for monitoring indoor activities}. In
  \bibinfo{booktitle}{\emph{Proceedings of the 18th Conference on Embedded
  Networked Sensor Systems}}. \bibinfo{pages}{729--730}.
\newblock


\bibitem[Modarres et~al\mbox{.}(2018)]%
        {modarres2018towards}
\bibfield{author}{\bibinfo{person}{Ceena Modarres}, \bibinfo{person}{Mark
  Ibrahim}, \bibinfo{person}{Melissa Louie}, {and} \bibinfo{person}{John
  Paisley}.} \bibinfo{year}{2018}\natexlab{}.
\newblock \showarticletitle{Towards explainable deep learning for credit
  lending: A case study}.
\newblock \bibinfo{journal}{\emph{arXiv}} (\bibinfo{year}{2018}).
\newblock


\bibitem[Monserrate(2022)]%
        {monserrate2022cloud}
\bibfield{author}{\bibinfo{person}{Steven~Gonzalez Monserrate}.}
  \bibinfo{year}{2022}\natexlab{}.
\newblock \showarticletitle{The cloud is material: On the environmental impacts
  of computation and data storage}.
\newblock \bibinfo{journal}{\emph{MIT Case Studies in Social and Ethical
  Responsibilities of Computing}} \bibinfo{number}{Winter 2022}
  (\bibinfo{date}{Jan} \bibinfo{year}{2022}).
\newblock
\urldef\tempurl%
\url{https://doi.org/10.21428/2c646de5.031d4553}
\showDOI{\tempurl}


\bibitem[Movva et~al\mbox{.}(2023)]%
        {movva2023large}
\bibfield{author}{\bibinfo{person}{Rajiv Movva}, \bibinfo{person}{Sidhika
  Balachandar}, \bibinfo{person}{Kenny Peng}, \bibinfo{person}{Gabriel
  Agostini}, \bibinfo{person}{Nikhil Garg}, {and} \bibinfo{person}{Emma
  Pierson}.} \bibinfo{year}{2023}\natexlab{}.
\newblock \showarticletitle{Large language models shape and are shaped by
  society: A survey of arXiv publication patterns}.
\newblock \bibinfo{journal}{\emph{arXiv preprint arXiv:2307.10700}}
  (\bibinfo{year}{2023}).
\newblock


\bibitem[Murshed et~al\mbox{.}(2021)]%
        {murshed2021machine}
\bibfield{author}{\bibinfo{person}{MG~Sarwar Murshed},
  \bibinfo{person}{Christopher Murphy}, \bibinfo{person}{Daqing Hou},
  \bibinfo{person}{Nazar Khan}, \bibinfo{person}{Ganesh Ananthanarayanan},
  {and} \bibinfo{person}{Faraz Hussain}.} \bibinfo{year}{2021}\natexlab{}.
\newblock \showarticletitle{Machine learning at the network edge: A survey}.
\newblock \bibinfo{journal}{\emph{Comput. Surveys}} \bibinfo{volume}{54},
  \bibinfo{number}{8} (\bibinfo{year}{2021}), \bibinfo{pages}{1--37}.
\newblock
\urldef\tempurl%
\url{https://doi.org/10.1145/3469029}
\showDOI{\tempurl}


\bibitem[NVIDIA(2023)]%
        {nvidia2023docs}
\bibfield{author}{\bibinfo{person}{NVIDIA}.} \bibinfo{year}{2023}\natexlab{}.
\newblock \bibinfo{title}{NVIDIA deep learning TensorRT documentation -
  optimizing TensorRT performance}.
\newblock
\newblock
\urldef\tempurl%
\url{https://docs.nvidia.com/deeplearning/tensorrt/developer-guide/index.html}
\showURL{%
\tempurl}


\bibitem[Ogueji et~al\mbox{.}(2022)]%
        {ogueji2022intriguing}
\bibfield{author}{\bibinfo{person}{Kelechi Ogueji},
  \bibinfo{person}{Orevaoghene Ahia}, \bibinfo{person}{Gbemileke Onilude},
  \bibinfo{person}{Sebastian Gehrmann}, \bibinfo{person}{Sara Hooker}, {and}
  \bibinfo{person}{Julia Kreutzer}.} \bibinfo{year}{2022}\natexlab{}.
\newblock \showarticletitle{Intriguing properties of compression on
  multilingual models}.
\newblock \bibinfo{journal}{\emph{arXiv preprint arXiv:2211.02738}}
  (\bibinfo{year}{2022}).
\newblock


\bibitem[Pascanu et~al\mbox{.}(2013)]%
        {pascanu2013difficulty}
\bibfield{author}{\bibinfo{person}{Razvan Pascanu}, \bibinfo{person}{Tomas
  Mikolov}, {and} \bibinfo{person}{Yoshua Bengio}.}
  \bibinfo{year}{2013}\natexlab{}.
\newblock \showarticletitle{On the difficulty of training recurrent neural
  networks}. In \bibinfo{booktitle}{\emph{International Conference on Machine
  Learning}}. PMLR, \bibinfo{pages}{1310--1318}.
\newblock


\bibitem[Passi and Jackson(2018)]%
        {passi2018trust}
\bibfield{author}{\bibinfo{person}{Samir Passi} {and} \bibinfo{person}{Steven~J
  Jackson}.} \bibinfo{year}{2018}\natexlab{}.
\newblock \showarticletitle{Trust in data science: Collaboration, translation,
  and accountability in corporate data science projects}.
\newblock \bibinfo{journal}{\emph{Proceedings of the ACM on Human-Computer
  Interaction}} \bibinfo{volume}{2}, \bibinfo{number}{CSCW}
  (\bibinfo{year}{2018}), \bibinfo{pages}{1--28}.
\newblock


\bibitem[Patel et~al\mbox{.}(2008)]%
        {patel2008investigating}
\bibfield{author}{\bibinfo{person}{Kayur Patel}, \bibinfo{person}{James
  Fogarty}, \bibinfo{person}{James~A Landay}, {and} \bibinfo{person}{Beverly
  Harrison}.} \bibinfo{year}{2008}\natexlab{}.
\newblock \showarticletitle{Investigating statistical machine learning as a
  tool for software development}. In \bibinfo{booktitle}{\emph{Proceedings of
  the SIGCHI Conference on Human Factors in Computing Systems}}.
  \bibinfo{pages}{667--676}.
\newblock
\urldef\tempurl%
\url{https://doi.org/10.1145/1357054.1357160}
\showDOI{\tempurl}


\bibitem[Piorkowski et~al\mbox{.}(2021)]%
        {piorkowski2021ai}
\bibfield{author}{\bibinfo{person}{David Piorkowski}, \bibinfo{person}{Soya
  Park}, \bibinfo{person}{April~Yi Wang}, \bibinfo{person}{Dakuo Wang},
  \bibinfo{person}{Michael Muller}, {and} \bibinfo{person}{Felix Portnoy}.}
  \bibinfo{year}{2021}\natexlab{}.
\newblock \showarticletitle{How ai developers overcome communication challenges
  in a multidisciplinary team: A case study}.
\newblock \bibinfo{journal}{\emph{Proceedings of the ACM on Human-Computer
  Interaction}} \bibinfo{volume}{5}, \bibinfo{number}{CSCW1}
  (\bibinfo{year}{2021}), \bibinfo{pages}{1--25}.
\newblock


\bibitem[Polino et~al\mbox{.}(2018)]%
        {polino2018model}
\bibfield{author}{\bibinfo{person}{Antonio Polino}, \bibinfo{person}{Razvan
  Pascanu}, {and} \bibinfo{person}{Dan Alistarh}.}
  \bibinfo{year}{2018}\natexlab{}.
\newblock \showarticletitle{Model compression via distillation and
  quantization}.
\newblock \bibinfo{journal}{\emph{arXiv}} (\bibinfo{year}{2018}).
\newblock
\showeprint[arXiv]{1802.05668}


\bibitem[PyTorch(2018)]%
        {pytorch2023quantization}
\bibfield{author}{\bibinfo{person}{PyTorch}.} \bibinfo{year}{2018}\natexlab{}.
\newblock \bibinfo{booktitle}{\emph{Quantization}}.
\newblock
\urldef\tempurl%
\url{https://pytorch.org/docs/stable/quantization.html}
\showURL{%
\tempurl}


\bibitem[PyTorch(2019)]%
        {pytorch2023sparsity}
\bibfield{author}{\bibinfo{person}{PyTorch}.} \bibinfo{year}{2019}\natexlab{}.
\newblock \bibinfo{booktitle}{\emph{Sparisty}}.
\newblock
\urldef\tempurl%
\url{https://pytorch.org/docs/stable/sparse.html}
\showURL{%
\tempurl}


\bibitem[PyTorch(2023)]%
        {pytorch2023examples}
\bibfield{author}{\bibinfo{person}{PyTorch}.} \bibinfo{year}{2023}\natexlab{}.
\newblock \bibinfo{booktitle}{\emph{PyTorch Examples}}.
\newblock
\urldef\tempurl%
\url{https://pytorch.org/tutorials/}
\showURL{%
\tempurl}


\bibitem[Rakova et~al\mbox{.}(2021)]%
        {rakova2021responsible}
\bibfield{author}{\bibinfo{person}{Bogdana Rakova}, \bibinfo{person}{Jingying
  Yang}, \bibinfo{person}{Henriette Cramer}, {and} \bibinfo{person}{Rumman
  Chowdhury}.} \bibinfo{year}{2021}\natexlab{}.
\newblock \showarticletitle{Where responsible AI meets reality: Practitioner
  perspectives on enablers for shifting organizational practices}.
\newblock \bibinfo{journal}{\emph{Proceedings of the ACM on Human-Computer
  Interaction}} \bibinfo{volume}{5}, \bibinfo{number}{CSCW1}
  (\bibinfo{year}{2021}), \bibinfo{pages}{1--23}.
\newblock


\bibitem[Robertson et~al\mbox{.}(2021)]%
        {robertson2021modeling}
\bibfield{author}{\bibinfo{person}{Samantha Robertson}, \bibinfo{person}{Tonya
  Nguyen}, {and} \bibinfo{person}{Niloufar Salehi}.}
  \bibinfo{year}{2021}\natexlab{}.
\newblock \showarticletitle{Modeling assumptions clash with the real world:
  Transparency, equity, and community challenges for student assignment
  algorithms}. In \bibinfo{booktitle}{\emph{Proceedings of the 2021 CHI
  Conference on Human Factors in Computing Systems}}. \bibinfo{pages}{1--14}.
\newblock


\bibitem[Sambasivan et~al\mbox{.}(2021)]%
        {sambasivan2021everyone}
\bibfield{author}{\bibinfo{person}{Nithya Sambasivan}, \bibinfo{person}{Shivani
  Kapania}, \bibinfo{person}{Hannah Highfill}, \bibinfo{person}{Diana Akrong},
  \bibinfo{person}{Praveen Paritosh}, {and} \bibinfo{person}{Lora~M Aroyo}.}
  \bibinfo{year}{2021}\natexlab{}.
\newblock \showarticletitle{``Everyone wants to do the model work, not the data
  work'': Data cascades in high-stakes AI}. In
  \bibinfo{booktitle}{\emph{Proceedings of the 2021 CHI Conference on Human
  Factors in Computing Systems}}. \bibinfo{pages}{1--15}.
\newblock
\urldef\tempurl%
\url{https://doi.org/10.1145/3411764.3445518}
\showDOI{\tempurl}


\bibitem[Sandler et~al\mbox{.}(2018)]%
        {sandler2018mobilenetv2}
\bibfield{author}{\bibinfo{person}{Mark Sandler}, \bibinfo{person}{Andrew
  Howard}, \bibinfo{person}{Menglong Zhu}, \bibinfo{person}{Andrey Zhmoginov},
  {and} \bibinfo{person}{Liang-Chieh Chen}.} \bibinfo{year}{2018}\natexlab{}.
\newblock \showarticletitle{Mobilenetv2: Inverted residuals and linear
  bottlenecks}. In \bibinfo{booktitle}{\emph{Proceedings of the IEEE Conference
  on Computer Vision and pattern Recognition}}. \bibinfo{pages}{4510--4520}.
\newblock
\urldef\tempurl%
\url{https://doi.org/10.1109/cvpr.2018.00474}
\showDOI{\tempurl}


\bibitem[Schein(1990)]%
        {schein1990organizational}
\bibfield{author}{\bibinfo{person}{Edgar~H Schein}.}
  \bibinfo{year}{1990}\natexlab{}.
\newblock \bibinfo{booktitle}{\emph{Organizational culture}}.
  Vol.~\bibinfo{volume}{45}.
\newblock \bibinfo{publisher}{American Psychological Association}.
\newblock


\bibitem[Sehgal and Kehtarnavaz(2019)]%
        {sehgal2019guidelines}
\bibfield{author}{\bibinfo{person}{Abhishek Sehgal} {and}
  \bibinfo{person}{Nasser Kehtarnavaz}.} \bibinfo{year}{2019}\natexlab{}.
\newblock \showarticletitle{Guidelines and benchmarks for deployment of deep
  learning models on smartphones as real-time apps}.
\newblock \bibinfo{journal}{\emph{Machine Learning and Knowledge Extraction}}
  \bibinfo{volume}{1}, \bibinfo{number}{1} (\bibinfo{year}{2019}),
  \bibinfo{pages}{450--465}.
\newblock


\bibitem[Smilkov et~al\mbox{.}(2016)]%
        {smilkov2017direct}
\bibfield{author}{\bibinfo{person}{Daniel Smilkov}, \bibinfo{person}{Shan
  Carter}, \bibinfo{person}{D Sculley}, \bibinfo{person}{Fernanda~B Viegas},
  {and} \bibinfo{person}{Martin Wattenberg}.} \bibinfo{year}{2016}\natexlab{}.
\newblock \showarticletitle{Direct-manipulation visualization of deep
  networks}. In \bibinfo{booktitle}{\emph{ICML Workshop on Vis for Deep
  Learning}}.
\newblock


\bibitem[Solaiman et~al\mbox{.}(2023)]%
        {solaiman2023evaluating}
\bibfield{author}{\bibinfo{person}{Irene Solaiman}, \bibinfo{person}{Zeerak
  Talat}, \bibinfo{person}{William Agnew}, \bibinfo{person}{Lama Ahmad},
  \bibinfo{person}{Dylan Baker}, \bibinfo{person}{Su~Lin Blodgett},
  \bibinfo{person}{Hal Daum{\'e}~III}, \bibinfo{person}{Jesse Dodge},
  \bibinfo{person}{Ellie Evans}, \bibinfo{person}{Sara Hooker},
  {et~al\mbox{.}}} \bibinfo{year}{2023}\natexlab{}.
\newblock \showarticletitle{Evaluating the social impact of generative AI
  systems in systems and cociety}.
\newblock \bibinfo{journal}{\emph{arXiv preprint arXiv:2306.05949}}
  (\bibinfo{year}{2023}).
\newblock


\bibitem[Stanford(2023)]%
        {stanford2023ai}
\bibfield{author}{\bibinfo{person}{Stanford}.} \bibinfo{year}{2023}\natexlab{}.
\newblock \bibinfo{title}{The AI index report: Measuring trends in artificial
  intelligence}.
\newblock
\newblock
\urldef\tempurl%
\url{https://aiindex.stanford.edu/report/}
\showURL{%
\tempurl}


\bibitem[Tan and Le(2019)]%
        {tan2019efficientnet}
\bibfield{author}{\bibinfo{person}{Mingxing Tan} {and} \bibinfo{person}{Quoc
  Le}.} \bibinfo{year}{2019}\natexlab{}.
\newblock \showarticletitle{Efficientnet: Rethinking model scaling for
  convolutional neural networks}. In \bibinfo{booktitle}{\emph{International
  Conference on Machine Learning}}. PMLR, \bibinfo{pages}{6105--6114}.
\newblock
\showeprint[arXiv]{1905.11946}


\bibitem[TensorFlow(2018)]%
        {tf2018introducing}
\bibfield{author}{\bibinfo{person}{TensorFlow}.}
  \bibinfo{year}{2018}\natexlab{}.
\newblock \bibinfo{booktitle}{\emph{Introducing the Model Optimization Toolkit
  for TensorFlow}}.
\newblock
\urldef\tempurl%
\url{https://blog.tensorflow.org/2018/09/introducing-model-optimization-toolkit.html}
\showURL{%
\tempurl}


\bibitem[TensorFlow(2020)]%
        {tf2020quantization}
\bibfield{author}{\bibinfo{person}{TensorFlow}.}
  \bibinfo{year}{2020}\natexlab{}.
\newblock \bibinfo{booktitle}{\emph{Quantization aware training with TensorFlow
  Model Optimization Toolkit - performance with accuracy}}.
\newblock
\urldef\tempurl%
\url{https://blog.tensorflow.org/2020/04/quantization-aware-training-with-tensorflow-model-optimization-toolkit.html}
\showURL{%
\tempurl}


\bibitem[Thomas(2003)]%
        {thomas2003general}
\bibfield{author}{\bibinfo{person}{David~R Thomas}.}
  \bibinfo{year}{2003}\natexlab{}.
\newblock \showarticletitle{A general inductive approach for qualitative data
  analysis}.
\newblock \bibinfo{journal}{\emph{American Journal of Evaluation}}
  \bibinfo{volume}{27}, \bibinfo{number}{2} (\bibinfo{year}{2003}),
  \bibinfo{pages}{237--246}.
\newblock


\bibitem[Toma{\v{s}}ev et~al\mbox{.}(2020)]%
        {tomavsev2020ai}
\bibfield{author}{\bibinfo{person}{Nenad Toma{\v{s}}ev},
  \bibinfo{person}{Julien Cornebise}, \bibinfo{person}{Frank Hutter},
  \bibinfo{person}{Shakir Mohamed}, \bibinfo{person}{Angela Picciariello},
  \bibinfo{person}{Bec Connelly}, \bibinfo{person}{Danielle~CM Belgrave},
  \bibinfo{person}{Daphne Ezer}, \bibinfo{person}{Fanny Cachat van~der Haert},
  \bibinfo{person}{Frank Mugisha}, {et~al\mbox{.}}}
  \bibinfo{year}{2020}\natexlab{}.
\newblock \showarticletitle{AI for social good: Unlocking the opportunity for
  positive impact}.
\newblock \bibinfo{journal}{\emph{Nature Communications}} \bibinfo{volume}{11},
  \bibinfo{number}{1} (\bibinfo{year}{2020}), \bibinfo{pages}{2468}.
\newblock


\bibitem[Touvron et~al\mbox{.}(2021)]%
        {touvron2021training}
\bibfield{author}{\bibinfo{person}{Hugo Touvron}, \bibinfo{person}{Matthieu
  Cord}, \bibinfo{person}{Matthijs Douze}, \bibinfo{person}{Francisco Massa},
  \bibinfo{person}{Alexandre Sablayrolles}, {and} \bibinfo{person}{Herv{\'e}
  J{\'e}gou}.} \bibinfo{year}{2021}\natexlab{}.
\newblock \showarticletitle{Training data-efficient image transformers \&
  distillation through attention}. In \bibinfo{booktitle}{\emph{International
  Conference on Machine Learning}}. PMLR, \bibinfo{pages}{10347--10357}.
\newblock


\bibitem[Treviso et~al\mbox{.}(2023)]%
        {treviso2023efficient}
\bibfield{author}{\bibinfo{person}{Marcos Treviso}, \bibinfo{person}{Ji-Ung
  Lee}, \bibinfo{person}{Tianchu Ji}, \bibinfo{person}{Betty~van Aken},
  \bibinfo{person}{Qingqing Cao}, \bibinfo{person}{Manuel~R Ciosici},
  \bibinfo{person}{Michael Hassid}, \bibinfo{person}{Kenneth Heafield},
  \bibinfo{person}{Sara Hooker}, \bibinfo{person}{Colin Raffel},
  {et~al\mbox{.}}} \bibinfo{year}{2023}\natexlab{}.
\newblock \showarticletitle{Efficient methods for natural language processing:
  A survey}.
\newblock \bibinfo{journal}{\emph{Transactions of the Association for
  Computational Linguistics}}  \bibinfo{volume}{11} (\bibinfo{year}{2023}),
  \bibinfo{pages}{826--860}.
\newblock


\bibitem[Vasu et~al\mbox{.}(2023a)]%
        {vasu2023fastvit}
\bibfield{author}{\bibinfo{person}{Pavan Kumar~Anasosalu Vasu},
  \bibinfo{person}{James Gabriel}, \bibinfo{person}{Jeff Zhu},
  \bibinfo{person}{Oncel Tuzel}, {and} \bibinfo{person}{Anurag Ranjan}.}
  \bibinfo{year}{2023}\natexlab{a}.
\newblock \showarticletitle{FastViT: A fast hybrid vision transformer using
  structural reparameterization}.
\newblock \bibinfo{journal}{\emph{arXiv preprint arXiv:2303.14189}}
  (\bibinfo{year}{2023}).
\newblock


\bibitem[Vasu et~al\mbox{.}(2023b)]%
        {vasu2022mobileone}
\bibfield{author}{\bibinfo{person}{Pavan Kumar~Anasosalu Vasu},
  \bibinfo{person}{James Gabriel}, \bibinfo{person}{Jeff Zhu},
  \bibinfo{person}{Oncel Tuzel}, {and} \bibinfo{person}{Anurag Ranjan}.}
  \bibinfo{year}{2023}\natexlab{b}.
\newblock \showarticletitle{An improved one millisecond mobile backbone}.
\newblock  (\bibinfo{year}{2023}).
\newblock


\bibitem[Villalobos et~al\mbox{.}(2022)]%
        {villalobos2022machine}
\bibfield{author}{\bibinfo{person}{Pablo Villalobos}, \bibinfo{person}{Jaime
  Sevilla}, \bibinfo{person}{Tamay Besiroglu}, \bibinfo{person}{Lennart Heim},
  \bibinfo{person}{Anson Ho}, {and} \bibinfo{person}{Marius Hobbhahn}.}
  \bibinfo{year}{2022}\natexlab{}.
\newblock \bibinfo{title}{Machine learning model sizes and the parameter gap}.
\newblock
\newblock
\showeprint[arxiv]{2207.02852}~[cs.LG]


\bibitem[Wang et~al\mbox{.}(2021a)]%
        {wang2021autods}
\bibfield{author}{\bibinfo{person}{Dakuo Wang}, \bibinfo{person}{Josh Andres},
  \bibinfo{person}{Justin~D Weisz}, \bibinfo{person}{Erick Oduor}, {and}
  \bibinfo{person}{Casey Dugan}.} \bibinfo{year}{2021}\natexlab{a}.
\newblock \showarticletitle{Autods: Towards human-centered automation of data
  science}. In \bibinfo{booktitle}{\emph{Proceedings of the 2021 CHI Conference
  on Human Factors in Computing Systems}}. \bibinfo{pages}{1--12}.
\newblock


\bibitem[Wang et~al\mbox{.}(2021b)]%
        {wang2021much}
\bibfield{author}{\bibinfo{person}{Dakuo Wang}, \bibinfo{person}{Q~Vera Liao},
  \bibinfo{person}{Yunfeng Zhang}, \bibinfo{person}{Udayan Khurana},
  \bibinfo{person}{Horst Samulowitz}, \bibinfo{person}{Soya Park},
  \bibinfo{person}{Michael Muller}, {and} \bibinfo{person}{Lisa Amini}.}
  \bibinfo{year}{2021}\natexlab{b}.
\newblock \showarticletitle{How much automation does a data scientist want?}
\newblock \bibinfo{journal}{\emph{arXiv}} (\bibinfo{year}{2021}).
\newblock


\bibitem[Wang et~al\mbox{.}(2019)]%
        {wang2019human}
\bibfield{author}{\bibinfo{person}{Dakuo Wang}, \bibinfo{person}{Justin~D
  Weisz}, \bibinfo{person}{Michael Muller}, \bibinfo{person}{Parikshit Ram},
  \bibinfo{person}{Werner Geyer}, \bibinfo{person}{Casey Dugan},
  \bibinfo{person}{Yla Tausczik}, \bibinfo{person}{Horst Samulowitz}, {and}
  \bibinfo{person}{Alexander Gray}.} \bibinfo{year}{2019}\natexlab{}.
\newblock \showarticletitle{Human-AI collaboration in data science: Exploring
  data scientists' perceptions of automated AI}.
\newblock \bibinfo{journal}{\emph{Proceedings of the ACM on Human-Computer
  Interaction}} \bibinfo{volume}{3}, \bibinfo{number}{CSCW}
  (\bibinfo{year}{2019}), \bibinfo{pages}{1--24}.
\newblock


\bibitem[Wang et~al\mbox{.}(2023)]%
        {wang2023genie}
\bibfield{author}{\bibinfo{person}{Yanfei Wang}, \bibinfo{person}{Zhiwen Yu},
  \bibinfo{person}{Sicong Liu}, \bibinfo{person}{Zimu Zhou}, {and}
  \bibinfo{person}{Bin Guo}.} \bibinfo{year}{2023}\natexlab{}.
\newblock \showarticletitle{Genie in the model: Automatic generation of
  human-in-the-loop deep neural networks for mobile applications}.
\newblock \bibinfo{journal}{\emph{Proceedings of the ACM on Interactive,
  Mobile, Wearable and Ubiquitous Technologies}} \bibinfo{volume}{7},
  \bibinfo{number}{1} (\bibinfo{year}{2023}), \bibinfo{pages}{1--29}.
\newblock


\bibitem[Wang et~al\mbox{.}(2021c)]%
        {wang2021cnnexplainer}
\bibfield{author}{\bibinfo{person}{Zijie~J. Wang}, \bibinfo{person}{Robert
  Turko}, \bibinfo{person}{Omar Shaikh}, \bibinfo{person}{Haekyu Park},
  \bibinfo{person}{Nilaksh Das}, \bibinfo{person}{Fred Hohman},
  \bibinfo{person}{Minsuk Kahng}, {and} \bibinfo{person}{Duen Horng~(Polo)
  Chau}.} \bibinfo{year}{2021}\natexlab{c}.
\newblock \showarticletitle{CNN explainer: Learning convolutional neural
  networks with interactive visualization}. In \bibinfo{booktitle}{\emph{IEEE
  Transactions on Visualization and Computer Graphics}}.
  \bibinfo{publisher}{IEEE}.
\newblock
\urldef\tempurl%
\url{https://doi.org/10.1109/TVCG.2020.3030418}
\showDOI{\tempurl}


\bibitem[Warden and Situnayake(2019)]%
        {warden2019tinyml}
\bibfield{author}{\bibinfo{person}{Pete Warden} {and} \bibinfo{person}{Daniel
  Situnayake}.} \bibinfo{year}{2019}\natexlab{}.
\newblock \bibinfo{booktitle}{\emph{Tinyml: Machine learning with tensorflow
  lite on arduino and ultra-low-power microcontrollers}}.
\newblock \bibinfo{publisher}{O'Reilly Media}.
\newblock


\bibitem[Welsh et~al\mbox{.}(2023)]%
        {welsh2023dnikit}
\bibfield{author}{\bibinfo{person}{Megan~Maher Welsh}, \bibinfo{person}{David
  Koski}, \bibinfo{person}{Miguel Sarabia}, \bibinfo{person}{Niv Sivakumar},
  \bibinfo{person}{Ian Arawjo}, \bibinfo{person}{Aparna Joshi},
  \bibinfo{person}{Moussa Doumbouya}, \bibinfo{person}{Luca Suau,
  Xavierand~Zappella}, {and} \bibinfo{person}{Nicholas Apostoloff}.}
  \bibinfo{year}{2023}\natexlab{}.
\newblock \bibinfo{booktitle}{\emph{Data and Network Introspection Kit}}.
\newblock
\urldef\tempurl%
\url{https://github.com/apple/dnikit}
\showURL{%
\tempurl}


\bibitem[Windl et~al\mbox{.}(2022)]%
        {windl2022not}
\bibfield{author}{\bibinfo{person}{Maximiliane Windl},
  \bibinfo{person}{Sebastian~S Feger}, \bibinfo{person}{Lara Zijlstra},
  \bibinfo{person}{Albrecht Schmidt}, {and} \bibinfo{person}{Pawel~W Wozniak}.}
  \bibinfo{year}{2022}\natexlab{}.
\newblock \showarticletitle{'It is not always discovery time': Four pragmatic
  approaches in designing AI systems}. In \bibinfo{booktitle}{\emph{CHI
  Conference on Human Factors in Computing Systems}}. \bibinfo{pages}{1--12}.
\newblock


\bibitem[Wirth and Hipp(2000)]%
        {wirth2000crisp}
\bibfield{author}{\bibinfo{person}{R{\"u}diger Wirth} {and}
  \bibinfo{person}{Jochen Hipp}.} \bibinfo{year}{2000}\natexlab{}.
\newblock \showarticletitle{CRISP-DM: Towards a standard process model for data
  mining}. In \bibinfo{booktitle}{\emph{Proceedings of the 4th international
  conference on the practical applications of knowledge discovery and data
  mining}}, Vol.~\bibinfo{volume}{1}. Manchester, \bibinfo{pages}{29--39}.
\newblock


\bibitem[Wright et~al\mbox{.}(2020)]%
        {wright2020comparative}
\bibfield{author}{\bibinfo{person}{Austin~P. Wright}, \bibinfo{person}{Zijie~J.
  Wang}, \bibinfo{person}{Haekyu Park}, \bibinfo{person}{Grace Guo},
  \bibinfo{person}{Fabian Sperrle}, \bibinfo{person}{Mennatallah El-Assady},
  \bibinfo{person}{Alex Endert}, \bibinfo{person}{Daniel Keim}, {and}
  \bibinfo{person}{Duen~Horng Chau}.} \bibinfo{year}{2020}\natexlab{}.
\newblock \showarticletitle{A comparative analysis of industry human-AI
  interaction guidelines}.
\newblock \bibinfo{journal}{\emph{Workshop on Trust and Expertise in Visual
  Analytics at IEEE VIS}} (\bibinfo{year}{2020}).
\newblock


\bibitem[Wu et~al\mbox{.}(2018)]%
        {wu2018deep}
\bibfield{author}{\bibinfo{person}{Junru Wu}, \bibinfo{person}{Yue Wang},
  \bibinfo{person}{Zhenyu Wu}, \bibinfo{person}{Zhangyang Wang},
  \bibinfo{person}{Ashok Veeraraghavan}, {and} \bibinfo{person}{Yingyan Lin}.}
  \bibinfo{year}{2018}\natexlab{}.
\newblock \showarticletitle{Deep k-means: Re-training and parameter sharing
  with harder cluster assignments for compressing deep convolutions}. In
  \bibinfo{booktitle}{\emph{International Conference on Machine Learning}}.
  PMLR, \bibinfo{pages}{5363--5372}.
\newblock


\bibitem[Xu and McAuley(2023)]%
        {xu2023survey}
\bibfield{author}{\bibinfo{person}{Canwen Xu} {and} \bibinfo{person}{Julian
  McAuley}.} \bibinfo{year}{2023}\natexlab{}.
\newblock \showarticletitle{A survey on model compression and acceleration for
  pretrained language models}. In \bibinfo{booktitle}{\emph{Proceedings of the
  AAAI Conference on Artificial Intelligence}}, Vol.~\bibinfo{volume}{37}.
  \bibinfo{pages}{10566--10575}.
\newblock


\bibitem[Xu et~al\mbox{.}(2022)]%
        {xu2022enabling}
\bibfield{author}{\bibinfo{person}{Xuhai Xu}, \bibinfo{person}{Jun Gong},
  \bibinfo{person}{Carolina Brum}, \bibinfo{person}{Lilian Liang},
  \bibinfo{person}{Bongsoo Suh}, \bibinfo{person}{Shivam~Kumar Gupta},
  \bibinfo{person}{Yash Agarwal}, \bibinfo{person}{Laurence Lindsey},
  \bibinfo{person}{Runchang Kang}, \bibinfo{person}{Behrooz Shahsavari},
  {et~al\mbox{.}}} \bibinfo{year}{2022}\natexlab{}.
\newblock \showarticletitle{Enabling hand gesture customization on wrist-worn
  devices}. In \bibinfo{booktitle}{\emph{Proceedings of the 2022 CHI Conference
  on Human Factors in Computing Systems}}. \bibinfo{pages}{1--19}.
\newblock


\bibitem[Yang(2018)]%
        {yang2018machine}
\bibfield{author}{\bibinfo{person}{Qian Yang}.}
  \bibinfo{year}{2018}\natexlab{}.
\newblock \showarticletitle{Machine learning as a UX design material: How can
  we imagine beyond automation, recommenders, and reminders?}. In
  \bibinfo{booktitle}{\emph{AAAI Spring Symposium Series}}.
\newblock


\bibitem[Yang et~al\mbox{.}(2018)]%
        {yang2018investigating}
\bibfield{author}{\bibinfo{person}{Qian Yang}, \bibinfo{person}{Alex Scuito},
  \bibinfo{person}{John Zimmerman}, \bibinfo{person}{Jodi Forlizzi}, {and}
  \bibinfo{person}{Aaron Steinfeld}.} \bibinfo{year}{2018}\natexlab{}.
\newblock \showarticletitle{Investigating how experienced UX designers
  effectively work with machine learning}. In
  \bibinfo{booktitle}{\emph{Proceedings of the 2018 Designing Interactive
  Systems Conference}}. \bibinfo{pages}{585--596}.
\newblock


\bibitem[Yao et~al\mbox{.}(2021)]%
        {yao2021context}
\bibfield{author}{\bibinfo{person}{Dixi Yao}, \bibinfo{person}{Liyao Xiang},
  \bibinfo{person}{Zifan Wang}, \bibinfo{person}{Jiayu Xu},
  \bibinfo{person}{Chao Li}, {and} \bibinfo{person}{Xinbing Wang}.}
  \bibinfo{year}{2021}\natexlab{}.
\newblock \showarticletitle{Context-aware compilation of dnn training pipelines
  across edge and cloud}.
\newblock \bibinfo{journal}{\emph{Proceedings of the ACM on Interactive,
  Mobile, Wearable and Ubiquitous Technologies}} \bibinfo{volume}{5},
  \bibinfo{number}{4} (\bibinfo{year}{2021}), \bibinfo{pages}{1--27}.
\newblock


\bibitem[Yildirim et~al\mbox{.}(2022)]%
        {yildirim2022experienced}
\bibfield{author}{\bibinfo{person}{Nur Yildirim}, \bibinfo{person}{Alex Kass},
  \bibinfo{person}{Teresa Tung}, \bibinfo{person}{Connor Upton},
  \bibinfo{person}{Donnacha Costello}, \bibinfo{person}{Robert Giusti},
  \bibinfo{person}{Sinem Lacin}, \bibinfo{person}{Sara Lovic},
  \bibinfo{person}{James~M O'Neill}, \bibinfo{person}{Rudi~O'Reilly Meehan},
  {et~al\mbox{.}}} \bibinfo{year}{2022}\natexlab{}.
\newblock \showarticletitle{How experienced designers of enterprise
  applications engage AI as a design material}. In
  \bibinfo{booktitle}{\emph{CHI Conference on Human Factors in Computing
  Systems}}. \bibinfo{pages}{1--13}.
\newblock


\bibitem[Zamzam et~al\mbox{.}(2019)]%
        {zamzam2019resource}
\bibfield{author}{\bibinfo{person}{Marwa Zamzam}, \bibinfo{person}{Tallal
  Elshabrawy}, {and} \bibinfo{person}{Mohamed Ashour}.}
  \bibinfo{year}{2019}\natexlab{}.
\newblock \showarticletitle{Resource management using machine learning in
  mobile edge computing: A survey}. In \bibinfo{booktitle}{\emph{2019 Ninth
  International Conference on Intelligent Computing and Information Systems}}.
  IEEE, \bibinfo{pages}{112--117}.
\newblock
\urldef\tempurl%
\url{https://doi.org/10.1109/icicis46948.2019.9014733}
\showDOI{\tempurl}


\bibitem[Zdanowska and Taylor(2022)]%
        {zdanowska2022study}
\bibfield{author}{\bibinfo{person}{Sabah Zdanowska} {and}
  \bibinfo{person}{Alex~S Taylor}.} \bibinfo{year}{2022}\natexlab{}.
\newblock \showarticletitle{A study of UX practitioners roles in designing
  real-world, enterprise ML systems}. In \bibinfo{booktitle}{\emph{CHI
  Conference on Human Factors in Computing Systems}}. \bibinfo{pages}{1--15}.
\newblock


\bibitem[Zhang et~al\mbox{.}(2020)]%
        {zhang2020data}
\bibfield{author}{\bibinfo{person}{Amy~X Zhang}, \bibinfo{person}{Michael
  Muller}, {and} \bibinfo{person}{Dakuo Wang}.}
  \bibinfo{year}{2020}\natexlab{}.
\newblock \showarticletitle{How do data science workers collaborate? Roles,
  workflows, and tools}.
\newblock \bibinfo{journal}{\emph{Proceedings of the ACM on Human-Computer
  Interaction}} \bibinfo{volume}{4}, \bibinfo{number}{CSCW1}
  (\bibinfo{year}{2020}), \bibinfo{pages}{1--23}.
\newblock


\bibitem[Zhang et~al\mbox{.}(2018)]%
        {zhang2018shufflenet}
\bibfield{author}{\bibinfo{person}{Xiangyu Zhang}, \bibinfo{person}{Xinyu
  Zhou}, \bibinfo{person}{Mengxiao Lin}, {and} \bibinfo{person}{Jian Sun}.}
  \bibinfo{year}{2018}\natexlab{}.
\newblock \showarticletitle{Shufflenet: An extremely efficient convolutional
  neural network for mobile devices}. In \bibinfo{booktitle}{\emph{Proceedings
  of the IEEE Conference on Computer Vision and Pattern Recognition}}.
  \bibinfo{pages}{6848--6856}.
\newblock
\urldef\tempurl%
\url{https://doi.org/10.1109/cvpr.2018.00716}
\showDOI{\tempurl}


\bibitem[Zhao et~al\mbox{.}(2022)]%
        {zhao2022survey}
\bibfield{author}{\bibinfo{person}{Tianming Zhao}, \bibinfo{person}{Yucheng
  Xie}, \bibinfo{person}{Yan Wang}, \bibinfo{person}{Jerry Cheng},
  \bibinfo{person}{Xiaonan Guo}, \bibinfo{person}{Bin Hu}, {and}
  \bibinfo{person}{Yingying Chen}.} \bibinfo{year}{2022}\natexlab{}.
\newblock \showarticletitle{A survey of deep learning on mobile devices:
  Applications, optimizations, challenges, and research opportunities}.
\newblock \bibinfo{journal}{\emph{Proc. IEEE}} \bibinfo{volume}{110},
  \bibinfo{number}{3} (\bibinfo{year}{2022}), \bibinfo{pages}{334--354}.
\newblock
\urldef\tempurl%
\url{https://doi.org/10.1109/jproc.2022.3153408}
\showDOI{\tempurl}


\bibitem[Zhao et~al\mbox{.}(2023)]%
        {zhao2023survey}
\bibfield{author}{\bibinfo{person}{Wayne~Xin Zhao}, \bibinfo{person}{Kun Zhou},
  \bibinfo{person}{Junyi Li}, \bibinfo{person}{Tianyi Tang},
  \bibinfo{person}{Xiaolei Wang}, \bibinfo{person}{Yupeng Hou},
  \bibinfo{person}{Yingqian Min}, \bibinfo{person}{Beichen Zhang},
  \bibinfo{person}{Junjie Zhang}, \bibinfo{person}{Zican Dong},
  {et~al\mbox{.}}} \bibinfo{year}{2023}\natexlab{}.
\newblock \showarticletitle{A survey of large language models}.
\newblock \bibinfo{journal}{\emph{arXiv preprint arXiv:2303.18223}}
  (\bibinfo{year}{2023}).
\newblock


\bibitem[Zhao et~al\mbox{.}(2016)]%
        {zhao2016low}
\bibfield{author}{\bibinfo{person}{Yong Zhao}, \bibinfo{person}{Jinyu Li},
  {and} \bibinfo{person}{Yifan Gong}.} \bibinfo{year}{2016}\natexlab{}.
\newblock \showarticletitle{Low-rank plus diagonal adaptation for deep neural
  networks}. In \bibinfo{booktitle}{\emph{IEEE International Conference on
  Acoustics, Speech and Signal Processing}}. IEEE, \bibinfo{pages}{5005--5009}.
\newblock


\bibitem[Zhen et~al\mbox{.}(2021)]%
        {zhen2021fast}
\bibfield{author}{\bibinfo{person}{Peining Zhen}, \bibinfo{person}{Hai-Bao
  Chen}, \bibinfo{person}{Yuan Cheng}, \bibinfo{person}{Zhigang Ji},
  \bibinfo{person}{Bin Liu}, {and} \bibinfo{person}{Hao Yu}.}
  \bibinfo{year}{2021}\natexlab{}.
\newblock \showarticletitle{Fast video facial expression recognition by a
  deeply tensor-compressed LSTM neural network for mobile devices}.
\newblock \bibinfo{journal}{\emph{ACM Transactions on Internet of Things}}
  \bibinfo{volume}{2}, \bibinfo{number}{4} (\bibinfo{year}{2021}),
  \bibinfo{pages}{1--26}.
\newblock


\bibitem[Zhou et~al\mbox{.}(2023)]%
        {zhou2023comprehensive}
\bibfield{author}{\bibinfo{person}{Ce Zhou}, \bibinfo{person}{Qian Li},
  \bibinfo{person}{Chen Li}, \bibinfo{person}{Jun Yu}, \bibinfo{person}{Yixin
  Liu}, \bibinfo{person}{Guangjing Wang}, \bibinfo{person}{Kai Zhang},
  \bibinfo{person}{Cheng Ji}, \bibinfo{person}{Qiben Yan},
  \bibinfo{person}{Lifang He}, {et~al\mbox{.}}}
  \bibinfo{year}{2023}\natexlab{}.
\newblock \showarticletitle{A comprehensive survey on pretrained foundation
  models: A history from BERT to ChatGPT}.
\newblock \bibinfo{journal}{\emph{arXiv preprint arXiv:2302.09419}}
  (\bibinfo{year}{2023}).
\newblock


\bibitem[Zhou et~al\mbox{.}(2019)]%
        {zhou2019edge}
\bibfield{author}{\bibinfo{person}{Zhi Zhou}, \bibinfo{person}{Xu Chen},
  \bibinfo{person}{En Li}, \bibinfo{person}{Liekang Zeng}, \bibinfo{person}{Ke
  Luo}, {and} \bibinfo{person}{Junshan Zhang}.}
  \bibinfo{year}{2019}\natexlab{}.
\newblock \showarticletitle{Edge intelligence: Paving the last mile of
  artificial intelligence with edge computing}.
\newblock \bibinfo{journal}{\emph{Proc. IEEE}} \bibinfo{volume}{107},
  \bibinfo{number}{8} (\bibinfo{year}{2019}), \bibinfo{pages}{1738--1762}.
\newblock
\urldef\tempurl%
\url{https://doi.org/10.1109/jproc.2019.2918951}
\showDOI{\tempurl}


\bibitem[Zhu et~al\mbox{.}(2021)]%
        {zhu2021dynamic}
\bibfield{author}{\bibinfo{person}{Mingjian Zhu}, \bibinfo{person}{Kai Han},
  \bibinfo{person}{Enhua Wu}, \bibinfo{person}{Qiulin Zhang},
  \bibinfo{person}{Ying Nie}, \bibinfo{person}{Zhenzhong Lan}, {and}
  \bibinfo{person}{Yunhe Wang}.} \bibinfo{year}{2021}\natexlab{}.
\newblock \showarticletitle{Dynamic resolution network}.
\newblock \bibinfo{journal}{\emph{Advances in Neural Information Processing
  Systems}}  \bibinfo{volume}{34} (\bibinfo{year}{2021}),
  \bibinfo{pages}{27319--27330}.
\newblock
\showeprint[arXiv]{2106.02898}


\end{thebibliography}

\appendix
\section{Interview Questions}
\label{appendix:questions}

The list of questions prepared for each interview participant.

\vspace{4mm}\hrule\vspace{4mm}

\begin{itshape}
\noindent\textbf{Background ML Information}
\begin{enumerate}
    \item[Q1.] What is your team?
    \item[Q2.] What is your role?
    \item[Q3.] How many years of ML experience do you have?
    \item[Q4.] How many years of efficient ML experience do you have?
\end{enumerate}

\noindent\textbf{General Compression Questions}
\begin{enumerate}
    \item[Q5.] Describe the overall use case for model compression in your work. What are your main motivations to use model compression? Why do you care about model compression?
    \item[Q6.] Model compression details:
    \begin{enumerate}[leftmargin=2em]
        \item[Q6.1] Which model compression techniques do you use, and why? Are there trade offs or preferences?
        \item[Q6.2] Do you do compression as a final step or as a part of the training process?
        \item[Q6.3] In your experience, are there any pitfalls people applying compression should be aware of?
        \item[Q6.4] How did you figure out your budgets and how did you satisfy them?
    \end{enumerate}
    \item[Q7.] Do you compare compressed models against baseline models? If so, how, and what do you look for, \eg power and performance, other metrics, or user experience changes?
\end{enumerate}

\noindent\textbf{Compression Tooling Questions}
\begin{enumerate}
    \item[Q8.] What tools or visualizations do you use in your compression work? Please be specific, \eg specific charts, views, or metrics. How do you use these tools?
    \item[Q9.] What do you like about these tools?
    \item[Q10.] What features or future tools would help you conduct better model compression work?
\end{enumerate}

\end{itshape}

\end{document}